\pdfoutput=1
\documentclass[12pt,a4paper]{article}            
 \usepackage[skins,theorems]{tcolorbox}
\tcbset{highlight math style={enhanced,
  colframe=red,colback=white,arc=0pt,boxrule=1pt}}
  \usepackage[bookmarksopen, bookmarksnumbered, bookmarksopenlevel=2]{hyperref}
  \usepackage{tikz}
  \usepackage{tikz-3dplot}
 \usetikzlibrary{calc}
 \usetikzlibrary{decorations} %
 \usepackage[UKenglish]{babel}
 \usepackage[toc,page]{appendix}
 \usepackage{amsmath}
 \usepackage{amssymb}
 \usepackage{graphicx}
 \usepackage{hhline}
 \usepackage{arydshln}
 \usepackage[bf]{caption}
\usepackage{cite}
\usepackage[vcentermath]{youngtab}
\usepackage{geometry}
\usepackage{slashed}
\usepackage{color}
\usepackage{stackrel}
\usepackage{tikz-cd} 
\usepackage{cancel}

\usepackage{tikz}
\usetikzlibrary{shadows} 
\usepackage[framemethod=tikz]{mdframed}

\global\mdfdefinestyle{myboxstyle}{%
  shadow=true,
  linecolor=black,
  shadowcolor=black,
  shadowsize=6pt,
  nobreak=false,
  innertopmargin=10pt,
  innerbottommargin=10pt,
  leftmargin=5pt,
  rightmargin=5pt,
  needspace=1cm,
  skipabove=10pt,
  skipbelow=15pt,
  middlelinewidth=1pt,
  afterlastframe={\vspace{5pt}},
  aftersingleframe={\vspace{5pt}},
  tikzsetting={%
draw=black,
very thick} }

\newmdenv[style=myboxstyle]{whitebox} \newmdenv[style=myboxstyle,backgroundcolor=black!20]{graybox}

\newmdenv[style=myboxstyle,nobreak=true]{blockwhitebox}
\newmdenv[style=myboxstyle,backgroundcolor=black!20,nobreak=true]{blockgraybox}
\newmdenv[nobreak=true,hidealllines=true]{blockbox}

\usepackage{empheq}
\usepackage{arydshln}

\newcommand{\bqa}{\begin{eqnarray}}
\newcommand{\eqa}{\end{eqnarray}}

\newcommand{\nn}{\nonumber}

\def\del{\partial}

\hypersetup{
    pdftitle={},
    pdfauthor={},
    pdfsubject={}
}
\numberwithin{equation}{section}
\numberwithin{table}{section}

\makeatletter

\newcommand{\be}{\begin{equation}}
\newcommand{\ee}{\end{equation}}
\newcommand{\beq}{\begin{equation}}
\newcommand{\eeq}{\end{equation}}
\newcommand{\ba}{\begin{aligned}}
\newcommand{\ea}{\end{aligned}}

\newcommand{\bea}{\begin{eqnarray}}
\newcommand{\eea}{\end{eqnarray}}

\newcommand{\cO}{\mathcal{O}}
\newcommand{\cT}{\mathcal{T}}

\newcommand{\cK}{\mathcal{K}}

\newcommand{\cW}{\mathcal{W}}

\newcommand{\cI}{\mathcal{I}}

\newcommand{\cV}{\mathcal{V}}

\newcommand{\R}{\text{Re}}

\newcommand\bi{\begin{itemize}}
\newcommand\ei{\end{itemize}}

\renewcommand{\a}{{\alpha}}

\newcommand{\g}{{\gamma}}
\newcommand{\m}{{\mu}}
\newcommand{\n}{{\nu}}
\renewcommand{\r}{{\rho}}
\renewcommand{\k}{{\kappa}}
\renewcommand{\l}{{\lambda}}

\def\Re{\mathop{\mathrm{Re}}\nolimits}

\def\unit{{1\kern-.65ex {\rm l}}}
\def\1{{1\kern-.65ex {\rm l}}}

\def\bbP{\mathbb{P}}

\def\bbF{{\mathbb{F}}}

\def\bbP{{\mathbb{P}}}

\def\bbZ{{\mathbb{Z}}}

\newcount\hour \newcount\minute
\hour=\time \divide \hour by 60
\minute=\time
\def\now{%
\ifnum \hour<13
  \ifnum \hour=0 \advance \hour by 12 \number\hour:\else \number\hour:\fi%
     \ifnum \minute<10 0\fi%
     \number\minute%
\ A.M.%
\else \advance \hour by -12 \number\hour:%
  \ifnum \minute<10 0\fi%
  \number\minute%
  \ P.M.%
\fi%
}

\def\fnote#1#2{\begingroup\def\thefootnote{#1}\footnote{#2}
     \addtocounter{footnote}{-1}\endgroup}

\makeatother

\begin{document}


\begin{flushright}
{\tt\normalsize CTPU-PTC-20-24}\\
{\tt\normalsize IFT-UAM/CSIC-20-148}\\
{\tt\normalsize MITP/20-064}\\
{\tt\normalsize ZMP-HH/20-21}
\end{flushright}

\vskip 15 pt
\begin{center}
{\large \bf
Quantum Corrections in 4d N=1 Infinite Distance Limits \\ \vspace{2mm} and the Weak Gravity Conjecture
} 

\vskip 7 mm

Daniel Klaewer${}^{1}$, Seung-Joo Lee${}^{2}$, 
 Timo Weigand${}^{1,3}$, and Max Wiesner${}^{4,5}$

\vskip 7 mm

\small ${}^{1}${\it PRISMA${}^+$ Cluster of Excellence and Mainz Institute for Theoretical Physics, \\
Johannes Gutenberg-Universit\"at, 55099 Mainz, Germany} \\[3 mm]

\small ${}^{2}${\it Center for Theoretical Physics of the Universe, \\ Institute for Basic Science, Daejeon 34051, South Korea} \\[3 mm]

\small ${}^{3}${\it II. Institut f\"ur Theoretische Physik, Universit\"at Hamburg, \\  Luruper Chaussee 149, 22607 Hamburg, Germany } \\[3 mm]

{\it Zentrum f\"ur Mathematische Physik, Universit\"at Hamburg, \\ Bundesstrasse 55, 20146 Hamburg, Germany  }   \\[3 mm]

\small ${}^{4}${\it Instituto de F\'isica Te\'orica UAM-CSIC, Cantoblanco, 28049 Madrid, Spain}  \\[3 mm]

\small ${}^{5}${\it Departamento de F\'isica Te\'orica, Universidad Aut\'onoma de Madrid, 28049 Madrid, Spain}

\fnote{}{klaewer at uni-mainz.de, seungjoolee  at ibs.re.kr, timo.weigand at desy.de, max.wiesner at uam.es}

\end{center}

\vskip 3mm

\begin{abstract}

We study quantum corrections in four-dimensional theories with $N=1$ supersymmetry
in the context of Quantum Gravity Conjectures.
According to the Emergent String Conjecture, 
infinite distance limits in quantum gravity either lead to decompactification of the theory or result in a weakly coupled string theory. 
We verify this conjecture in the framework of $N=1$ supersymmetric F-theory compactifications to four dimensions
 including perturbative $\alpha'$ as well as non-perturbative corrections. 
After proving uniqueness of the emergent critical string at the classical level,
we show that quantum corrections obstruct precisely those limits in which the scale of the emergent critical string  would lie parametrically below the Kaluza-Klein scale.
Limits in which the tension of the asymptotically tensionless string sits at the Kaluza-Klein scale, by contrast, are not obstructed.

In the second part of the paper we study the effect of quantum corrections 
for the Weak Gravity Conjecture away from the strict weak coupling limit. 
We propose that gauge threshold corrections and mass renormalisation effects modify the super-extremality bound in four dimensions. For the infinite distance limits in F-theory
 the classical super-extremality bound is generically satisfied by a 
 sublattice of states in the tower of excitations of an emergent heterotic string.
By matching the F-theory $\alpha'$-corrections to gauge threshold corrections of the dual heterotic theory we predict 
how the masses of this tower must be renormalised
in order for the Weak Gravity Conjecture to hold at the quantum level.

\end{abstract}

\vfill

\thispagestyle{empty}
\setcounter{page}{0}

\setcounter{page}{1}
\newpage

\tableofcontents

\section{Introduction}
Recently there has been much activity aiming to distinguish effective field theories (EFTs) that can be consistently completed to a theory of gravity from those that do not allow for such a completion and hence lie in the so-called {\it Swampland} \cite{Vafa:2005ui}.
This has led to the formulation of a growing web of general conjectures about the nature of any theory of quantum  gravity that lie at the heart of the \textit{Swampland Program}  (for reviews see \cite{Brennan:2017rbf,Palti:2019pca}). Since string theory combines both quantum gravitational and field theoretic dynamics it provides an ideal arena to test these quantum gravity conjectures in explicit setups. In this context the asymptotic boundaries of the moduli space of string compactifications are of particular interest as here the structure of the moduli space simplifies considerably allowing for precise statements. 

If these asymptotic regions lie at infinite distance in moduli space the Swampland Distance Conjecture (SDC) \cite{Ooguri:2006in} asserts that a tower of states has to become asymptotically massless in the vicinity of the infinite distance point. 
According to the Emergent String Conjecture \cite{Lee:2019oct}, an infinite distance limit in {\it any} quantum gravity theory is either an effective decompactification limit or a limit in which a unique critical string
becomes weakly coupled and tensionless compared to the Planck scale. 
The asymptotic physics crucially depends on the parametric behaviour of the string tension, $T_{\rm str}$, with respect to the Kaluza-Klein (KK) scale (in less than ten dimensions) \cite{Lee:2019oct}:
\bea  \label{phases}
T_{\rm str}   \succ   M^2_{\rm KK}     && {\rm decompactification \, \,  limit}   \nn \\
T_{\rm str}   \sim M^2_{\rm KK}       && {\rm  emergent \, \, string \, \, limit}   \\
T_{\rm str}  \prec M^2_{\rm KK}     && {\rm pathological}   \nn 
\eea
In the first case, the physics is dominated by a KK tower leading to decompactification.
In the emergent string limit, by contrast, the weakly coupled string excitations and the KK tower are parametrically comparable.\footnote{Here the symbols $\succ$, $\sim$, $\prec$ refer to parametric behaviour in the infinite distance limit. In particular $\sim$ also refers to cases where for example the string tension and the KK scale are just separated by finite volume effects. If this suppression is large, but finite in the infinite distance limit, we speak of a numerical suppression.}
The theory hence asymptotes to a weakly coupled string theory without changing the number of dimensions, even if it is not formulated as a string theory away from the boundaries of its moduli space.
The third possibility will be discussed further below.
Examples of the first type include limits in which the parametrically dominant tower of states consists of BPS particles due to branes wrapped on certain non-contractible cycles as studied in \cite{Grimm:2018ohb,Grimm:2018cpv,Corvilain:2018lgw,Font:2019cxq,Grimm:2019wtx,Gendler:2020dfp}.
Emergent string limits, on the other hand, arise when solitonic strings due to wrapped branes become weakly coupled and acquire a tension comparable with $M^2_{\rm KK}$ \cite{Lee:2018urn, Lee:2018spm, Lee:2019jan,Lee:2019apr,Lee:2019oct}.
The special role played by tensionless strings to understand the Swampland Distance Conjecture in four dimensions has also been stressed in \cite{Lanza:2020qmt}, which studies in particular the backreaction of such strings on the fields in an ${N}=1$ supersymmetric effective action.

Since an infinite distance point implies a weak-coupling limit for a gauge symmetry \cite{Grimm:2018ohb,Lee:2018urn,Heidenreich:2020ptx}, the Weak Gravity Conjecture (WGC) \cite{ArkaniHamed:2006dz} requires that among the infinite tower of light states there should be charged states whose charge-to-mass ratio exceeds that of an extremal black hole. In a stronger version of the WGC these charged states are even expected to populate a sublattice of the entire charge lattice \cite{Cheung:2014vva, Heidenreich:2015nta, Heidenreich:2016aqi,Montero:2016tif} or at least an infinite subtower of states \cite{Andriolo:2018lvp}. Alternatively one can interpret this condition as the statement that there should exist a tower of states for which the net force is repulsive \cite{Palti:2017elp}, \cite{Heidenreich:2019zkl, Lee:2018spm}.

\begin{figure}
  \center
  \includegraphics[width=0.55\textwidth]{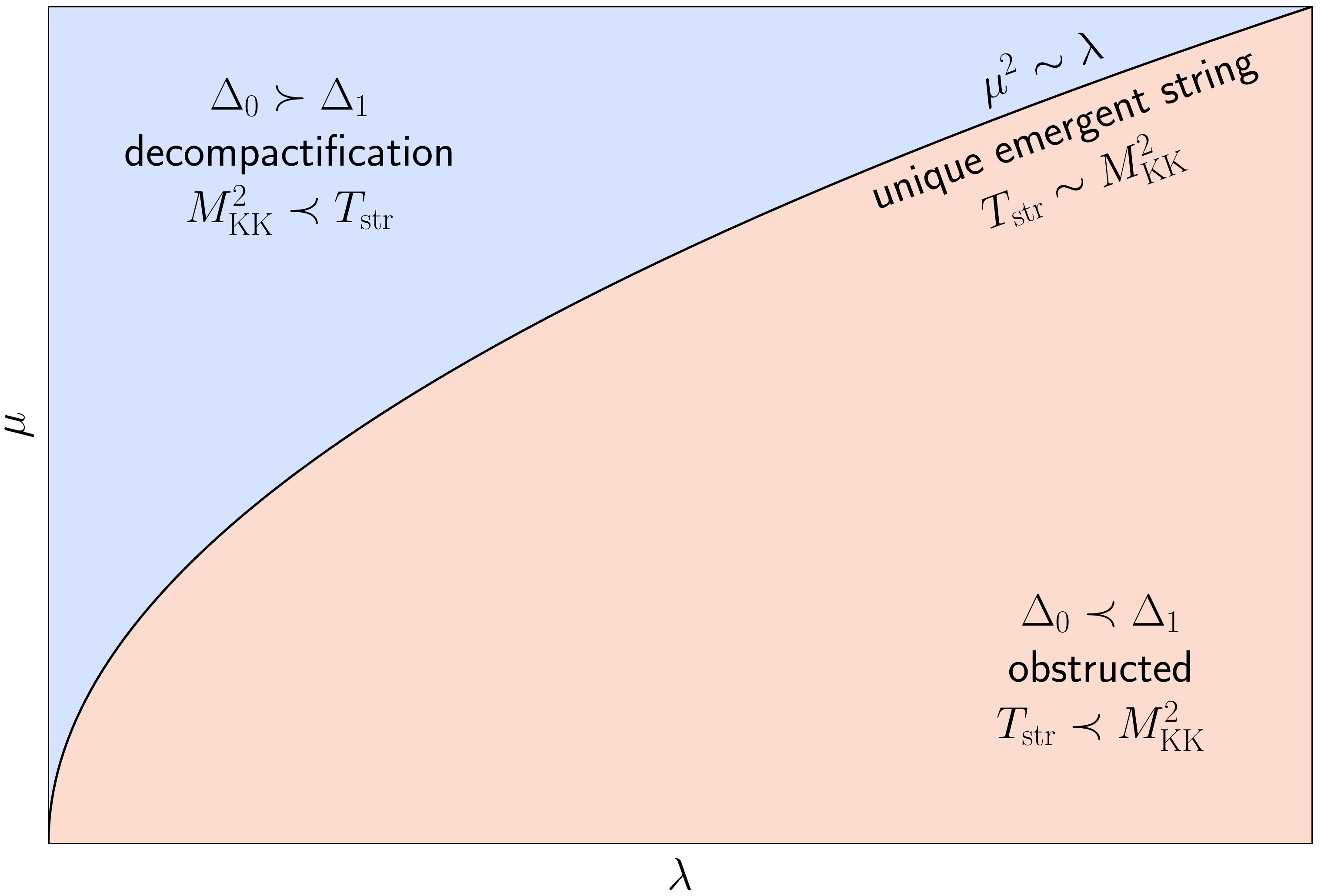}
  \caption{Infinite distance limits of the F-theory geometry $C_0\to B_3\to B_2$ (for the class where $C_0$ is a rational curve): The quantity $\lambda\to\infty$ parametrises a finite volume infinite distance limit in which the volumes $\mathcal{V}_{C_0}\to 0$ and $\mathcal{V}_{B_2}\to\infty$. The parameter $\mu$ describes an additional homogeneous rescaling $\mathcal{V}_{B_3}\sim\mu^3$. The plane is separated by the codimension one locus $\mu^2\sim\lambda$ at which a unique emergent heterotic obtained by wrapping a D3 brane on $C_0$ becomes light at the same rate as the KK scale. The region where $\mu^2/\lambda\to 0$ is obstructed by $\alpha'$-corrections, whereas the opposite regime $\mu^2/\lambda\to\infty$ is a decompactification limit where $M_{\rm KK}^2/T_{\rm str}\to0$. Note that $\Delta_0$ and $\Delta_1$ dualise to holomorphic and non-holomorphic threshold corrections in the heterotic frame, see Section \ref{sec_SecFHet}.}
  \label{fig:mu_lambda_plane}
\end{figure}

Consistency with the Emergent String Conjecture requires that in any infinite distance limit which does not correspond to an effective decompactification,
the states satisfying the (sub-) lattice WGC must be the excitations of emergent weakly coupled strings.
This was confirmed in non-perturbative compactifications with eight supercharges in 
six dimensions  \cite{Lee:2018urn, Lee:2018spm, Lee:2019apr} and with four supercharges in four dimensions \cite{Lee:2019jan}.
However, in compactifications to four dimensions quantum corrections have to be taken into account which can in principle obstruct classical infinite distance limits. This  has been shown for $N=2$ theories in the vector multiplet \cite{Lee:2019oct} and the hypermultiplet moduli space \cite{Marchesano:2019ifh, Baume:2019sry}. The issue of quantum corrections in four-dimensional setups with $N=1$ supersymmetry has been addressed in \cite{Cicoli:2018tcq,Gonzalo:2018guu,Xu:2020nlh}. 

In particular the results of \cite{Baume:2019sry} show that (non-)perturbative quantum corrections and tensionless strings go hand-in-hand: The quantum corrections precisely obstruct geometric limits in which the tension of the emergent, critical string would naively drop parametrically below the KK-scale. This would lead to a pathological phase of the theory 
 with a genuine critical four-dimensional string which is not expected to exist on general grounds (see (\ref{phases})).

\begin{figure}\center
  \includegraphics[width=\textwidth]{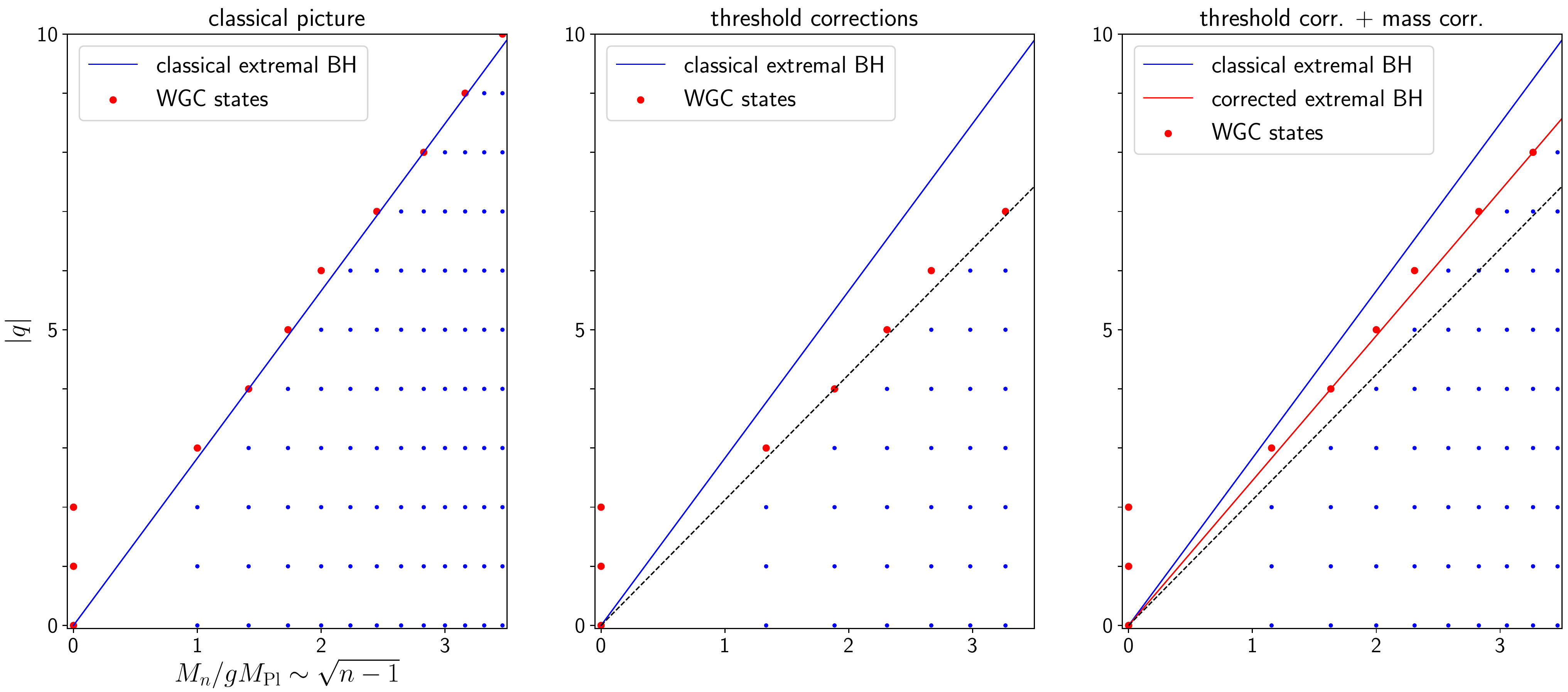}
  \caption{Left: Neglecting corrections away from the infinite distance point the (sub-)lattice WGC is fulfilled by a (super-)extremal tower of heterotic string excitations. Center: The gauge coupling is corrected away from the infinite distance point such that the former WGC states do not marginally lie above the classical bound (here, they lie below, as in the example of Section \ref{sec_WGC1loop}.). Right: Taking into account corrections to the relation $M_n^2\sim n T_{\rm str}$ the tower of formerly extremal states satisfies the repulsive force condition $|F_{\rm Coulomb}|\geq|F_{\rm Grav}|+|F_{\rm Yuk}|$ although it is still sub-extremal with respect to the classical extremality bound. We expect corrections to black hole solutions to be such that (super-)extremality of the string states is restored (red line).}
  \label{fig:WGC_relation_corrections}
\end{figure}

 In the present work, we discuss the effect of quantum corrections for the swampland conjectures in four-dimensional theories with ${N}=1$ supersymmetry\footnote{Quantum log corrections to swampland conjectures have been studied in~\cite{Blumenhagen:2019vgj,Blumenhagen:2020dea}.}. 
 We will find a conspiracy of quantum effects to obstruct precisely the pathological limits in (\ref{phases}), while both decompactification limits and emergent string limits
 are unobstructed (see Figure \ref{fig:mu_lambda_plane}). As our second main result we will see how the precise charge-to-mass ratio appearing in the WGC is modified at the quantum level as a consequence of gauge threshold corrections (cf. Figure \ref{fig:WGC_relation_corrections}).

 To obtain these results, we reconsider the setup of \cite{Lee:2019jan} and investigate the role of perturbative $\alpha'$-corrections in infinite distance limits in the K\"ahler moduli space of F-theory compactifications on Calabi--Yau fourfolds.\footnote{For a discussion of infinite distance limits in the complex structure moduli space of Calabi--Yau fourfolds in the context of the swampland conjectures, see \cite{Grimm:2019ixq}.} Our starting point for this investigation are the leading perturbative  corrections  at order $(\alpha')^2$ in the form computed in \cite{Grimm:2013gma, Grimm:2013bha,Grimm:2017pid,Weissenbacher:2019mef, Weissenbacher:2020cyf}.
 The first part of this paper is thus dedicated to investigating whether such quantum corrections obstruct certain classical infinite distance limits similar to what happens in $N=2$ setups. 

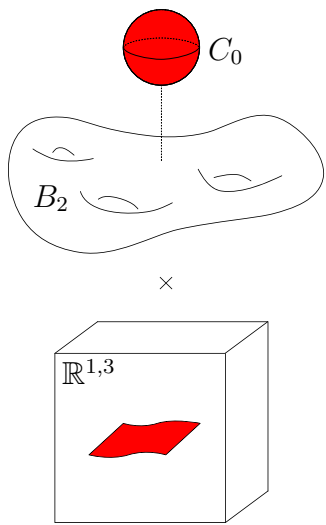
\begin{figure}
\center
  \begin{tikzpicture}
      \node at (0,0) {\includegraphics[width=0.25\textwidth]{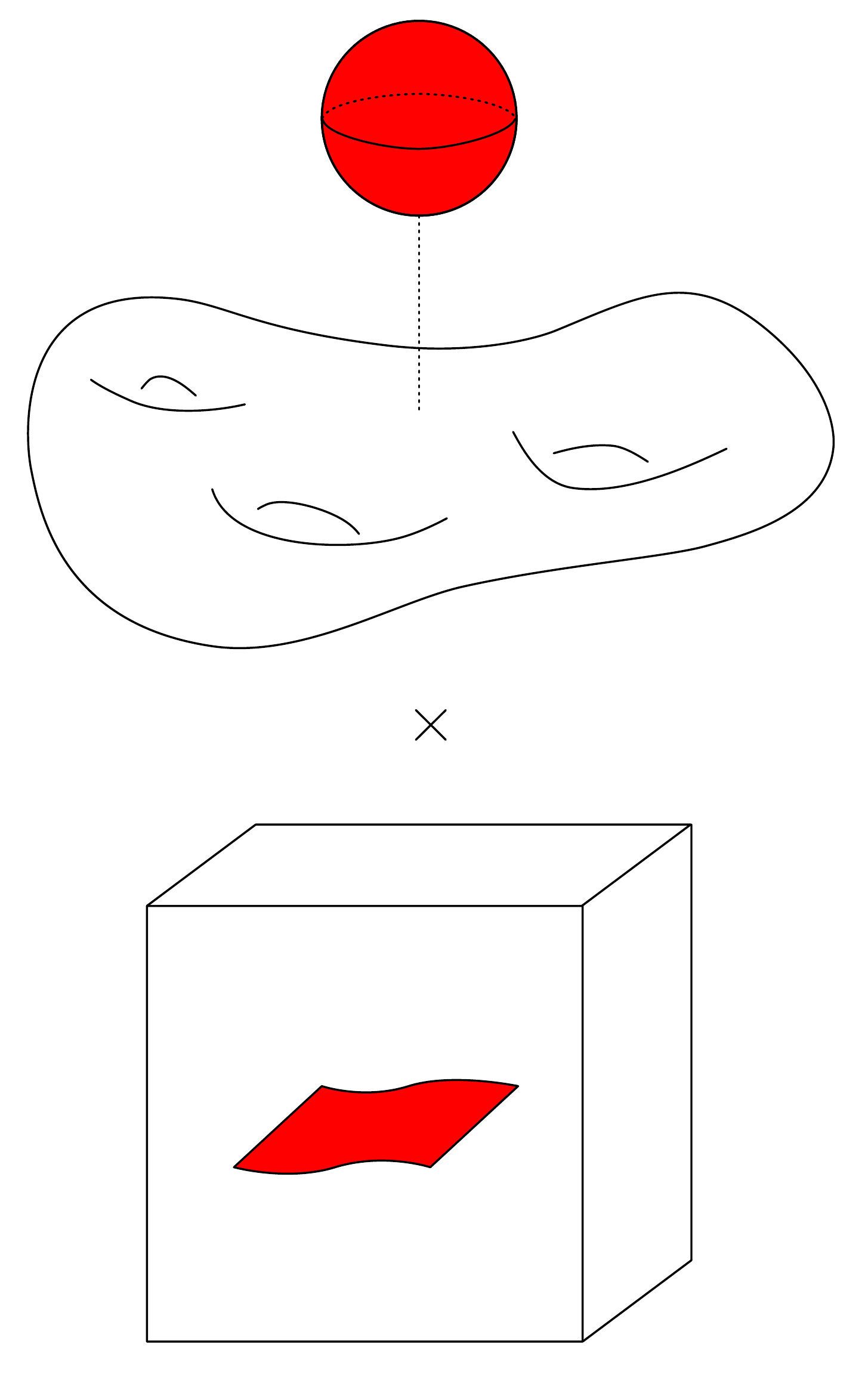}};
      \node at (-1,-1.4) {$\mathbb{R}^{1,3}$};
      \node at (-1.5,0.9) {$B_2$};
      \node at (0.8,2.85) {$C_0$};
    \end{tikzpicture}
  \caption{We consider F-theory compactifications on $\mathcal{E}_\tau\to Y_4\to B_3$. As in \cite{Lee:2019jan}, infinite distance limits with emergent strings feature a fibration structure $C_0\to B_3\to B_2$, such that the fiber $C_0$ shrinks relative to $B_2$. The role of the emergent string is played by a D3 brane wrapped on $C_0$. The curve $C_0$ can either be a torus leading to a Type II string, or a rational curve leading to a heterotic string.}
  \label{fig:fibration}
\end{figure}

Our main interest is in infinite distance limits that do not correspond to decompactification limits and thus reduce to a weakly-coupled string theory as displayed in Figure~\ref{fig:fibration}. To make contact with the WGC we further need a gauge theory to become weakly-coupled at the infinite distance point. As analysed in \cite{Lee:2019jan} such limits can only be attained if the base of the elliptically-fibered Calabi-Yau fourfold  admits a rational fibration. The emergent string then corresponds to a D3-brane wrapped on the rational fiber and is dual to a critical heterotic string. In the case of a $U(1)$ gauge theory, a sublattice of the string excitations captured by the elliptic genus classically satisfy the WGC \cite{Lee:2019jan}. Since also the asymptotic geometry of the moduli space receives perturbative corrections a particularly interesting question is whether this classical statement remains true even after the inclusion of quantum effects. 

Quantum corrections should also be a necessary ingredient when analysing the swampland conjectures upon  moving slightly away from the infinite distance points towards the interior of the moduli space. The general expectation is that swampland conditions such as the WGC should be still valid in the interior of the moduli space. However, quantities such as the charge-to-mass ratio of extremal black holes  receive corrections encoded in the corrections to the F-theory moduli space geometry slightly away from the strict emergent string limit. Analysing the structure of the F-theory corrections then allows us to also make a statement about the WGC slightly inside the quantum moduli space.

In the following, we give an overview and summary of the key results of this paper:
The first part of the paper revisits the geometry of infinite distance limits in the K\"ahler moduli space of F-theory compactifications on Calabi-Yau fourfolds \cite{Lee:2019jan}. Such limits can lead to the appearance of asymptotically tensionless heterotic or Type II strings. In this case the F-theory base is fibered by a curve such that the string is realised by a wrapped D3 brane. The situation is summarized in Figures \ref{fig:mu_lambda_plane} and \ref{fig:fibration}.

Regarding the classical geometry of the emergent string limits we present a uniqueness result in Section~\ref{sec_Uniqueness}. We find that any infinite distance limit in which more than one non-contractible curve shrinks to zero size at the fastest rate is necessarily a decompactification limit. In this case the tower of states predicted by the distance conjecture is the Kaluza-Klein tower and not the tower of excitations associated with the string from wrapping D3-branes on the shrinking curves, which asymptotically disappear from the spectrum. Any limit in which this is not the case can only feature a \emph{unique} such shrinking curve corresponding to a unique emergent critical heterotic or Type II string. See also Figure \ref{fig:limit_modulispace} for illustration.

In section~\ref{sec_corr} we show that even the limits with a unique emergent string are severely constrained by the leading perturbative $\alpha'^2$-corrections to the F-theory effective action,
as computed in \cite{Grimm:2013gma, Grimm:2013bha,Weissenbacher:2019mef, Weissenbacher:2020cyf}. These generically present an obstruction to
what is called pathological limits in (\ref{phases}), i.e. to
 those limits in which the tension of the emergent string asymptotically vanishes when compared to the KK scale.
 The parametric hierarchy between the two scales translates to the shrinking of divisors vertical with respect to the fibration by $C_0$, leading to a loss of control. This censorship by $\alpha'$-corrections offers a tentative way out of the conundrum of a four-dimensional critical string theory that such limits would propose. 
At least for a subclass of such limits, we will understand the quantum corrections responsible for this shielding of the pathological infinite distance limits as follows:
They lead to a strong coupling transition at finite distance before the potentially problematic infinite distance regime can be entered. This effect is reflected not only in the perturbative $\alpha'$-corrections, but also in the non-perturbative effect of D3-brane instantons.
On the other hand, there are no quantum obstructions both for the decompactification limits and for the emergent string limits in which the emergent string scale is parametrically at the same order in the infinite distance parameter as the KK scale.
This result is intuitive because the infinite distance limits correspond to weak coupling limits, and hence the potential is expected to vanish asymptotically. It also fits with the general  asymptotic behaviour of the potential advocated in \cite{Ooguri:2018wrx}.

Combining our results, we can conclude the following:
\begin{whitebox}
  Infinite distance limits in the K\"ahler moduli space of F-theory on Calabi-Yau fourfolds correspond either to a pure decompactification limit or to a transition to a different duality frame determined by a {\bfseries unique} emergent string such that $T_{\rm str}\sim M_{\rm KK}^2$. Any attempt to decouple the string from the KK tower is faced with diverging $\alpha'$-corrections.
\end{whitebox}

In the second part of the paper we focus on those limits resulting in a weakly coupled heterotic string and elaborate on the interpretation of the heterotic string states as the constituents of the WGC tower. Section~\ref{sec_SecFHet} reviews the relevant aspects of F-theory/heterotic duality in four dimensions. 
To leading order in the infinite distance parameter, the gauge coupling is determined by the ratio of the volume of the compactification space $B_3$ and the shrinking fiber, $\frac{1}{g_{\rm YM}^2}\sim\mathcal{V}_{B_3}/\mathcal{V}_{C_0}$.
This relation receives subleading correction terms, called $\Delta_0$ in Figure~\ref{fig:mu_lambda_plane}, already at the level of classical geometry in F-theory as we move away from the strict infinite distance point,
 as well as contributions due to the perturbative $\alpha'$-corrections ($\Delta_1$ in Figure~\ref{fig:mu_lambda_plane}).
We will explain how, in the dual heterotic frame, $\Delta_0$ maps to the holomorphic gauge threshold corrections, whereas $\Delta_1$ maps to non-holomorphic loop corrections 
of $\frac{1}{g_{\rm YM}^2}$.


Section~\ref{sec_wgc} then studies the subleading modifications to the WGC due to these two effects.
To this end we invoke the interpretation of the WGC bound as the requirement that the net force between two test particles from the tower be repulsive \cite{Palti:2017elp, Heidenreich:2019zkl, Lee:2018spm}.  
To leading order, this is equivalent to imposing super-extremality with respect to an extremal black hole.
The repulsive-force bound depends both on $g^2_{\rm YM}$ and on the mass $M_k$ of the test particle of charge $q_k$.
We first show that for generic fluxes in F-theory, a sublattice of string excitations satisfies the resulting bound in the strict weak coupling limit. 
This is an extension of the results in~\cite{Lee:2019jan} which builds on recent observations in \cite{Lee:2020gvu,Lee:2020blx}  concerning the modular properties of the elliptic genus of strings in generic flux backgrounds.

Naively, the loop corrections to $g^2_{\rm YM}$ seem to imply that the states which satisfy the leading order WGC bound 
violate the quantum corrected bound. 
 A more careful investigation reveals that while the tension of the heterotic string (which is BPS) is uncorrected at the quantum level, the masses of its non-BPS excitations may receive corrections to the classical relation $M_n^2\sim n T_{\rm str}$. 
Hence the WGC predicts the form of these
 mass renormalisations if we impose that the same tower of states continues to satisfy the repulsive force condition (Figure~\ref{fig:WGC_relation_corrections}).
 Furthermore, if the equivalence between the repulsive-force condition and super-extremality survives  the loop corrections, 
  the repulsive-force bound
  predicts the charge-to-mass ratio for extremal black holes in the 1-loop corrected theory. 
  Note that the gauge loop corrections are not to be confused with higher derivative  $\alpha'$-corrections to the charge-to-mass ratio of extremal black holes, whose role 
  in the WGC has been noted early on in \cite{ArkaniHamed:2006dz,Kats:2006xp} and which have been analysed more recently e.g. in \cite{Cheung:2018cwt,Hamada:2018dde,Bellazzini:2019xts,Aalsma:2019ryi}.
 The WGC in situations without supersymmetry has been studied in \cite{Bonnefoy:2018tcp}.

The resulting corrected extremality bound satisfied by the tower of string excitations depends on the holomorphic and non-holomorphic gauge threshold corrections $\Delta_0$ and $\Delta_1$ as 

\begin{whitebox}
  \begin{equation}  \label{correctedboundWGC1}
    \frac{g_{\rm YM}^2 q_k^2M_{\rm Pl}^2}{M_k^2}\geq \frac{g_{\rm YM}^2 Q_{\rm BH}^2M_{\rm Pl}^2}{M_{\rm BH}^2}=1-\frac12 (\Delta_0+\Delta_1) \,.
  \end{equation}
\end{whitebox}

We conjecture that (\ref{correctedboundWGC1}) holds generally for a four-dimensional ${N}=1$ quantum gravity theory including 1-loop corrections.

\section{Uniqueness results for classical infinite distance limits in 4d \texorpdfstring{${N=1}$}{N=1} F-theory}   \label{sec_Uniqueness}

This section is devoted to the classical infinite distance limits in the K\"ahler moduli space of F-theory compactifications to four dimensions with ${N}=1$ supersymmetry.
A hallmark of such limits is that, apart from a potential homogeneous rescaling of the volume, the base $B_3$ of the F-theory elliptic fourfold $Y_4$ must itself exhibit a rational or genus-one fibration $C_0\to B_3\to B_2$ whose generic fiber shrinks in the limit compared to the two-dimensional base $B_2$.
A D3-brane wrapping this shrinking fiber then gives rise to a (relatively) tensionless fundamental heterotic or Type II string, respectively, signalling the transition to a dual, weakly coupled frame emerging in the infinite distance limit.

While this general structure has been uncovered already 
in \cite{Lee:2019jan}, 
our new insights in this paper are two-fold: First, we prove {\it uniqueness} of the fastest shrinking fibral curves of $B_3$ already at the level of classical K\"ahler geometry, using similar methods as in \cite{Lee:2019oct}.
This is crucial in order for the resulting asymptotically tensionless string  to play the role of a dual fundamental string.
Second, we point out that some of these classically attainable limits seemingly give rise to tensionless critical strings at a scale parametrically below the Kaluza-Klein scale. As we will see in Section \ref{sec_corr}, this pathological feature is remedied once quantum corrections are taken into account.

\subsection{Systematics and general picture}

We begin with a general overview of the classical infinite distance limits and a summary of our new results.
The technical details are then elaborated on in Sections \ref{subsec_JAclassical} and \ref{subsec_JBclassical}.

Consider the base space $B_3$ of an elliptically fibered Calabi-Yau fourfold $Y_4$ on which we compactify F-theory to four dimensions.
Neglecting, for the time being, quantum corrections we focus in this section on the classical K\"ahler geometry of $B_3$.
The infinite distance limits in the K\"ahler moduli space can be parametrised in terms of the K\"ahler form 
\be
J = \sum_\a {v}^\a J_\a \,,
\ee 
where $J_\a$ denote the generators of the K\"ahler cone.
In an infinite distance limit, one or more of the K\"ahler parameters ${v}^\a$ scale to infinity \cite{Lee:2019jan,Lee:2019oct}.

\begin{figure}
  \centering
    \includegraphics[width=0.5\textwidth]{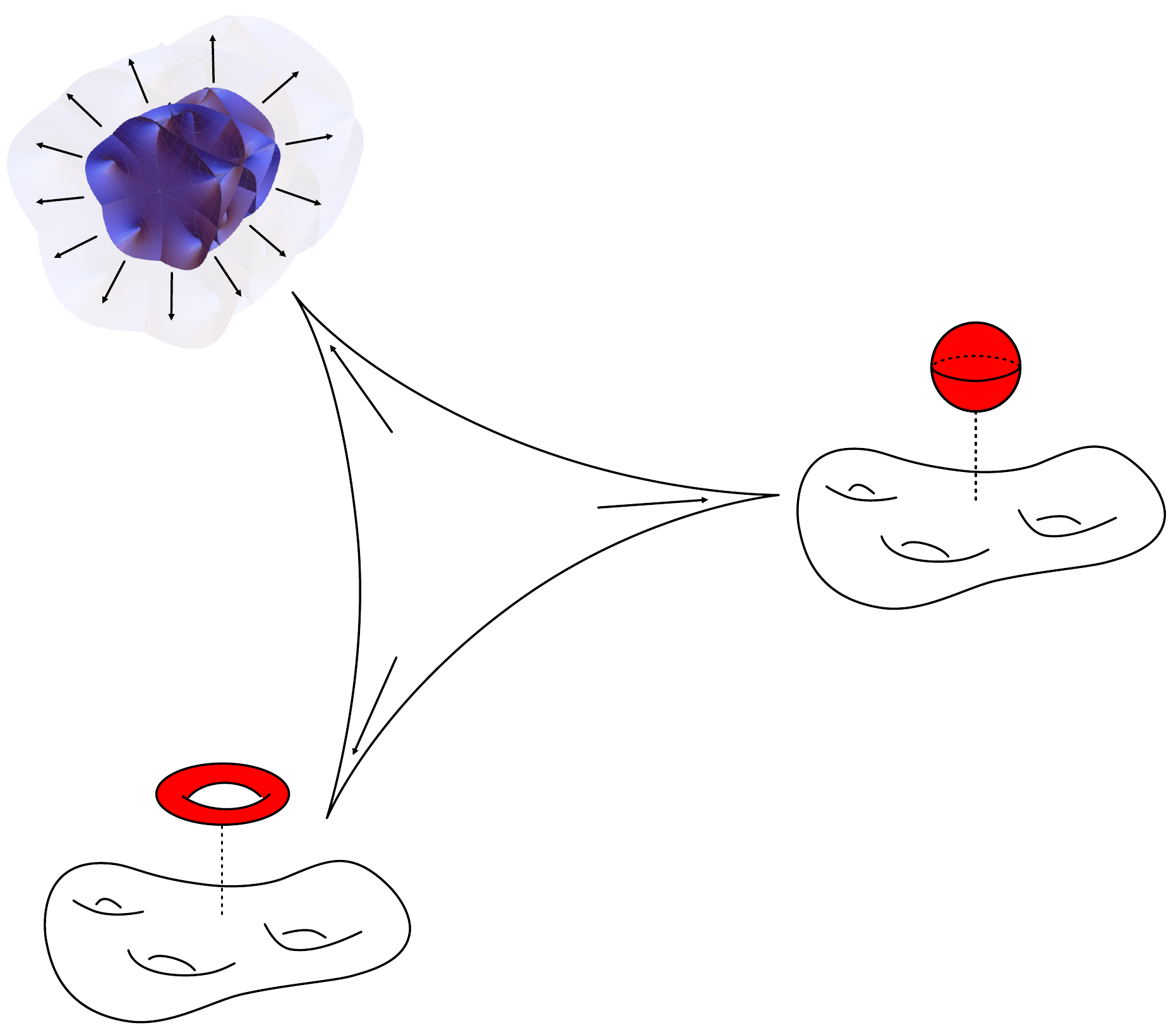}
  \caption{Shown are the possible infinite distance limits in the K\"ahler moduli space of F-theory compactified to four dimensions with ${N}=1$ supersymmetry.
  These are either decompactification limits (upper left corner) or limits in which a unique heterotic (upper right corner) or Type II string (lower left corner) becomes tensionless (at the fastest rate) with respect to the four-dimensional Planck scale.}
  \label{fig:limit_modulispace}
\end{figure}

Generically this implies that the volume of $B_3$, measured in ten-dimensional string units,
\be
{\cal V}_{B_3} =  \frac{1}{6} \int_{B_3} J^3 \,,
\ee
diverges. If this is the case, we can introduce the homogeneously rescaled K\"ahler form $J'$ defined by
\bea   \label{JvsJp}
J =  \mu   \, J'  
\eea
such that the rescaled volume, 
\be   \label{VvsVp}
 {\cal V}'_{B_3} =  \,   \frac{1}{6} \int_{B_3} (J')^3 =\, \mu^{-3}\,  {\cal V}_{B_3}  \,,
\ee
stays finite in the infinite distance limit with $\mu \to \infty$.\footnote{Of course, if the volume ${\cal V}_{B_3}$ does not diverge in the infinite limit, we can simply set $\mu =1$.}

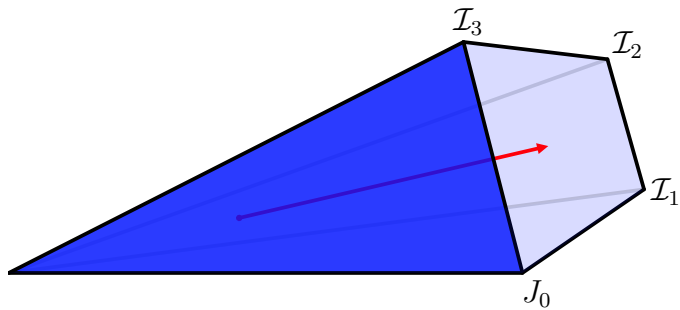
\begin{figure}
   \centering
   \begin{tikzpicture}
    \node at (0,0) {\includegraphics[width=0.5\textwidth]{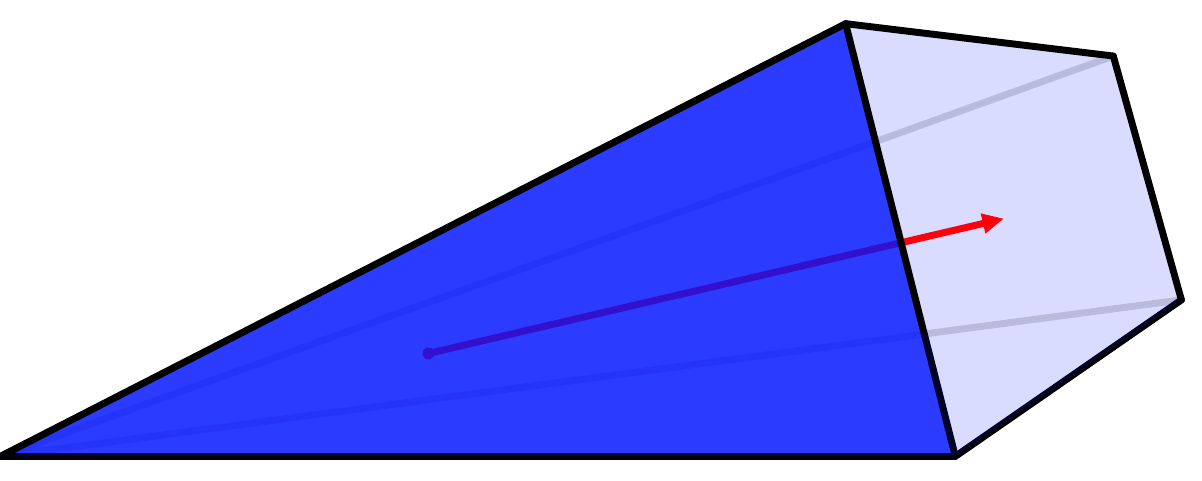}};
    \node at (1.8,1.8) {$\mathcal{I}_3$};
     \node at (3.9,1.5) {$\mathcal{I}_2$};
     \node at (4.4,-0.45) {$\mathcal{I}_1$};
     \node at (2.7,-1.8) {$J_0$};
   \end{tikzpicture}
  \caption{The existence of a finite volume limit in the K\"ahler cone of $B_3$ where one of the K\"ahler moduli $v^{\prime 0}\to\infty$ leads to a natural classification of the remaining K\"ahler cone generators into three different sets $\cI _{1\dots3}$ according to their intersection properties with $J_0^2$ and $J_0$ (see~\cite{Lee:2019jan} for details). Limits of $J$-class A only allow for generators of type $\cI _1$ and $\cI _3$, whereas limits of $J$-class B can only feature generators of type $\cI _2$.}
   \label{fig:limit_general}
 \end{figure}

There are now two possibilities \cite{Lee:2019oct}. The first option is that the rescaled K\"ahler form $J'$ does not undergo any further infinite distance limit, i.e. the rescaled K\"ahler parameters $v^{\prime\a}$
stay finite. In this situation, the infinite distance limit of $J$ corresponds to a straightforward decompactification limit, in which the ratio of the Kaluza-Klein (KK) scale to the F-theory/Type IIB string scale $M_{\rm IIB}$ tends to zero as
\be
\frac{M_{\rm KK}}{M_{\rm IIB}} \sim ({\cal V}_{B_3})^{-1/6}  \sim  \mu^{-1/2}  \to  0    \,.
\ee
Importantly, the KK tower is the {\it only} infinite tower of asymptotically massless states as measured in units of $M_{\rm IIB}$. This follows from the assumption that the rescaled K\"ahler form $J'$ does not by itself undergo any further infinite distance limit.
The only effect of the limit is therefore the overall rescaling of  ${\cal V}_{B_3}$.

The second, more interesting option is that the original limit for $J$ implies a residual infinite distance limit in the rescaled K\"ahler moduli space whose K\"ahler metric is determined by $J'$.
Such limits correspond to {\it finite volume infinite distance limits} (with respect to the rescaled K\"ahler form $J'$) because by construction the volume in the rescaled metric, ${\cal V}'_{B_3}$, does not scale to infinity.
They fall into two classes, dubbed {\it $J$-class  A} and {\it $J$-class B} in \cite{Lee:2019jan} (see Figure~\ref{fig:limit_general}). The two types of residual limits are distinguished by the precise scaling behaviour of the K\"ahler parameters $v^{\prime\a}$ of $J'$, as will be described in detail in Section \ref{subsec_JAclassical} and \ref{subsec_JBclassical}, respectively.
The complete classification of infinite distance limits is then obtained by superimposing one of the two limits for $J'$ with the overall scaling (\ref{JvsJp}). This parallels the classical situation for Calabi-Yau threefolds \cite{Lee:2019oct}.

Let us first focus exclusively on the classical finite volume limits (with respect to $J'$).
A preliminary analysis of the properties of the two types of finite volume limits has appeared in \cite{Lee:2019jan}.
The general intuition is that if one or several of the K\"ahler parameters $v^{\prime\a}$ diverge, a curve or divisor on $B_3$ must become asymptotically large.
In order for the total volume of $B_3$  to remain finite (with respect to $J'$),  the volume form must have the structure of a product of the volume of the large cycle with an asymptotically small cycle, at least to leading order.
The more accurate statement is that $B_3$ must admit a fibration whose generic fiber is either an asymptotically vanishing curve or surface.

If $B_3$ admits a fibration by a curve, then this curve must be either a rational curve, i.e. a $\mathbb P^1$, or a genus-one curve. This follows from the requirement that the anti-canonical class $\bar K_{B_3}$ of $B_3$ must be effective in order for an elliptically fibered Calabi-Yau $Y_4$ over $B_3$ to exist.
The significance of such fibrations is that a D3-brane wrapping the generic fiber gives rise to a solitonic string corresponding to a critical heterotic string (in the case of a $\mathbb P^1$ fiber \cite{Lee:2018urn,Lee:2019jan}) or to a critical Type II string (in the case of a genus-one fiber \cite{Lee:2019apr,Lee:2019oct}). As the fiber shrinks (at least in the metric determined by $J'$) these strings become asymptotically light and weakly coupled. Their excitations yield a tower of light, weakly coupled states. This tower of states competes with the KK modes from large curves or divisors on $B_3$, which are always present: Indeed, as noted above, if the volume of $B_3$ stays finite (with respect to the K\"ahler form $J'$), some curve or divisor becomes large. The asymptotic physics of the infinite distance limit depends on which 
of the towers of {\it weakly coupled}, asymptotically massless states dominate in the limit, i.e. which towers of states become light at the fastest rate. 
If the KK tower is parametrically lighter than the string tower, the limit describes a decompactification limit, while if the mass scales of the two towers are comparable, we enter the regime dominated by a light string in an effectively four-dimensional theory.

Generically, $B_3$ may admit several rational or genus-one fibrations. The important question is then if there is always a {\it unique} such fibration whose fiber vanishes (in the metric $J'$) at the fastest rate and how this rate compares to
the KK scale (again defined, for the time being, with respect to $J'$).
In the remainder of this section we will provide a very sharp answer to this question within the range of {\it classical} finite volume infinite distance limits, 
using similar techniques as in \cite{Lee:2019oct}:

\begin{whitebox}
For any finite volume infinite distance limit of $J$-class A or B, precisely one of the following three possibilities occurs (all volumes are measured with respect to $J'$):
\begin{enumerate}
\item The base $B_3$ exhibits a unique $\mathbb P^1$-fibration whose generic fiber shrinks at a rate faster than the rate at which any other non-contractible curve on $B_3$ vanishes. 
 A D3-brane wrapping this rational fiber gives rise to an asymptotically tensionless heterotic string. 
\item The base $B_3$ exhibits a  unique $T^2$-fibration whose generic fiber shrinks at a rate faster than the rate at which any other non-contractible curve on $B_3$ vanishes.
A D3-brane wrapping this torus fiber gives rise to an asymptotically tensionless fundamental Type II string. 
\item There exists no {\it unique} rational or genus-one fibration within $B_3$ whose fiber volume vanishes at a rate faster than the rate of any other non-contractible curve. However, the KK scale is parametrically smaller than the string scale from any wrapped D3-brane on a non-contractible 2-cycle. 
\end{enumerate}
\end{whitebox}

Having established the behaviour of the finite volume infinite distance limits for $J'$, the physics of the full infinite distance limit requires taking into account also the overall scaling $\mu$.
In particular, this modifies the relation between the scales associated with the solitonic fundamental strings (if present) and the KK scale, both of which have to be measured with respect to the full K\"ahler form $J$, rather than $J'$.
As we will see, the first two types of limits for $J'$ above translate into either
a decompactification limit for $J$ or to a limit in which the duality frame changes to that of the respective weakly coupled fundamental string
which becomes light. In the latter case, one speaks of an emergent string limit.
Which of the two scenarios occurs depends on the precise comparison between the parameter $\mu$ appearing in (\ref{JvsJp}) and the parameter controlling the volume of the vanishing fiber, that we will denote as $\lambda$ later.
The third possibility above remains a decompactification limit.
All this is illustrated in Figure \ref{fig:limit_modulispace}.

The uniqueness results summarised above exclude the exotic possibility that there are two or several weakly coupled fundamental strings emerging in the limit:
According to our findings, if several fundamental strings become light at the same rate, then their string scale is always parametrically above the KK scale, i.e. the limit is a decompactification limit. This resolves potential inconsistencies which would follow from the coexistence of several weakly coupled fundamental strings at or below the KK scale.
Interestingly, consistency with the emergent string picture is obtained in this respect already within classical geometry.
Nonetheless, another conundrum remains:
The first two types of classical limits above allow for situations in which the emergent string scale is parametrically suppressed with respect to the KK scale, rather than sitting at the same parametric scale. Such a behaviour would point to a new, genuinely four-dimensional weakly coupled string theory. As we will establish in Section \ref{sec_corr}, the geometric regime in which this would occur is afflicted by diverging $\alpha'$-corrections and must hence be ruled out.

\subsection{Classical infinite distance limits of \texorpdfstring{$J$}{J}-class A}  \label{subsec_JAclassical}

\begin{figure}
  \centering
  \begin{tikzpicture}
  \node at (0,0) {\includegraphics[width=0.5\textwidth]{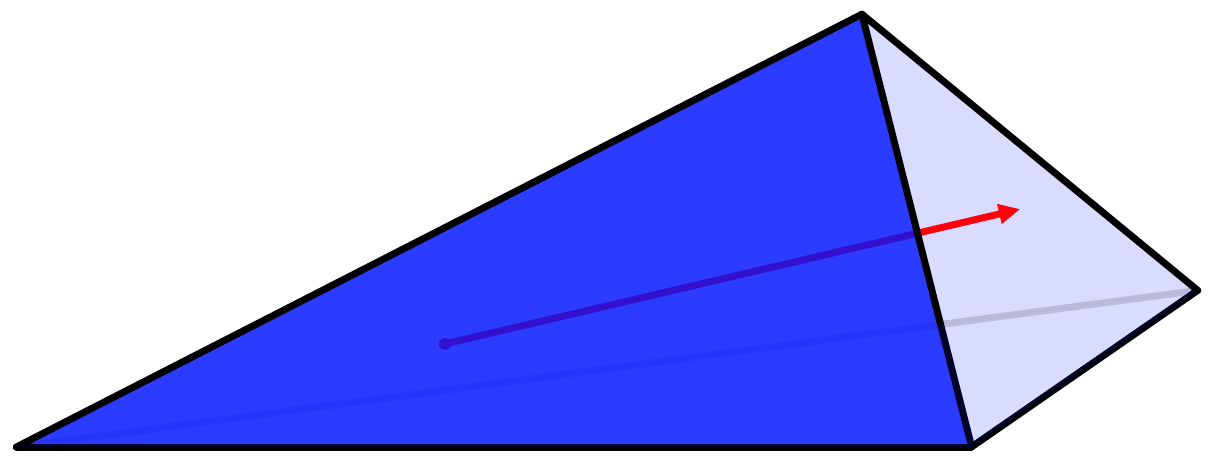}};
  \node at (1.9,1.8) {$\mathcal{I}_3$};
  \node at (4.5,-0.45) {$\mathcal{I}_1$};
  \node at (2.8,-1.75) {$J_0$};
  \end{tikzpicture}
  \caption{Besides $J_0$, the $J$-class A limits feature two other sets of K\"ahler cone generators, distinguished by their intersection properties with $J_0^2$.}
  \label{fig:JclassA}
\end{figure}
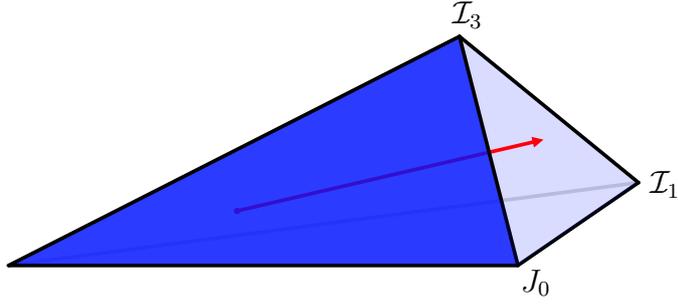

We now show the claims summarised in the previous section for the first type of infinite distance limits, based on the classical finite volume limits of $J$-class A  for the rescaled K\"ahler form $J'$.
Recall first the defining characteristics of the limit \cite{Lee:2019jan}.
\paragraph{ \bf $J$-class A limit for $J'$:}
The K\"ahler form $J' = \sum_\alpha v'^\alpha J_\alpha$ can be parametrised as
\begin{equation}   \label{JclassA-1}
    J'=\lambda J_0+\sum_{\mu\in\mathcal{I}_1}\frac{a^\mu}{\lambda^2}J_\mu+\sum_{r\in\mathcal{I}_3}b^{\prime r} J_r\,, \qquad \text{with}\quad a^\mu \precsim 1,\;\; b'^r \precsim \lambda \,, 
\end{equation}
where $\lambda \to \infty$ in the limit. As depicted in figure~\ref{fig:JclassA}, the K\"ahler cone generators are grouped into a distinguished generator $J_0$ and two further types of generators,  the set $\{J_\mu\}$ with  $\mu \in \mathcal{I}_1$ and $\{J_r\}$ with  $r \in \mathcal{I}_3$, defined by
 the intersection numbers
\bea   \label{JclassA-2}
 J_0^3&=& 0  \,,  \nn  \\
J_0^2\cdot J_\mu   &\neq& 0   \qquad \forall \mu \in \mathcal{I}_1\,,   \\
J_0^2\cdot J_r&=&J_0\cdot J_r\cdot J_s=   J_r\cdot J_s   \cdot J_t = 0 \,  \qquad \forall \,  r,s,t \in \mathcal{I}_3     \nn  \,.
\eea 
For later purposes we also define the finite parameter vector 
\be
\tilde b^{m}=(1,b^{\prime r}/\lambda)    \,,
\ee
where the index $m$ runs over $\{0\}\cup\cI_3$.

As detailed in \cite{Lee:2019jan}, the existence of the K\"ahler class generator $J_0$ with $J_0^3 = 0$ and $J_0^2 \neq 0$ implies that the base $B_3$ admits the structure of a fibration whose generic fiber is the curve in class 
\be
C_0 = J_0 \cdot J_0  \,.
\ee
Since $\bar K_{B_3}$ is effective and $J_0$ is a generator of the K\"ahler cone, $C_0 \cdot \bar K_{B_3}  = J_0 \cdot J_0 \cdot \bar K_{B_3}  \geq 0$.
In light of the adjunction formula,
\beq
2-2g = \int_{B_3} (\bar K_{B_3} - 2 J_0)  J_0^2  = \bar K_{B_3} \cdot C_0 \,,
\eeq
we are thus left with either of the two options: 
If $C_0 \cdot  \bar K_{B_3} =2$, the generic fiber $C_0$ is a $\mathbb P^1$, while if  $C_0 \cdot  \bar K_{B_3} =0$, it is a curve of genus one, i.e. a torus.

Classically, the volume of $C_0$, measured in the rescaled metric $J'$, tends to zero as 
\be
{\cal V}'_{C_0} \sim \lambda^{-2} \,.
\ee
At the same time any divisor ${\bf S}$ with $J_0 \cdot J_0 \cdot {\bf S} \neq 0$ expands as ${\cal V}'({\bf S}) \sim \lambda^2$. If one such divisor is a component of the discriminant of the elliptic fibration $Y_4$,
$C_0$ is necessarily a rational curve.\footnote{If by assumption $J_0 \cdot J_0 \cdot {\bf S} \neq 0$ and hence $J_0 \cdot J_0 \cdot {\bf S} > 0$, then  $J_0 \cdot J_0 \cdot \bar K_{B_3} > 0$ because  ${\bf S} \leq \Delta = 12 \bar K_{B_3}$ in order for ${\bf S}$ to be a component of the discriminant. See also the discussion at the beginning of Section \ref{sec_WGCrelQC_recap}.} In this case the infinite distance limit is a weak coupling limit for the non-abelian gauge group factor supported on the stack of 7-branes along the divisor ${\bf S}$. 
Similarly $C_0$ is a $\mathbb P^1$ if one of the expanding divisors corresponds to the  height-pairing associated with an abelian gauge group factor \cite{Lee:2018urn}.\footnote{Technically, this follows from the conjectural relation $b \leq n \bar K_{B_3}$ for some $n \in \mathbb N$ for the height-pairing divisor $b$ of a $U(1)$. }

Our first important new point, as mentioned above, is that if $B_3$ admits any other $\mathbb P^1$ or $T^2$-fibration whose generic fiber class $\tilde C_0$ is not identical to the class of $C_0$ (up to a potential scaling), then the volume of its generic fiber $\tilde C_0$  cannot vanish in the limit $\lambda \to \infty$. In this sense a $J$-class A limit admits a unique rational or genus-one fibration with asypmtotically vanishing fiber volume. 

To see this, let us suppose that $B_3$ admits another rational or genus-one fibration whose generic fiber is given by  a curve $\tilde C_0$ different from $C_0$.
  This implies that there exists a nef divisor $\tilde D \in \overline{\cK(B_3)}$, with $\overline{\cK(B_3)}$ the closure of the K\"ahler cone of $B_3$, such that\footnote{To obtain such $\tilde D$  we just pick a nef divisor of the base of the fibration with a positive self-intersection and pull it back to define $\tilde D$. }
\bea 
\tilde D^2 &=& \tilde n \, \tilde C_0\,, \quad \tilde n \in \mathbb N \,,  \label{D2}  \\
\tilde D^3 &=& 0 \,. \label{D3}
\eea
Then, using the intersection properties of the K\"ahler cone generators, one can indeed show that ${\cal V}'_{\tilde D^2}$ cannot vanish asymptotically in the limit $\lambda \to \infty$.
The proof is presented in Appendix \ref{App_JclassA}.

The full infinite distance limit, including the overall scaling $\mu$ in (\ref{JvsJp}), can then be expressed as 
\be \label{combinedJclassA}   
J = \mu \Big(\lambda J_0+\sum_{\mu\in\mathcal{I}_1}\frac{a^\mu}{\lambda^2}J_\mu+\sum_{r\in\mathcal{I}_3} b^{\prime r} J_r\Big)   \,, 
\ee
where we write
\be
\mu = \lambda^{\frac{1}{2} + x} \,, \qquad \text{with}\quad x\geq -\frac12\,.
\ee
Here $x$ is a free parameter, which, at the classical level,
 controls the
asymptotic ratio of the Kaluza-Klein scale and the scale of the solitonic fundamental string excitations:
 The tension of the solitonic string from a D3-brane wrapping the curve $C_0$ scales like
\bea  \label{MsolTypeA}
\frac{M^2_{\rm sol}}{M^2_{\rm IIB}} \sim   {\cal V}_{C_0}  \sim \frac{\mu}{\lambda^2} \sim \lambda^{-3/2 + x}   \,.
\eea
The Kaluza-Klein scale, on the other hand, is set by the volume of the expanding divisors (whose class contains one or more of the generators $J_\mu$) on $B_3$ as
\bea   \label{MKKTypeA}
\frac{M^2_{\rm KK}}{M^2_{\rm IIB}}  \sim (\mu \lambda)^{-1}   \sim \lambda^{-3/2 - x}   \,.   
\eea

Assume first that the curve $C_0$ is a rational curve such that a D3-brane wrapping $C_0$ gives rise to a heterotic string.
We can identify three different regimes:

\begin{enumerate}
\item
If $x > 0$, the ratio 
\be
\frac{M^2_{\rm KK}}{M^2_{\rm het}}   \sim \lambda^{-2 |x|} \to 0 \,.
\ee
This indicates that the limit corresponds to a straightforward decompactification limit.
\item
If $x = 0$, 
\be
\frac{M^2_{\rm het}}{M^2_{\rm IIB}} \sim \frac{M^2_{\rm KK}}{M^2_{\rm IIB}} \sim \lambda^{-3/2} \to 0   \,.
\ee
In this limit, the heterotic string takes over the role of the new fundamental string in a dual description, and the new heterotic string scale sits parametrically at the same scale as the Kaluza-Klein scale.
This corresponds to an effectively four-dimensional weak coupling limit in the dual heterotic frame, i.e. to an emergent heterotic string.
For later purposes, let us summarise this limit as
\begin{equation}   \label{JclassAhetrelabelded}
\boxed{
\text{$J$-class A, rescaled:}  \qquad J= \tilde\lambda J_0+ \sum_{r\in \mathcal{I}_3} b^r J_r       + \sum_{\alpha\in \mathcal{I}_1}\frac{a^\mu}{\tilde\lambda}J_\mu \,,  \quad  \tilde\lambda \to \infty    \,,}
\end{equation}
where the intersection numbers satisfy the relation (\ref{JclassA-2}) and the parameters $a^\mu$ and ${b^r/\tilde\lambda\equiv b^{\prime r}/\lambda}$ are finite in the limit  $\tilde\lambda \to \infty$.
The $\mathbb P^1$-fiber class is given by $C_0 = J_0 \cdot J_0$ and we have relabelled $\tilde \lambda = \mu\lambda =  \lambda^{3/2}$ in terms of the parameter $\lambda$ appearing in (\ref{JclassA-1}). This incorporates the rescaling by $\mu = \lambda^{\frac{1}{2}}$ in (\ref{combinedJclassA}).

\item
If $x < 0$, then 
\be
\frac{M^2_{\rm het}}{M^2_{\rm KK}} \sim \lambda^{-2|x|} \to 0 \,.
\ee
In this limit, the dual heterotic string would become parametrically lighter than the Kaluza-Klein scale.
Such behaviour would point to a completely new type of intrinsically four-dimensional theory, rather than to a weakly coupled compactification of the dual heterotic string.
However, as we will show in Section \ref{sec_alphapcorr}, this regime is out of computational control due to the appearance of unsuppressed $\alpha'$-corrections.
Note that this includes in particular the case $x=-\frac{1}{2}$ in which $\mu$ does not scale to infinity at all, as in a fully finite volume limit. 
\end{enumerate}

Most of this discussion applies equally well to situations in which the fastest shrinking curve $C_0$ is a genus-one fiber, rather than a $\mathbb P^1$ fiber, within $B_3$.
In this case, the solitonic string obtained by wrapping a D3-brane along $C_0$ is a fundamental Type II string \cite{Lee:2019apr,Lee:2019oct}, rather than a heterotic string. 
The physics of the limit is most easily determined by performing two T-dualities along the directions of the toroidal fiber. 
The T-duality maps
\bea
{\cal V}_{C_0}    &\longrightarrow&     \widehat{\cal V}_{C_0}   = {\cal V}^{-1}_{C_0}  \sim  \lambda^{3/2-x}    \,,   \\
g_{\rm IIB}       &\longrightarrow& \hat g_{\rm IIB}    =  {\cal V}^{-1}_{C_0}  \, g_{\rm IIB}    \sim  \lambda^{3/2-x}   \,
\eea
and the string associated with a D3-brane  wrapped on $C_0$ becomes an unwrapped D1-string. Its tower of string excitations sits at the mass scale
\be
\frac{M^2_{\rm D1}}{M^2_{\rm IIB}} \sim   \hat g^{-1}_{\rm IIB}   \sim \lambda^{-3/2+x}  \,. 
\ee
There are now two types of Kaluza-Klein states, associated with either the large T-dualised curve $C_0$  
or with the large divisors on $B_3$, whose volume does not change by the T-duality:
\be
\frac{M^2_{{\rm KK}; C_0}}{M^2_{\rm IIB}}  \sim  \lambda^{-3/2+x}     \,, \qquad   \frac{M^2_{{\rm KK}; B_3}}{M^2_{\rm IIB}}  \sim \lambda^{-3/2 - x}     \,.
\ee
This leaves us with the following behaviour:
\begin{enumerate}
\item
If $x > 0$, then the KK scale is set by $M^2_{{\rm KK}; B_3}$, and the system undergoes a decompactification limit because
\be
\frac{M^2_{{\rm KK}; B_3}}{M^2_{\rm D1}}   \sim \lambda^{-2 |x|} \to 0 \,.
\ee
\item
If $x=0$,  the D1-string scale and the KK scale agree parametrically and we enter the regime of a dual four-dimensional compactification with the role of the fundamental string played by the D1-string.
\item
If $x <0$, the KK scale is set by $M^2_{{\rm KK}; C_0}$, which again parametrically coincides with $M^2_{\rm D1}$. The pathological behaviour observed for the heterotic string in the regime $x<0$ is therefore avoided.
\end{enumerate}

\subsection{Classical infinite distance limits of \texorpdfstring{$J$}{J}-class B}   \label{subsec_JBclassical}

\begin{figure}
  \centering
  \begin{tikzpicture}
  \node at (0,0) {\includegraphics[width=0.5\textwidth]{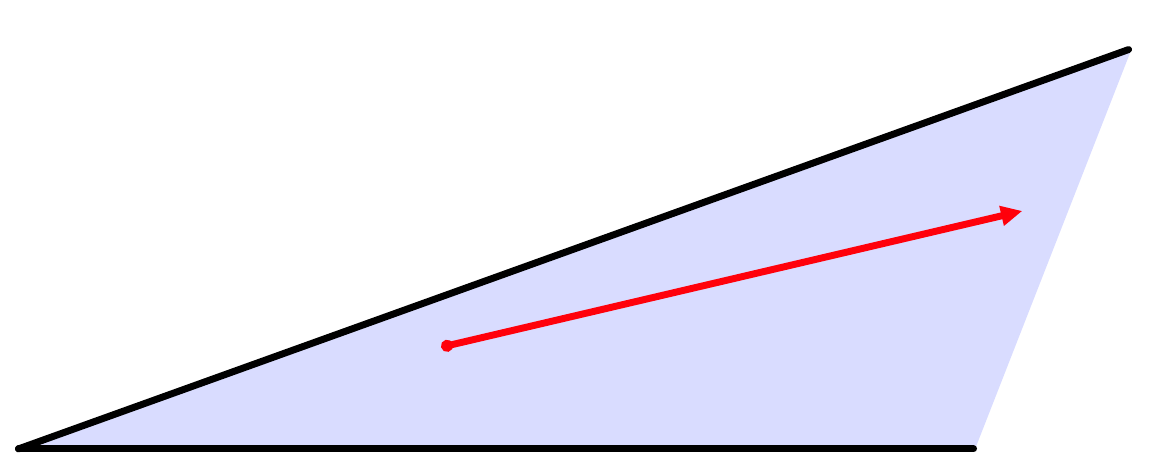}};
  \node at (4.4,1.3) {$\mathcal{I}_2$};
  \node at (3.25,-1.7) {$J_0$};
  \end{tikzpicture}
  \caption{In limits of $J$-class B all the remaining K\"ahler cone generators except for $J_0$ belong to the set $\cI _2$ as defined in~\cite{Lee:2019jan}. }
  \label{fig:JclassB}
\end{figure}

The second possibility for a classical finite volume limit at infinite distance corresponds to a 
\paragraph{\bf $J$-class B limit for $J'$  \cite{Lee:2019jan}:}
The K\"ahler form $J'$ can be parametrised as (Figure~\ref{fig:JclassB})
\be \label{Jprime-classB}
J' =  \lambda J_0 + \sum_{\mu \in {\cal I}_2} b^{\prime\mu} J_\mu  \,,    \qquad \text{with}\quad b^{\prime\mu} \precsim \lambda \,,
\ee
subject to the intersection properties 
\begin{equation}     \label{Jprime-classB-int}
\begin{split}   
J_0^3 &= 0 \,,   \cr
J_0^2 \cdot J_\mu &= 0   \qquad \forall \mu \in {\cal I}_2  \,,
\end{split}
\end{equation}
and such that for each $J_{\mu}$ there exists at least one  $J_{\nu}$ with $J_0 \cdot J_{\mu} \cdot J_{\nu} \neq 0$.

To obtain the full infinite distance limit we must again include the overall scaling factor $\mu$ in (\ref{JvsJp}), which is a priori independent of the parameter $\lambda$.
Before doing this, 
let us first investigate the properties of the finite volume limit (\ref{Jprime-classB}) for $J'$ itself, as this has not been fully analysed in \cite{Lee:2019jan}.

The Kaluza-Klein scale, measured with respect  to $J'$, is set by the curve with the largest volume, which is given by
\be
{\cal V}'_{C_{\rm max}} \sim   \lambda   \,.
\ee
Hence, ignoring for now the overall scaling $\mu$, we find
\be \label{Mkkprimed}
\frac{M^{\prime 2}_{{\rm KK};C_{\rm max}}}{M^2_{\rm IIB}} \sim \lambda^{-1}  \,. 
\ee
This serves as an upper bound for the Kaluza-Klein scale of the full limit including the overall sale factor $\mu \succsim 1$.

There are now two possibilities:
 A priori 
 in order for a limit (\ref{Jprime-classB}) to exist, the base $B_3$ need not admit a fibration by a (rational or genus-one) curve $C_0$.\footnote{As an example, $B_3$ can admit a fibration by a $\mathbb P^2$ surface over $\mathbb P^1$, with the volume of the base $\mathbb P^1$ scaling to infinity in the limit while the overall volume of $B_3$ remains finite.}
 In this case there do not arise any tensionless critical strings, and hence, the limit is 
 a decompactification limit as signalled by the light KK tower.
 The second possibility is that 
 there does exist a shrinking rational or genus-one fiber $C_0$, and hence, a nef divisor $D \in \overline{\cK(B_3)}$ such that
\bea\label{D2-B}
D^2 &=& n \, C_0 \,, \quad n\in \mathbb N \,, \\ \label{D3-B}
D^3 &=&0 \,. 
\eea 
If the volume of the generic fiber $C_0$ on $B_3$ scales as 
\be \label{VprimeC0B}
{\cal V}'_{C_{0}} \sim   \lambda^{-\gamma}     \,,\qquad\text{with}\quad \gamma  \geq 1   \,,
\ee
then the fundamental string obtained by wrapping a D3-brane along $C_0$ leads to a tower of states at or below the Kaluza-Klein scale (\ref{Mkkprimed}).\footnote{Note that eventually we need to take into account also the scaling behaviour of the overall parameter $\mu$. 
However, as we will see below, as long as $\mu \succsim 1$ the bound for the string tower to lie at or below the actual Kaluza-Klein scale continues to be given by $\gamma  \geq 1$.}
Alternatively, if  $\gamma < 1$, the limit corresponds to a decompactification limit because the Kaluza-Klein tower is the parametrically leading tower of states. In the latter case, no further work is required.

Let us therefore assume that a fibral curve $C_0$ with the scaling behaviour (\ref{VprimeC0B}) exists. We may further suppose that the fiber $C_0$ has its volume vanishing at a fastest rate, in case $B_3$ admits other fibrations whose fibers also shrink. Our first task is to show that such a fastest-shrinking curve $C_0$ is unique, to the extent that no other curve fiber $C_0'$ vanishes at the same rate as $C_0$. 
One can now expand the divisor $D$ in (\ref{D2-B}) in terms of the generators of the K\"ahler cone and deduce from  (\ref{VprimeC0B}) various constraints on the K\"ahler parameters (\ref{Jprime-classB}).
This analysis, which is performed in Appendix \ref{App_JclassB}, establishes that the parameter $\gamma \geq 1$ must in fact lie in the regime
\be  \label{gammaregime}
1 \leq \gamma \leq 2  \,.
\ee 
Furthermore, we show there that any other fibral curve $C_0'$, if it exists and shrinks in the limit, must necessarily shrink at a rate $\lambda^{-\gamma'}$ with  $\gamma' < \gamma$.
This establishes that the fundamental string which becomes light parametrically at the same rate as the Kaluza-Klein scale (or, in principle, faster than the Kaluza-Klein scale) is unique.

We are now in a position to combine these results about the limit for the rescaled K\"ahler form $J'$ with the overall scaling $\mu$. 
The full infinite distance limit takes the form
\be   \label{combinedJclassB}
J =  \mu (\lambda J_0 + \sum_{\mu \in {\cal I}_2} b^{\prime \mu} J_\mu) .
\ee
This leads to
\be
{\cal V}_{B_3} = \mu^3 \,, \qquad {\cal V}_{C_0} = \mu \lambda^{-\gamma} \,, \qquad   {\cal V}_{C_{\rm max}} = \mu \lambda  \,,
\ee
where we are assuming again the existence of a fibral curve $C_0$ as in (\ref{VprimeC0B}) and where 
$C_{\rm max}$ continues to denote the curve class with the maximal volume.

Consider first the case where $C_0$ is a rational fiber such that a D3-brane wrapping $C_0$ leads to an asymptotically light heterotic string.
The Kaluza-Klein scale $M_{\rm KK}$ and the heterotic string scale $M_{\rm het}$, set by ${\cal V}_{C_{\rm max}}$ and ${\cal V}_{C_0}$, respectively, compare as follows:
\be
\frac{M^2_{{\rm KK}}}{M^2_{\rm IIB}}    \sim (\mu \lambda)^{-1}   \,,  \qquad   \qquad
\frac{M^2_{\rm het}}{M^2_{\rm IIB}}  \sim   \mu \lambda^{-\gamma} \,.  
\ee
If we parametrise 
\be \label{muparalimitB}
\mu = \lambda^{\frac{1}{2} + x}   \,, \qquad \text{with}\quad x\geq -\frac12\,,
\ee
this leads to 
\be
M^2_{\rm het}  = M^2_{\rm KK}  \lambda^{2-\gamma + 2 x} \,.
\ee

There are now three different regimes for the parameter $x$:

\begin{enumerate}
\item
If $x > 0$,  we find that 
\bea
\frac{M^2_{\rm KK}}{M^2_{\rm het}} \sim \lambda^{\gamma -2 - 2 |x|}    \rightarrow 0  \,,  
\eea
where the asymptotic behaviour follows from our previous result that $\gamma \leq 2$, as asserted in (\ref{gammaregime}). Hence this regime corresponds to a decompactification limit.
\item
If $x=0$ and if there indeed exists a rational fibral curve $C_0$ whose scaling exponent $\gamma$ saturates the maximal possible value $\gamma=2$, then the infinite distance limit takes us to a dual heterotic frame with an emergent asymptotically weakly coupled heterotic string satisfying
\bea
M^2_{\rm het}  \sim M^2_{\rm KK}    \,.
\eea
This describes a dual weakly coupled effectively four-dimensional heterotic compactification.
If no such fibral curve with $\gamma=2$ exists, the limit is again a straightforward decompactification limit. 
For later purposes, let us rewrite the limit which does give rise to a weakly coupled heterotic string ($\gamma=2$) as
\be  \label{JclassBhetrelabelded} \boxed{
\text{$J$-class B, rescaled:}  \qquad J = \tilde \lambda J_0    + \sum_{\hat \mu \in \mathcal{\hat I}_{ 2}}   b^{\hat \mu} J_{\hat \mu}     +  \sum_{\check \mu \in \mathcal{\check I}_{ 2}}    \frac{\tilde b^{\check \mu}}{\tilde \lambda} J_{\check \mu}   \,,  \quad  \tilde\lambda \to \infty   \,. }   
\ee
The parameters $\frac{b^{\hat \mu}}{\tilde \lambda}$ and  $\tilde b^{\check \mu}$ remain finite in the limit and the generators satisfy the relations (\ref{Jprime-classB-int}). The $\mathbb P^1$-fiber class takes the form $C_0 = D \cdot D$ for $D$ given as a linear combination involving all $J_{\hat \mu}$ (this, in fact, defines the subset $\mathcal{\hat I}_{ 2}$) and possibly also $J_0$.  The remaining generators $J_{\check \mu}$ have the property that $D \cdot D \cdot J_{\check \mu} > 0$. For more details on the intersection numbers, see Appendix  \ref{App_JclassB}. Note the similarity to the limit (\ref{JclassAhetrelabelded}).

\item If $x < 0$, then
\bea
\frac{M^2_{\rm KK}}{M^2_{\rm het}} \sim \lambda^{\gamma -2 + 2 |x|}    \,.  
\eea
In principle, according to the purely classical analysis so far, this regime  could give rise to situations where $M^2_{\rm het}/M^2_{\rm KK} \to 0$ (for $\gamma > 2 -2 |x|$).\footnote{The minimal value for $x$ compatible with $\mu \succsim 1$ is $x=-\frac{1}{2}$. In this case $M^2_{\rm het}/M^2_{\rm KK}  \precsim 1$  requires $\gamma \geq 1$. This confirms that restricting to the regime $\gamma \geq 1$ in search for an emergent fundamental string whose excitations tower lies at or (in principle) below the Kaluza-Klein scale was indeed justified even in the full limit including the overall scaling $\mu$.}
However, we will show in the next section that this regime is out of computational control due to the appearance of perturbative $\alpha'$-corrections.
\end{enumerate}

So far we have assumed that $C_0$ is a rational curve such that the emergent string from a wrapped D3-brane is a fundamental heterotic string.
Let us briefly turn to the possibility that $C_0$ is a genus-one fiber, associated with an emergent Type II fundamental string.
As in the previous section, it is convenient to perform two T-dualities along the toroidal fiber $C_0$. After the T-dualities the fiber expands and the Type IIB string coupling is rescaled correspondingly:
\be
\widehat{\cal V}_{C_0} \sim \lambda^{-\frac{1}{2} - x  + \gamma}    \,, \qquad \quad \frac{1}{\hat g_{\rm IIB}} \sim   \lambda^{\frac{1}{2} + x - \gamma}   \,,
\ee
where we are using the parametrisation (\ref{muparalimitB}).
The new light string after the T-duality corresponds to the unwrapped D1-string, and the Kaluza-Klein scale is set by the volume of either the expanding curve ${C_0}$ or of the curve $C_{\rm max}$, whose volume scaling is unaffected by the T-duality.
As a result
\bea
\frac{M^2_{{\rm D1}}}{M^2_{\rm IIB}}    \sim   \lambda^{\frac{1}{2} + x - \gamma}    \,, \qquad \frac{M^2_{{\rm KK}}}{M^2_{\rm IIB}}    \sim   {\rm min}(\lambda^{\frac{1}{2} + x - \gamma} ,\lambda^{-\frac{3}{2} - x}    )   \,.
\eea

This leads to the following three regimes:
\begin{enumerate}
\item
If $x > 0$, then
\bea
\frac{M^2_{\rm KK}}{M^2_{\rm D1}} \sim \lambda^{\gamma -2  - 2 |x|}    \rightarrow 0  \,,  
\eea
resulting in a decompactification limit.
\item
If $x=0$, then $M^2_{\rm D1} \sim M^2_{\rm KK}$ if $\gamma=2$, leading to an effectively four-dimensional limit with the role of the fundamental string played by the D1-string. If $\gamma <2$, on the other hand, we encounter again a decompactification limit of the original Type IIB theory.
\item
If $x<0$, then  $M^2_{\rm KK}   \precsim M^2_{\rm D1}$, where the precise behaviour depends on the relation between $x$ and $\gamma$. Interestingly, this regime does no longer contain a pathological situation in which the D1-string tower would lie parametrically below the Kaluza-Klein tower.
\end{enumerate}

\section{Quantum corrected infinite distance limits in 4d \texorpdfstring{${N=1}$}{N=1} F-theory}   \label{sec_corr}

The analysis of the previous section has been purely classical on the F-theory side. At the same time, for compactifications with ${N}=1$ supersymmetry in four dimensions, quantum corrections are expected to arise.
The goal of this section is to investigate how such corrections modify the structure of infinite distance limits. 

In fact, we have already observed a conundrum that arises at the purely classical level: For certain ratios of the K\"ahler parameters we might enter a regime where the mass scale associated with a solitonic critical heterotic string becomes parametrically smaller than the KK scale. This would point to a new type of weakly coupled inherently four-dimensional string theory, in the sense that the higher-dimensional KK scale can be decoupled parametrically from the string scale in four dimensions. 
From all we know about critical string theory, such a phase of string theory is unlikely to exist.

We will see that this happens precisely in a regime  which is inflicted by parametrically large $\alpha'$-corrections which render the geometric analysis unreliable. 
This establishes consistency of the infinite distance limits. At the same time, no obstructions are found to the decompactification or emergent string limits as classified in (\ref{phases}).

\subsection{Overview}
Before we dive into the details let us briefly summarize the different classes of perturbative and non-perturbative corrections that can correct the four-dimensional effective action of $N=1$ compactifications of F-theory: 
\begin{enumerate}
    \item The four-dimensional K\"ahler potential $K^F$ receives perturbative $\alpha'$-corrections. We consider the form of corrections as calculated in \cite{Grimm:2013gma, Grimm:2013bha} and more recently refined in \cite{Weissenbacher:2019mef, Weissenbacher:2020cyf}.\footnote{Different approaches for determining $\alpha'$-corrections in F-theory have been developed in \cite{GarciaEtxebarria:2012zm,Minasian:2015bxa}.}
They are obtained by including the eight-derivative terms into the action $S_{11}$ of 11-dimensional supergravity. Subsequent dimensional reduction yields the $\alpha'$-corrected 3d $N=2$ effective action of M-theory on  a Calabi-Yau fourfold $Y_4$. The higher-derivative terms translate into corrections of the 3d $N=2$ K\"ahler potential and coordinates. Uplifting to F-theory leads to corrections of the corresponding four-dimensional $N=1$ quantities.    As anticipated above, we will see that these corrections reinstate consistency of the infinite distance limits.

    \item In addition there are non-perturbative corrections due to e.g. D3-instantons in F-theory. In the related context of four-dimensional $N=2$ compactifications of Type IIB string theory these play an important role in obstructing infinite distance limits in the hypermultiplet moduli space for which $M_{\rm KK}\succ M_\text{sol.}$ \cite{Marchesano:2019ifh, Baume:2019sry}. In certain cases these non-perturbative corrections to the moduli space geometry can give a non-zero quantum volume to curves that were classically required to shrink. While the $N=2$ analysis of \cite{Marchesano:2019ifh,Baume:2019sry} mainly considered D-instanton effects on the moduli space geometry, in our $N=1$ F-theory set-up the effect of D3-instantons on the K\"ahler potential is not under computational control. Luckily, we will see in Section \ref{sec_otherqc} that the effect of such instantons is sufficiently suppressed, at least for the infinite distance limits 
    leading to decompactification or to a weakly coupled emergent string.

    \item In four-dimensional $N=1$ theories non-perturbative effects such as D-brane instantons can also correct the superpotential $W$. In F-theory, D3-brane instantons wrapping divisors of the base $B_3$ of the Calabi-Yau fourfold $Y_4$ can contribute to the superpotential in case they exhibit a suitable amount of fermionic zero-modes. The F-theory superpotential gives rise to an F-term scalar potential $V_F$ such that in principle a diverging non-perturbative superpotential can yield a divergence in the scalar potential. As a consequence, the point in moduli space where the superpotential diverges cannot be reached dynamically. In the context of the swampland distance conjecture this has been analysed in \cite{Gonzalo:2018guu} and we will revisit this issue in the present context in Section \ref{sec_otherqc}. 
While we find no obstructions to the decompactification and emergent string limits, we provide additional evidence that the problematic limit of a heterotic string at a scale below the KK scale 
is dynamically shielded and one runs into a strong coupling transition at finite distance in moduli space before reaching the infinite distance limit.

\end{enumerate}

\subsection{\texorpdfstring{$\alpha'$}{alpha'}-corrections in the K\"ahler potential}    \label{sec_alphapcorr}

The first type of corrections which will play an important role are perturbative $\alpha'$-corrections to the K\"ahler potential that arise from higher curvature corrections in M-theory.
To begin, we first note that the infinite distance limits  should be reformulated as a genuine limit in the F-theory moduli space. Analogously to Type IIB orientifolds the four-dimensional ${N}=1$ chiral multiplet moduli space relevant for the discussion here is classically spanned by 
\begin{align}\label{DefModuli}
    \text{Re}\,T^{\rm cl.}_\alpha = \frac{1}{2} \int_{D_\alpha} J\wedge J\,, \qquad S=\frac{1}{g_s}+i C_0\,, 
\end{align}
where the $D_\alpha$ are a basis of divisors on the base $B_3$ of the elliptically fibered Calabi-Yau fourfold $Y_4$ and $J$ is, as before, the K\"ahler form on the base. 
We have also added the IIB axio-dilaton to the list of relevant chiral multiplets even though we do not consider parametric rescalings for this modulus in the following. 

Let us focus on perturbative $\alpha'$-corrections to these moduli. The leading ${\cal O}(\alpha^{\prime 2})$  corrections have been calculated  in \cite{Grimm:2013bha,Grimm:2013gma,Weissenbacher:2019mef} by dimensionally reducing higher derivative terms in the 11-dimensional M-theory effective action on a Calabi--Yau fourfold and subsequently lifting to F-theory.  For completeness we display the corrections in the M-theory effective action in Appendix \ref{App_Lalpha}. After the uplift to F-theory, the $\alpha'$-corrections affect both the four-dimensional K\"ahler potential and the K\"ahler coordinates \eqref{DefModuli}. The correction to the former is given by \cite{Weissenbacher:2019mef}
\begin{align}\label{KFdef}
	K^F = -2 \log (\cV_{B_3})      \
\end{align}
in terms of the quantum corrected base volume\footnote{
The corrections considered here are leading compared to further $\alpha'^3$-corrections to the K\"ahler potential, which in Type IIB orientifolds read $K=-2\log(\mathcal{V}_{B_3}+\xi)$ with $\xi = - \chi(X_3) \zeta(3)/(2(2 \pi)^3 g_s^{3/2})$~\cite{Becker:2002nn}. 
While these can be relevant at small $g_s$, we are here taking a parametric limit in the K\"ahler moduli without changing $g_s$ and hence the $(\alpha')^2$ corrections are the dominant corrections.}
\begin{align}\label{VolB3def}	
\cV_{B_3}  = 
{\mathcal V}_{B_3}^{(0)}  +\alpha^2\left( (\tilde \kappa_1+\tilde \kappa_2) \mathcal{Z} +\tilde \kappa_2 \mathcal{T} \right) \,,
\end{align}
and the K\"ahler coordinates are \cite{Weissenbacher:2019mef}
\begin{align}\label{ReTmain}
    \text{Re}\, T_\alpha= \frac{K_\alpha}{2} + \alpha^2\left((\kappa_3+\kappa_5)\frac{K_\alpha\mathcal{Z}}{2 {\mathcal V}_{B_3}^{(0)}} + \kappa_5 \frac{K_\alpha \mathcal{T}}{2 {\mathcal V}_{B_3}^{(0)}}+ \tilde{\mathcal{Z}_\alpha}  \log \mathcal{V}_{B_3}^{(0)}+\kappa_6 \mathcal{T}_\alpha + \kappa_7 {\cal Z}_\alpha \right)\,.
\end{align}
Here ${\mathcal V}_{B_3}^{(0)}$ is the classical volume of $B_3$ computed from the K\"ahler form $J = v^\alpha J_\alpha$ as ${{\mathcal V}_{B_3}^{(0)} = \frac{1}{6} \int_{B_3} J^3}$, and we define
\bea   \label{Kalphabeta}
K_{\alpha} = \int_{B_3} J_\alpha \wedge J \wedge J \,, \qquad K_{\alpha \beta} = \int_{B_3} J_\alpha \wedge J_\beta \wedge J   \,.
\eea 
The $\alpha'$-corrections are encoded in the quantities
\begin{align}\label{chidef}
    \mathcal{Z}_\alpha=\int_{Y_4} c_3(Y_4)\wedge \pi^\ast(J_\alpha) \,,\qquad \mathcal{Z}=  \mathcal{Z}_\alpha v^\alpha = \int_{Y_4} c_3(Y_4)\wedge \pi^\ast(J) \,,
\end{align}
and $\mathcal{T}_\alpha$, which is given, for smooth Weierstrass models, by
\begin{align} \label{Talpha}
 \mathcal{T}_\alpha &= -18(1+\alpha_2) \frac{1}{\text{Re}\,T_\alpha^\text{cl.}}\int_{D_\alpha} c_1(B_3) \wedge J \int_{D_\alpha} J_\alpha \wedge J\,,    \\
 \mathcal{T} &= \mathcal{T}_\alpha   v^\alpha   \,.
\end{align}
Furthermore 
\be
\begin{split}   \label{tildeZdef}
\tilde{\mathcal{Z}_\alpha}  &= \kappa'_{4,\alpha} \mathcal{Z}_\alpha    \qquad \quad (\text{no sum over} \, \alpha)   \cr
\tilde{\mathcal{Z}} &= \tilde{\mathcal{Z}_\alpha}   v^\alpha   
\end{split}
\ee
and
\begin{align}
    \alpha^2=\frac{1}{3^2 2^{13}} \,.
\end{align}
The parameters $\kappa_i$ and  $\tilde \kappa_i=\frac{3}{2}\kappa_i$ are a priori unfixed numerical constants which are inherited from the uncertainties of the K\"ahler potential and the K\"ahler coordinates in the underlying three-dimensional M-theory compactification. 
By comparison with the one-modulus case, \cite{Weissenbacher:2019mef} fixed 
\begin{align}
    \kappa_1+\kappa_2=512\,, 
    \qquad 
    4\kappa_3+4\kappa_5+\kappa_6=0
\end{align}
and furthermore comparison with the results of \cite{Grimm:2013gma} (in which $\mathcal{T}_\alpha$ is effectively set to zero) requires
\begin{align}\label{k3plusk5}
    \kappa_3+\kappa_5=768\,. 
\end{align}

On the other hand, $\kappa_{4,\alpha}'$, which appear in the four-dimensional K\"ahler coordinates (\ref{ReTmain}) in F-theory, may differ from their counterparts in M-theory through 1-loop corrections to the F-theory lift \cite{Grimm:2013gma,Weissenbacher:2019mef}. Compared to \cite{Weissenbacher:2019mef} we allow for these renormalised couplings, which appear in (\ref{ReTmain}) via (\ref{tildeZdef}), to depend on the specific K\"ahler coordinate $T_\alpha$ in question. This important point will be explained further in Section \ref{subsec_classical_het-F_duality}, where we will argue that some of these renormalised couplings $\kappa_{4,\alpha}'$ must in fact vanish by consistency with F-theory/heterotic duality.

The chiral multiplets are dual to linear multiplets whose scalar component is the real saxion $L^\alpha$. These are related to the $\text{Re}\,T_\alpha$ via 
\begin{align}
    L^\alpha=-\frac{\partial K^F}{\partial \text{Re}\,T_\alpha}=-\frac{\partial K^F}{\partial v^\beta}\frac{\partial v^\beta}{\partial \text{Re}\, T_\alpha} =-\frac{\partial K^F}{\partial v^\beta}   K^{\alpha \beta} \,,
\end{align}
where $K^{\alpha \beta}$ is the inverse of the matrix $K_{\alpha \beta}$ defined in (\ref{Kalphabeta}).
Using (\ref{KFdef}) and (\ref{ReTmain}), we find
\begin{align}
\label{CorrectedLinearMultiplets}
\nonumber L^\alpha=\frac{v^\alpha}{ \mathcal{V}_{B_3}^{(0)}} + &\frac{\alpha^2}{ \mathcal{V}_{B_3}^{(0)}} \Bigg(\left(2\tilde \kappa_2 v^\gamma\partial_\beta \mathcal{T}_\gamma-\kappa_6 v^\gamma \partial_\gamma \mathcal{T}_\beta-3\kappa_5 v^\gamma\partial_\beta \mathcal{T}_\gamma
    \right)\mathcal{K}^{\beta \alpha}\\
    &+ \left(2\tilde \kappa_2-3\kappa_5\right)\mathcal{K}^{\beta \alpha} \mathcal{T}_\beta+\left(2(\tilde \kappa_1+\tilde \kappa_2)-3(\kappa_3+\kappa_5)\right) \mathcal{K}^{\beta \alpha}\mathcal{Z}_\beta \\
     & -\frac{v^\alpha }{\mathcal{V}_{B_3}^{(0)}}\left((\tilde \kappa_1+\tilde\kappa_2)-\frac{\kappa_3+\kappa_5}{2}\right)\mathcal{Z}-\frac{1}{2} \frac{v^\alpha }{\mathcal{V}_{B_3}^{(0)}}  \tilde{\mathcal{Z}} -\frac{v^\alpha }{\mathcal{V}_{B_3}^{(0)}}\left(\tilde \kappa_2-\frac{\kappa_5}{2}\right)\mathcal{T}\Bigg)\,. \nonumber 
\end{align}
Note that the $\alpha'$-corrections depend on $K^{\alpha \beta}$, which is  the inverse of $K_{\alpha \beta}$ defined in (\ref{Kalphabeta}), i.e. $K^{\alpha \beta}K_{\beta \gamma} = \delta^\alpha_\gamma$.
The geometric meaning of the linear multiplets $L^\a$ is that they parametrise the $\alpha'$-corrected volumes of the curves $C^\alpha$ dual to the divisors $D_\alpha$ via
\begin{equation}   \label{VCalphauantum}
{\cal V}_{C^\alpha}=\mathcal{V}_{B_3} \, L^\alpha  \,,  
\end{equation}
where the $\alpha'$-corrected volume $\mathcal{V}_{B_3}$  is given in (\ref{VolB3def}).

The above corrections at order $(\alpha')^2$  are the first of an infinite series of perturbative corrections.
Perturbative control requires that the correction terms are suppressed with respect to the purely classical contributions. 
This is equivalent to demanding that
\bea \label{cZcond1}
\frac{{\cal Z}}{\mathcal{V}^{(0)}_{B_3}}     \stackrel{!}{\ll} 1    \,,  \qquad \quad  \frac{\cT}{\mathcal{V}^{(0)}_{B_3}}   \stackrel{!}{\ll}  1     \,,
\eea
which in the cases considered here also ensures
\bea
  \frac{{\cal Z}_\alpha}{K_\alpha} \ll 1 \,,\qquad \quad \frac{\cT_\alpha}{K_\alpha} \ll 1\,,\qquad \forall \alpha\,.
\eea
Finally in the models that we consider the logarithmic correction to $\Re\, T_\alpha$ only appears for $\alpha\in \mathcal{I}_1 (\check{\mathcal{I}}_2)$ for limits of $J$-class A (B) as we will discuss around \eqref{kappa4rel} and therefore never spoils perturbative control in our limits.

Let us examine under which conditions the infinite distance limits analysed in the previous section satisfy this constraint.
For both types of limits (\ref{combinedJclassA}) and (\ref{combinedJclassB}) we see that 
\bea\label{chiJscaling}
\frac{{\cal Z}}{\mathcal{V}^{(0)}_{B_3}}   &\sim&  \frac{ \mu \lambda \, {\cal Z}_0 + \ldots}{\mu^3}    \sim     {\cal Z}_0  \, \lambda^{-2x}  + \ldots    \\
\frac{{\cal T}}{\mathcal{V}^{(0)}_{B_3}}   &\sim&  \frac{ \mu \lambda \, {\cal T}_0 + \ldots}{\mu^3}    \sim     {\cal T}_0  \, \lambda^{-2x}  + \ldots
\eea
where the omitted terms scale at strictly subleading order in $\lambda$ (for the $J$-class A limits (\ref{combinedJclassA})) or at most at the same order (for the $J$-class B limits (\ref{combinedJclassB})).
In (\ref{chiJscaling}) the index $\alpha =0$ refers to the K\"ahler class $J_0$ as defined in the respective limits and we recall that the parameter $x$ is defined via $\mu = \lambda^{1/2 + x}$.\footnote{Here, we assumed that $v^0$ is the only classical K\"ahler modulus that grows as $v^0\sim \lambda$ in our limit. However, in principle also the moduli $v^\alpha$ with $\alpha\in \mathcal{I}_3(\hat{\mathcal{I}}_2)$ for $J$-class A (B) are classically allowed to have the same $\lambda$ scaling as $v^0$. Since all topological data related to the corresponding $J_\alpha$ can be re-expressed in terms of those for $J_0$ it is enough to consider the quantities associated to the latter K\"ahler cone generator. So here and in the rest of this section we restrict to discussing $\mathcal{Z}_0$ and $\cT_0$.}

We can thus single out the topological invariant $\mathcal{Z}_0$ together with $\cT_0$ to indicate whether a chosen infinite distance limit exists or not. \\

\noindent {\bf Case 1:} ${\cal T}_0 \neq 0$ or ${\cal Z}_0 \neq 0$ \\

\noindent Perturbative control is lost at order $(\alpha')^2$ for infinite distance limits in which the parameter $x <0$
because (\ref{cZcond1}) is violated. 
Importantly, this excludes the pathological limits discussed in Section \ref{sec_Uniqueness} in which, at the classical level, a tensionless heterotic string seems to emerge whose mass scale is parametrically below the KK scale. Including $\alpha'$-corrections is therefore crucial to understand the asymptotic physics of the infinite distance limits and establishes its consistency.
Note that if ${\cal T}_0 \neq 0$ or ${\cal Z}_0 \neq 0$, this would likewise exclude the limits with $x < 0$ in which a Type II string becomes tensionless, even though in this case no inconsistency arises.
This in particular excludes infinite distance limits in the regime where $\cV_{B_3}$ is finite (where $x=-\frac{1}{2}$).\footnote{This is also in agreement with the analysis of \cite{Cicoli:2018tcq} of certain Large Volume Compactifications \cite{Balasubramanian:2005zx}, in which $\cV_{B_3}$ is stabilised at finite volume and where an interplay between the K\"ahler cone structure and potential $(\alpha')^2$ corrections renders the remaining moduli space compact.}

Before we provide additional evidence for this breakdown of the effective theory, let us also consider limits in different regimes of the parameter $x$.
For $x>0$ the perturbative corrections are manifestly parametrically suppressed in $\lambda$. According to the classification of the previous section, in this parameter regime  
the theory asymptotically decompactifies, at least partially, due to the parametrically leading tower of KK states. 

For $x=0$, on the other hand, the $\alpha^{\prime 2}$ corrections are not parametrically suppressed with $\lambda$. Their strength crucially depends on the chosen reference values for the K\"ahler parameters $v^\alpha$ that define the trajectory. To see this, we first observe that for $x=0$ limits as given in (\ref{JclassAhetrelabelded}) and (\ref{JclassBhetrelabelded}), respectively,   the F-theory K\"ahler moduli classically behave as
\begin{align}   \label{ReTdisc}
    \text{Re}\, T^\text{cl.}_0 \sim \text{const.}\,,\qquad  \text{Re}\,T ^\text{cl.}_\mu \sim \tilde\lambda^2 \,,\qquad \mu\in \mathcal{I}_1 \, \, \text {or } \, {\mathcal{I}}_{\check 2}\,. 
\end{align}
Looking at the corrections to the F-theory K\"ahler coordinate
\begin{align}
    \text{Re}\,T_0\simeq \text{Re}\,T_0^\text{cl.} \left(1+\frac{1}{96}\frac{2\cW_0}{ \text{Re}\,T_0^\text{cl.}}\right)+\kappa_6\mathcal{Z}_0+\kappa_7 \mathcal{T}_0= \text{Re}\,T_0^\text{cl.}+ \text{const.}\,,
\end{align}
where
\bea \label{W0def}
  \cW_0\equiv \mathcal{Z}_0 +\frac{\kappa_5}{768} \cT_0  \,,
\eea 
we observe that for the correction to be negligible we should choose $\text{Re}\,T_0^\text{cl.} \gg 1$ appropriately, which then translates into a condition for the reference values of the K\"ahler parameters $v^\alpha$. 

To understand the significance of this consider, for definiteness, a limit on a rationally fibered base.
As will be discussed in greater detail in Section \ref{subsec_classical_het-F_duality}, a suitable linear combination of the $\text{Re}\,T_\mu$ is dual to the heterotic dilaton $\text{Re}\,S$, while $\text{Re}\,T_0$ corresponds to a K\"ahler modulus of the heterotic base $B_2$ that enters quadratically in the volume form of $B_2$. Hence, the classical $x=0$ limit on the heterotic side corresponds to a weak coupling limit while keeping the volume of the base $B_2$ constant. In other words, our limit crucially depends on the point in heterotic K\"ahler moduli space that we choose as our starting point.

If we choose $\text{Re}\,T_0^\text{cl.}$ and thus the heterotic base K\"ahler modulus too small, we encounter the same problem as in the regime where $x<0$ case: On the F-theory side we cannot trust the geometric description due to the large $\alpha'$-corrections. On the heterotic side, in the limit of small K\"ahler moduli we also expect $\mathcal{O}(\alpha'^2)$ corrections as they have  been calculated in e.g. \cite{Anguelova:2010ed}. For $x=0$ we see that we should choose the heterotic base volume large enough such as we are in the regime where we can trust the classical heterotic/F-theory duality without having to worry about the $\alpha'$-corrections that might spoil the weak-coupling duality.

Let us now come back to the fate of the infinite distance limits in the regime $x <0$, focussing again on the heterotic limits by taking $B_3$ to be rationally fibered.
While the appearance of uncontrolled $\alpha'$-corrections is already sufficient to resolve the conundrum proposed by these classical infinite distance limits, it is desirable to better understand the physics as one enters the regime of large $\lambda$ for $x<0$. 
Answering this question requires full control of all perturbative and non-perturbative corrections, but we can nonetheless provide a hint 
already on the basis of the ${\cal O}(\alpha'^2)$ results for negative values of the combination $\cW_0$ defined in (\ref{W0def})
which appears in \eqref{ReTmain}. 

To this end note that classically the heterotic string coupling corresponds in F-theory to the volume of a section (i.e. of a copy of the base $B_2$) of the rational fibration.
This well-known relation will be reviewed in more detail in Section \ref{subsec_classical_het-F_duality}.
 For $\cW_0<0$, \eqref{ReTmain} shows that the volume $\text{Re}\, T_{B_2}$ vanishes at this order in $\alpha'$ along the trajectory once 
\be
\lambda_\text{max} \sim -\frac{96 \mathcal{V}^{(0)}_{B_3}}{\cW_0} > 0
\ee
is reached. Hence, instead of a weakly-coupled heterotic string at \textit{infinite} distance, at ${\cal O}\left(\alpha^{\prime 2}\right)$ we observe a strong coupling singularity for the solitonic string obtained by wrapping a D3-brane on the $\mathbb{P}^1$-fiber at \text{finite} distance in the F-theory moduli space. At the same time at this order in $\alpha'$ the corrected K\"ahler potential 
\be\label{Kahlerpot}
K^F=-2 \log {\cal V}_{B_3}\,,
\ee 
with ${\cal V}_{B_3}$ as in \eqref{VolB3def}, diverges in the vicinity of $\lambda_\text{max}$. Even though for a definite statement higher-order and non-perturbative corrections should be taken into account, this observation suggests that for $x<0$ and $\cW_0<0$ there is no tensionless and weakly-coupled heterotic string in the spectrum that we expected classically. In particular along the classical trajectory we are leaving the geometric F-theory phase around $\lambda_\text{max}$ and beyond that point the fate of the solitonic string, which is classically dual to the critical heterotic string, is unclear as we enter a strong-coupling phase for its associated string coupling. We thus see that for $\cW_0<0$ and $x<0$ the classical emergent string limit does not persist at the quantum level. 
We will provide more evidence for this picture in Section \ref{sec_otherqc} by analysing also the non-perturbative superpotential.

In Appendix \ref{app_T0Z0} we study examples of classes of geometries for which $\mathcal{W}_0<0$ is indeed fulfilled. In particular, this is the case in $J$-class B limits for a smooth Weierstrass model over a projective, smooth, and almost Fano base $B_3$. For more general geometries and in particular $J$-class A limits we cannot fix the overall sign of ${\cal W}_0$. One reason is that ${\cal T}_0$ depends on unknown parameters $\alpha_2$ and $\kappa_5$.  Still, for generic models with a rational fibration for $B_3$ that we consider in Appendix \ref{app_T0Z0} at least one of the quantities $\mathcal{Z}_0$ and $\mathcal{T}_0$ is non-vanishing. \\

\noindent {\bf  Case 2:} ${\cal T}_0 = 0$ and ${\cal Z}_0 = 0$ \\

\noindent  
If $B_3$ is a rational fibration,
then ${\cal T}_0={\cal Z}_0 = 0$ occurs at best in highly non-generic models (even though we cannot provide an example). 
In this case there would be no obstruction to taking even a problematic limit in the regime $x <0$, at leading order in $\lambda$. But at the same time there exist other 4-cycles $J_\alpha$,  $\alpha\ne 0$, in these non-generic models whose volume $\text{Re} \,T_\alpha$ vanishes in the classical infinite distance limit. By the same arguments, the vanishing of such four-cycles is then obstructed 
if the analogous combination
$\mathcal{W}_\alpha$ (rather than (\ref{W0def})), is non-zero. As a consequence, unless all $\mathcal{W}_\alpha$ vanish\footnote{Based on genericity arguments we conjecture that this cannot happen, but it would be interesting to prove or disprove this rigorously.}, the classical infinite distance limit is still obstructed at $\mathcal{O}(\alpha'^2)$ even though this obstruction does not occur at leading order in $\lambda$.

The situation is quite different if $B_3$ is elliptically fibered.  Here, we have ${\cal T}_0 =0$ and ${\cal Z}_0 =0$ already for every smooth Weierstrass model over $B_3$, as also discussed in  Appendix \ref{app_T0Z0}, and in fact ${\cal T}_\alpha =0$ and ${\cal Z}_\alpha =0$ for all vanishing divisors. Thus at least for these simple models, the characteristic integrals controlling the $\mathcal{O}\left(\alpha^{\prime 2}\right)$ corrections vanish. Therefore, at this order in the perturbative corrections there is no obstruction to taking the small fiber limit for any value of $x\ge -1/2$. Recall that the T-dual theory has a solitonic D1-string probing the T-dual base $\hat{B}_3$ in the large volume limit. In this duality frame, we do not expect any large $\alpha'$-corrections to occur. The analysis of the previous section showed that for $x<0$ the KK-scale is set by the modes on the torus fiber and parametrically coincides with the tension of the solitonic string. Hence, the absence of a quantum obstruction at the perturbative level to the limit with $x<0$ in the case of a $T^2$ fibration is not in contradiction with the general expectation that there should not be a genuinely four-dimensional critical string. 

Though the corrections at the order $\alpha'^2$ vanish, for the classical limit to persist even for $x<0$ one should also consider the higher order corrections that will scale as
\begin{align}\label{higherorder}
    \mathcal{O}(\alpha'^{2n})\sim \frac{1}{\left(\text{Re}\,T_0^\text{cl.}\right)^n}\sim \left(\frac{\mu^2}{\lambda}\right)^{\!-n}= \lambda^{-2nx}\,. 
\end{align}
If these vanish for all $n$ the classical limit indeed survives for $x<0$. It would be interesting to investigate this 
for the smooth Weierstrass models over an elliptically-fibered base $B_3$, for which the $\alpha'$-corrections are not required to establish consistency.

\subsection{Non-perturbative obstructions}\label{sec_otherqc}

So far we have focused on perturbative $\alpha'$-corrections to the K\"ahler potential and observed that in the case of a rational fibration they obstruct the classical F-theory limits with $x<0$. 
We have also seen that limits with $x=0$ are marginally allowed. However, since the fiber of the F-theory base $B_3$  shrinks to zero size also in such limits, the question about obstructions due to non-perturbative effects remains. This has both a kinematical and a dynamical aspect.

In this section we will argue that there appear no non-perturbative obstructions to taking an infinite distance limit corresponding to either a decompactification limit (the regime $x > 0$ in the notation Section \ref{sec_Uniqueness})
or to an emergent string limit (the regime $x=0$). 
The pathological limit $x<0$ (for rationally fibered base), on the other hand, in which we already lose  perturbative control, is also inflicted with non-perturbative corrections, and at least in certain cases we 
are able to interpret these as triggering a phase transition at finite distance in moduli space to a theory without a heterotic string.

Let us begin with potential kinematical obstructions.
From experience with dual Type IIA compactifications on Calabi-Yau spaces one might be worried if quantum corrections to the two-cycle volume as computed by mirror symmetry  could obstruct
a vanishing cycle volume limit. We will give two reasons why such corrections do not affect the analysis of our F-theory infinite distance limits.

To see this, compactify Type IIA string theory on the same Calabi-Yau fourfold $Y_4$ to two dimensions.
This theory is dual to the four-dimensional ${N}=1$ supersymmetric F-theory model on $Y_4$, compactified down to two dimensions on an additional $T^2$.
In the two-dimensional Type IIA setting non-perturbative effects due to worldsheet instantons are well-known to modify the notion of two-cycle volumes.
At the quantum level, it is the $\sigma$-model volume rather than the classical volume which governs the BPS masses of wrapped objects.
If the point of vanishing $\sigma$-model volume is not part of the full moduli space, this signals a quantum obstruction to contracting the cycle in the two-dimensional theory.

One might therefore wonder if in the related Type IIA compactification on $Y_4$ 
 the shrinking of the $\mathbb{P}^1$-fiber is obstructed in this manner.
The $\sigma$-model volume of a curve $C\in H_2(Y_4)$ can be calculated using fourfold mirror symmetry \cite{Mayr:1996sh}.
In Appendix \ref{app_QV} we study the resulting expressions for the Type IIA $\sigma$-model volume of the $\mathbb{P}^1$-fiber $C_0$ of $B_3$
in a prototypical example. 
We will see that the quantum corrections are exponentially suppressed with the volume of curves on the base $B_2$ of $B_3$, which tend to infinity 
in the limit as the volume of the $\mathbb{P}^1$-fiber shrinks.
Specifically we will find for the $\sigma$-model volume in the infinite distance limit of large $\mu$ and $\lambda$ that 
\bea
 \cV_{\rm IIA}(C_0) \sim \frac{\mu}{\lambda^2} + e^{-2 \pi \mu \lambda}    \,.
\eea
The fact that the non-perturbative corrections are exponentially suppressed is a special property of the curves which shrink at infinite distance in the moduli space.
Similar conclusions have already been obtained for infinite distance limits in Type IIA string theory on Calabi-Yau threefolds in \cite{Lee:2019oct}.

The upshot is that in the related two-dimensional Type IIA context, the shrinking of the rational fiber is not obstructed by worldsheet instanton corrections.
Even if we had found such an obstruction, this would still not prevent a vanishing tension limit for the wrapped string in the four-dimensional 
F-theory. The reason is that the worldsheet instantons modifying the classical volume in Type IIA are dual to the effect of D3-branes wrapping a curve on $B_3$ times the $T^2$ factor.
In the limit of infinite torus volume, corresponding to F-theory in four dimensions, this effect would be suppressed.

By contrast, D3-brane instantons wrapping four-cycles, rather than curves, on $B_3$ are known to correct both the K\"ahler potential and superpotential of the F-theory effective action.
The effect of an instanton wrapping a four-cycle $D_\alpha$ is exponentially suppressed by
\bea
e^{-S_{\rm inst}, \alpha} \sim e^{-{\rm Re}T_\alpha}   \,.
\eea
Depending on the zero-mode structure instantons may correct either the K\"ahler potential or lead to a superpotential of the form 
\begin{align}\label{superpot}
    W= \sum_{N,\alpha} A_N\; e^{-k_N^\alpha T_\alpha}\,,
\end{align}
where $A_N$ are moduli-dependent coefficient.
The question whether a significant contribution to the superpotential comes from the D3-instantons thus depends on the number of fermionic zero modes as well as the scaling of the instanton action. The first condition is rather difficult to answer in general and so in the following we will ask what happens \textit{if} D3-instantons with exactly the right number of fermionic zero modes exist. 
This is, in fact, expected in theories with genuine ${N}=1$ supersymmetry \cite{Palti:2020qlc}.

The second condition depends on the precise scaling of the K\"ahler moduli in \eqref{combinedJclassA}.
Let us consider the two different types of fibration separately again:

\subsubsection*{Rational Fibration}
In a decompactification limit for a rational fibration, i.e. for limits in the regime $x>0$, all D3-brane instanton contributions to the effective action are suppressed since 
\begin{align}
    \text{Re}\, T_\alpha \succsim \frac{\mu^2}{\lambda} = \lambda^{2x}\rightarrow \infty \, \qquad \forall \alpha\,.
\end{align}
This mirrors, in the dual heterotic frame, the parametric suppression of all non-perturbative corrections 
 in the large volume, weak-coupling limit of the heterotic string theory. 

For $x=0$, this is no longer the case as the classical action of  D3-instantons wrapping vertical divisors of $B_3$  scales as
\begin{align}
    \text{Re}\, T_\text{vert.} \sim \text{const.}\,,
\end{align}
which is sensitive to the $\alpha'$-corrections to  $\text{Re}\, T_\alpha$ discussed in Section \ref{sec_alphapcorr}. These D3-instantons map to heterotic world-sheet instantons wrapping 2-cycles in the heterotic base $B_2$. For $x=0$ the volume of these 2-cycles remains constant along the trajectory along which we take the infinite distance limit. 
The amount of suppression of these instantons depends on the starting point from where in moduli space we take the infinite distance limit, similar to the discussion around (\ref{ReTdisc}), rather than
on the limit as such. For a suitable choice of this starting point we can safely exclude a kinematical obstruction to the infinite distance limit due to such instantons.
What is more, the contribution of D3-instantons wrapping $B_2$, i.e. heterotic space-time instantons, is even exponentially suppressed with the inverse heterotic coupling
and hence negligible in the limit.

In particular, it is clear that there cannot be a dynamical obstruction of the infinite distance limit due to D3-brane instantons in the regime where $x=0$. To see this consider the F-term potential that derives from the superpotential
(\ref{superpot})
\begin{align}
    V_F=e^{K}\left[K^{\alpha \bar \beta} D_{T_\alpha} W \overline{D_{T_\beta} W} - 3 |W|^2\right]\,. 
\end{align}
In the limit with $x=0$, this term is suppressed due to the $e^K$ factor. Indeed, using \eqref{Kahlerpot}, $V_F$ asymptotically scales as 
\be\label{scalingVF}
V_F=\cV^{-1}_{B_3}\left[\text{const.}\right]\sim \mu^{-3} \sim \lambda^{-3/2}\rightarrow 0\,. \
\ee
Thus, asymptotically the scalar potential vanishes for precisely those infinite distance limits that are kinematically allowed and thus correspond to weak coupling limits also at the quantum level. This is in agreement with the general expectation that close to any infinite distance point the scalar potential has to vanish asymptotically as argued for in \cite{Ooguri:2018wrx}.

This should be contrasted with the situation for $x<0$ where we already saw that the geometric description breaks down at finite distance due to perturbative $\alpha'$-corrections. In this case we do not necessarily expect the scalar potential to vanish asymptotically as the classical infinite distance/weak coupling limit does not persist at the quantum level. At least for $\cW_0<0$, with $\cW_0$ given in \eqref{W0def}, we already argued that we enter a strong-coupling phase for the solitonic string obtained by wrapping a D3-brane on the $\bbP^1$-fiber of $B_3$. Since $\R\,T_\alpha$ also controls the D3-brane instanton action, the non-perturbative contribution to the superpotential \eqref{superpot} as well as all contributions to the K\"ahler potential are likewise affected by the quantum corrections. In particular for $\cW_0<0$ the correction is such that the action of \textit{all} D3-instantons becomes small at this order in $\alpha'$. This includes D3-instantons wrapping the base $B_2$ which are irrelevant in the limits with $x\ge 0$.  Interestingly these D3-instantons are magnetically dual to the D3-brane wrapped on the rational fiber, i.e. the heterotic string. Thus, instead of a limit where the solitonic string becomes tensionless, we enter a region where, at this order in $\alpha'$, the tension of the string remains finite but its magnetic dual -- the D3-instanton on $B_2$ -- acquires a vanishing action. In this regime all D3-instantons (with suitable fermionic zero-modes) contribute at $\mathcal{O}(1)$ to the superpotential:
\begin{align}
    W= \sum_{N,\alpha} A_N \exp\left\{-k_N^\alpha\left[ \text{Re}\,T_\alpha+ \text{Im}T_\alpha\right]\right\}\rightarrow \sum_{N,\alpha} A_N \,\mathcal{O}(1)\,,
\end{align}
where $\Re\,T_\alpha$ is given in \eqref{ReTmain}.

If $A_N\ne 0$, which for genuine $N=1$ models should be true for at least a subset of instantons  \cite{Palti:2020qlc}, a convergence of the superpotential in the region where the perturbative expansion breaks down cannot be guaranteed due to the small instanton action. The divergence in the superpotential might dynamically shield the transition point, but this point lies at finite distance corresponding to a strong coupling limit for the heterotic string instead of an infinite distance limit with a weakly coupled heterotic string. 

This observation can be better understood in simpler models that are related to six-dimensional $N=(1,0)$ (or four-dimensional $N=2$) F-theory models. Therefore consider F-theory on a Calabi-Yau threefold that is a smooth elliptic fibration over the Hirzebruch surface $\mathbb{F}_1$. The Mori cone generators of $\mathbb{F}_1$, the base $h$ and the fiber $f$, satisfy
\begin{align}
    h\cdot f=1\,,\qquad h\cdot h=-1\,,\qquad f\cdot f=0\,. 
\end{align}
The spectrum of solitonic strings in this theory consists of a critical heterotic string obtained by wrapping a D3-brane on $f$ (cf. \cite{Lee:2018urn}) and an E-string obtained by wrapping a D3-brane on $h$. As realised long ago by Witten \cite{Witten:1996qb} shrinking the F-theory base $h$ gives rise to a strong coupling singularity for the heterotic string. Furthermore, after shrinking $h\subset \bbF_1$ we enter a new phase corresponding to F-theory on $T^2\rightarrow \bbP^2$ which clearly has no heterotic string in its string spectrum. During this process the tension of the heterotic string remains constant. We thus see that strong coupling singularities associated to a vanishing tension of the E-string in this particular setup can give rise to a phase transition which removes the heterotic string from the spectrum.

We can now compactify this set-up on a torus to obtain a four-dimensional $N=2$ theory. When wrapped on the $T^2$ the six-dimensional E-string yields D3-instantons in four dimensions. As shown in \cite{Berglund:2005dm} if supersymmetry is further broken to $N=1$, corresponding to F-theory with rationally fibered base $B_3$, these D3-instantons generate a superpotential of the form 
\begin{align}\label{eq:superpotEstring}
    W=f(G,z)\times \left( Q \frac{E_4(q_0)}{\eta(q_0)^{12}} +\cO(Q^2)\right) \,.
\end{align}
Here, $Q=e^{-2\pi T_b}$ is the exponentiated complexified volume of the base $B_2$ dual to the heterotic dilaton and $q_0=e^{-2\pi T_0}$ is the exponentiated complexified volume of some vertical divisor of $B_3$. The pre-factor $f(G,z)$ depends on the choice of fluxes $G$ and the other geometric moduli $z$ of the system which we ignore here. The $q_0$-dependent contribution in \eqref{eq:superpotEstring} can be interpreted as a perturbative correction to the D3-instanton action \cite{Berglund:2005dm} and the origin of these D3-instantons in the E-string is reflected by the appearance of the E-string partition function.

Expanding the $\eta$-function for large $T_0$ we observe that the leading contribution to the superpotential due to the D3-instanton from the E-string behaves as 
\be\label{eq:Wleading}
W\sim f(G,z) \times e^{-2\pi (T_b - \frac{1}{2}T_0)} + \dots \,.
\ee
Accordingly, if $T_b$ is finite the effective D3-instanton action vanishes for some finite $\Re T_0$ and the convergence of the superpotential is not guaranteed any more. This is the four-dimensional avatar of the six-dimensional strong coupling singularity discussed above. 

To understand what happens at this strong coupling singularity in four dimensions, let us go back to the $N=2$ parent theory. In addition to the duality to the heterotic string used so far, F-theory compactified on $(T^2\rightarrow \bbF_1)\times T^2$ is also dual to Type IIA string theory compactified on $T^2\rightarrow \bbF_1$. In terms of heterotic variables, the volume $t_b$ of the base $\bbP^1$ of $\bbF_1$ is given by
\begin{align}
    t_b=S-\frac{1}{2} U -\frac{1}{2} T\,,
\end{align}
where $T$ and $U$ are the standard moduli of the heterotic torus. This implies  that the locus
\begin{align}
    t_b=0 \,,\qquad \text{i.e.} \qquad S=\frac{1}{2}\left(T+U\right)\,,
    \end{align}
 is the Type IIA analogue of the strong coupling singularity encountered around \eqref{eq:Wleading} upon identifying $(T+U)\leftrightarrow T_0$. Similarly to the six-dimensional F-theory model, blowing down the base of $\mathbb{F}_1$ allows us to enter the phase where Type IIA string theory is compactified on the Calabi-Yau threefold $T^2\rightarrow \mathbb{P}^2$. Since this threefold does not have a K3-fibration, the new phase of the four-dimensional theory does not have a heterotic string in its string spectrum. 

To sum up, in this simple example (of a limit  in the regime $x <0$) we have seen that 
\begin{enumerate}
    \item due to quantum corrections the D3-instanton action vanishes along a finite distance locus which can give rise to unsuppressed non-perturbative contributions to the superpotential,
    \item the vanishing of the instanton action is associated to a strong coupling singularity of the heterotic string,
    \item at least in the six-dimensional $N=(1,0)$ and the four-dimensional $N=2$ parent theories this strong coupling singularity is associated to a phase transition which (among other things) removes the heterotic string from the spectrum. 
\end{enumerate}

However, the superpotential does not diverge at every strong coupling singularity. A simple example is obtained by replacing $\bbF_1\rightarrow \bbF_2$. In this case, the six-dimensional string that is associated to the strong coupling singularity is not an E-string but a string associated to an $A_1$ orbifold singularity. This string in fact enjoys higher supersymmetry as it is already present in Type IIB models \cite{Witten:1996qb}. As was shown in \cite{Berglund:2005dm} after compactifying further to four dimensions and breaking to $N=1$ supersymmetry, the coefficient of the superpotential term similar to \eqref{eq:superpotEstring} vanishes identically such that the D3-instantons associated to the non-critical $A_1$-string do not contribute to the superpotential. 

It would be very interesting to see how the observations in these simple examples generalize to the more general case of infinite distance limits in the regime $x <0$. For us, the important message is that the region where perturbative control is lost can in certain cases be further obstructed by a diverging superpotential and that the solitonic heterotic string is not even necessarily stable along trajectories with $x<0$.

This is in contrast to the regime where $x\ge0$, for which we saw that the infinite distance limit is not obstructed by a non-perturbative superpotential.

\subsubsection*{Genus-one fibration}
If we consider a genus-one fibered $B_3$, all arguments apply in an analogous manner for limits in the regimes $x>0$
or $x=0$.
However, the case $x<0$ is different: Recall that here, at least for smooth Weierstrass models $Y_4$ over $B_3$, we found no kinematic obstruction at $\cO(\alpha'^2)$
to taking the vanishing fiber limit, in agreement with the fact that such limits would be perfectly well-defined.
This raises the question whether a non-perturbative superpotential can obstruct such a kinematically allowed\footnote{We reiterate from the previous section that for the limit to really be allowed perturbatively, also all higher order $\alpha'$ would have to vanish.} infinite distance emergent string limit. In the following, we give some preliminary arguments based on simple examples why we expect this not to be the case. A systematic analysis of the superpotential for genus-one fibrations in the limit of vanishing fibers is, however, left for future work.

If the superpotential diverges in the regime of a vanishing fiber volume, $\cV_{C_0}=0$, this should be caused by a tensionless fundamental Type IIB string, obtained by wrapping a D3-brane on $C_0$. This is similar to the diverging superpotential due to the E-string discussed in the heterotic set-up. However, while the E-string (and its generalisations) in four dimensions is a genuine $N=(0,2)$ string, the fundamental Type IIB string enjoys a higher, $N=(2,2)$ worldsheet supersymmetry. This corresponds to an $N=2$ space-time supersymmetry that is broken by the orientifold. Suppose that the four-dimensional model arises from a six-dimensional model which in turn undergoes an infinite distance limit of a vanishing $T^2$-fiber on the base $B_2$. In this case the fundamental Type IIB string already exists in six dimensions. Compactifying further to obtain a four-dimensional theory, the fundamental Type IIB string can either wrap the additional curve or not, yielding either a D3-instanton or staying as fundamental Type IIB string in the four-dimensional spectrum of strings.  Taking the volume of the $T^2$-fiber to zero causes the Type IIB string to become tensionless and at the same time the D3-instanton to acquire a vanishing action. This is only true since we saw that the $\alpha'$-corrections in this limit vanish and we can trust the classical geometry. We can therefore ask whether this D3-instanton associated to the fundamental Type IIB string can contribute to the superpotential. Following the intuition of the example of \cite{Berglund:2005dm} in the heterotic case, we would expect the contribution to the superpotential to be of the form 
\begin{align}\label{superpotentialIIB}
    W=f(G,z) \times b\left( q_0 Z_\text{ell.}(q_i) +  \dots \right)\,,
\end{align}
where $q_0=e^{-2\pi T_0}$ is the exponentiated volume of the cycle wrapped by the D3-instantons, $Z_\text{ell.}$ the six-dimensional elliptic genus of the Type IIB string and $b$ a coefficient multiplying the superpotential. We first notice that the six-dimensional elliptic genus $Z_\text{ell.}=\text{const.}$ \cite{Lee:2019oct} such that the instanton action $S\sim \Re{T_0}$ is not corrected. This complies with our previous results about the $\alpha'$-corrections. The question if a non-perturbative superpotential obstructs the infinite distance limit with $x<0$ then boils down to whether $b$ is vanishing or not. 

As an example take the base $B_3$ to be $dP_9\times \bbP^1$, where $dP_9$ is elliptically fibered over $\hat \bbP^1$. This base allows for both a $T^2$- and a $\bbP^1$-limit by either taking the elliptic fiber of $dP_9$ as the curve $C_0$ or viewing the $\bbP^1$-factor as the fiber of a trivial rational fibration. We are interested in the $T^2$-limit and the resulting duality to a D1-string. The non-perturbative superpotential due to D3-instantons for this model has been investigated starting with \cite{Donagi:1996yf} and is of the form
\begin{align}
    W= e^{-2\pi T_1}\frac{E_4(T_0)}{\eta^{12}(T_0)}\,.
\end{align}
Here classically $\Re{T_1}=\cV_{\bbP^1\times \sigma}$, with $\sigma$ the zero section of $dP_9$ and $\Re{T_0}=\cV_{T^2\times\bbP^1}$. Note that in deriving this expression, the effect of space-time filling D3-branes has not been taken into account \cite{Braun:2018fdp}.  If one of the D3-instantons intersects such a D3-brane, additional zero modes may modify or even cancel its contribution to the superpotential. Irrespective of this complication, we note that any of these instanton contributions is suppressed in the limit of vanishing $T^2$-fiber as $\Re{T_1}\rightarrow \infty$. In particular a contribution scaling like $e^{-2\pi T_0}$ as expected from \eqref{superpotentialIIB} is absent. This is similar to the case of a non-critical string arising in the case of a F-theory compactification involving $\bbF_2$ as discussed in the heterotic context above. For this string, which also preserves higher supersymmetry, the superpotential coefficient also vanishes.  

Note however, that again the effects of space-time filling D3-branes can in principle lift certain zero modes of D3-instantons making them contribute to the superpotential \cite{Palti:2020qlc}. A detailed study of this effect also in more general set-ups is left for future work.

At the level discussed here we conclude that in the case of a genus-one fibration, we do not expect the limit of vanishing fibral volume to be obstructed by a superpotential.

\subsection{Flux induced D-terms}   \label{sec_FluxedD}

So far we have considered compactifications without specifying the possible background fluxes.
In F-theory compactifications on fourfolds $Y_4$ such fluxes take values in $G \in H^{4}(Y_4)$.
Even though the effect of background fluxes is a priori not quantum, but arises already at the classical level,
we conclude this section by commenting on potential flux induced obstructions to taking the infinite distance limits.

Fluxes within the so-called primary horizontal subspace $H^{2,2}_{\rm hor}(Y_4)$ are known to induce a classical Gukov-Vafa-Witten superpotential \cite{Gukov:1999ya} which depends on the complex structure moduli
of $Y_4$. In the language of Type IIB compactifications with 7-branes, these include the Type IIB complex structure moduli, the 7-brane moduli and the axio-dilaton.
The infinite distance limits in the K\"ahler moduli space which form the subject of this work are therefore not obstructed by such F-term potentials.
Their effect on infinite distance limits in the complex structure moduli space of M-theory compactifications on fourfolds has been studied in \cite{Grimm:2019ixq}.

By contrast, four-form fluxes in the primary vertical subspace $H^{2,2}_{\rm vert}(Y_4)$ induce a D-term potential, which manifestly depends on the K\"ahler moduli.
In Type IIB language, such fluxes map to certain gauge fluxes on the 7-branes present in the background.
If we consider the gauge group on a 7-brane wrapping a divisor ${\bf S}$, the flux induced contribution to the D-term potential takes the  form (see e.g. \cite{Jockers:2005zy})
\bea  \label{VD}
V_D = \frac{g^2_{\rm YM}}{2} \xi^2    \,,  \qquad \xi = \frac{\int_{\bf S} J_{B_3} \wedge {\rm tr} F}{{\cal V}_{B_3}}  \,.
\eea
As it stands this formula holds for a stack of 7-branes wrapping a divisor ${\bf S}$ on $B_3$, with suitably embedded gauge flux $F \in H^{1,1}({\bf S})$.
More generally, it continues to apply to abelian gauge groups if we identify the  divisor ${\bf S}$ with the height-pairing of the rational section which is responsible 
for the appearance of an abelian gauge group factor in F-theory.
Such fluxes will play a central part in the context of proving the Weak Gravity Conjecture in Section \ref{sec_WGCpf}, which motivates us to investigate the consequences of their induced D-term
potential for the infinite distance limits.

The gauge coupling appearing in the D-term potential (\ref{VD}) depends on the volume of ${\bf S}$ in string units via
\bea
\frac{2 \pi}{g^2_{\rm YM}} = {\cal V}_{\bf S}
\eea
and is hence maximised for the parametrically smallest divisors. In the infinite distance limits under consideration, these are of the form ${\bf S} = J_0$. 
In this case
\bea
g^2_{\rm YM}   \precsim \frac{\lambda}{\mu^2}   \,,\qquad  \int_{\bf S} J \wedge F   \precsim \mu \lambda  \, ,
\eea
in both $J$-class A and B infinite distance limits. The scaling behaviour of $\int_{\bf S} J \wedge F$ depends on the concrete flux background but is bounded as indicated.
Since ${\cal V}_{B_3}  \sim \mu^3$,  the asymptotic D-term potential hence scales at worst as 
\bea
V_D   \precsim \left(\frac{\lambda}{\mu^2}\right)^3    \qquad \quad {\rm if} \quad  {\bf S} \sim J_0 \,.
\eea

For $ \mu  \precsim \lambda^{1/2}$, the flux-induced D-term potential could hence in principle diverge. This regime, however, has already been excluded based on our analysis of the perturbative $\alpha'$-corrections in Section \ref{sec_alphapcorr}. 
The only remaining potential obstruction to taking the infinite distance limit could therefore arise in the emergent string regime $ \mu \sim \lambda^{1/2}$ (cf. (\ref{JclassAhetrelabelded}) and (\ref{JclassBhetrelabelded}), in which the flux-induced contributions to the D-term potential 
remain (at worst) parametrically constant. 

However, even in this regime, the flux-induced potential may be cancelled by condensation of charged matter fields.
This, of course, is only possible for suitable signs of the massless field charges. 
In \cite{Aldazabal:2018nsj} it has been conjectured that such a cancellation of the D-term is in fact always possible in a consistent theory of quantum gravity.
We will assume this for the remainder of this paper. 

For the central application of the infinite distance limits in this paper, however, this assumption is not even required. Namely, to prove the Weak Gravity Conjecture in Section \ref{sec_wgc} we will be interested in configurations where the class of the gauge divisor ${\bf S}$ receives a contribution from one of K\"ahler generators of 
type $J_\mu$, $\mu \in \cI_1$ (in limits of $J$-class A, (\ref{combinedJclassA})) or $J_{\check\mu}$, $\check\mu \in \check{\cI}_2$ (for $J$-class B limits).
In this case, the divisor volume scales as  $\cV_{\bf S} \sim \mu^2 \lambda^2$ and hence 
\be
V_D   \precsim \frac{1}{(\mu\lambda)^6}   \,.
\ee
Clearly this potential can never obstruct the infinite distance limit of large $\mu$ and $\lambda$.

\section{F-theory/heterotic duality and quantum corrections}   \label{sec_SecFHet}

We have seen that for a rationally fibered F-theory base $B_3$ the infinite distance limits studied in the previous sections correspond to the weak-coupling limit of a critical heterotic string. In this section we therefore discuss some relevant aspects of F-theory/heterotic duality in order to understand the precise interpretation
of the F-theory limits in the dual heterotic frame. In particular we will elaborate on how both the classical F-theory geometry and the perturbative $\alpha'$-corrections encode
 perturbative loop corrections to the heterotic gauge coupling.
This fact will be of central importance in the context of the Weak Gravity Conjecture studied in Section \ref{sec_wgc}.

\subsection{Classical corrections to F-theory/heterotic duality}\label{subsec_classical_het-F_duality}

As is well-known  \cite{Morrison:1996na,Morrison:1996pp}, F-theory/heterotic duality in four dimensions can be thought of as adiabatically fibering the eight-dimensional duality of the heterotic string on $T^2$ and F-theory on K3 over a K\"ahler surface $B_2$. On the heterotic side this results in an elliptically fibered Calabi--Yau threefold 
\be
\rho: Z_3 \to B_2 \,.
\ee 
On the dual F-theory side one arrives at a K3-fibered Calabi-Yau fourfold $Y_4$ which, as always in F-theory, also allows for an elliptic fibration $\pi: Y_4 \rightarrow B_3$. As a consequence of $Y_4$ admitting both an elliptic and a K3-fibration, the base $B_3$ must be rationally fibered with projection
\be  \label{pfibration}
p: B_3 \rightarrow B_2 \,
\ee
and generic rational fiber $C_0$.
This means that $B_3$ is in general a blowup of a projectivisation $\mathbb{P}(\mathcal{O}\oplus \mathcal{T})$ of a rank 2 bundle $\mathcal{O}\oplus \mathcal{T}$ over the base $B_2$, which is either a Hirzebruch surface (or a blow-up in r points thereof), a del Pezzo surface or the Enriques surface $K3/\bbZ_2$.
 The group $H^{1,1}(B_3)$ of divisor classes is spanned by  a so-called exceptional section $S_-$ with the property that $S_- \cdot_{B_3} S_- = - S_- \cdot_{B_3} p^\ast(c_1({\cal T}))$, furthermore the 
 $p$-vertical divisors $p^\ast(C_\alpha)$, $\alpha = 1, \ldots, h^{1,1}(B_2)$ (where $C_\alpha$ is a basis of $H^{1,1}(B_2))$ as well as additional blowup divisors $E_a$. 
 Of course one can define several section divisors in addition to $S_-$ by taking linear combinations of $S_-$ with a pullback divisor. We will collectively denote these as $S_i$.
 Of particular relevance is the section 
 \be
 S_+ = S_- + p^\ast(c_1({\cal T}))
 \ee
 for the following reason:
 If we consider an F-theory model with a stack of 7-branes wrapping $S_-$ and $S_+$,  then the associated non-abelian gauge groups
 arise, on the dual heterotic side,
  from either of the two $E_8$ factors of the perturbative heterotic string.
  The twist of the fibration (\ref{pfibration}) is encoded in the class $c_1({\cal T}) \in H^{1,1}(B_2)$ which governs the asymmetric embedding of a gauge instanton
  into the two $E_8$ factors \cite{Friedman:1997yq}.
  
   For abelian gauge groups, the role of the divisor wrapped by the stack of 7-branes is played by the height-pairing
  of the associated rational section of $Y_4$ (see the reviews \cite{Weigand:2018rez,Cvetic:2018bni} for details and references), and the embedding of the abelian gauge group into both $E_8$ factors is more intricate \cite{Cvetic:2015uwu,Cvetic:2016ner}.
 
  In view of the defining intersection numbers, we can identify these three types of divisors with suitable linear combinations of the different types of K\"ahler cone generators:
\be
 \begin{aligned}
&J-{ \rm class \, \, A}:& \quad    &p^\ast(C_\alpha), E_a   &\longleftrightarrow&    \quad J_m\,,   \quad m \in \{0\}\cup \cI_3 \cr
&&                                   &                                S_i  &\longleftrightarrow& \quad  J_\mu   \,,  \quad \, \mu \in  \cI_1 \cr
&J-{ \rm class \, \, B}:& \quad    &p^\ast(C_\alpha), E_a   &\longleftrightarrow&    \quad J_m\,,   \quad m \in \{0\}\cup \hat{\cI}_2 \cr
&&                                   &                                S_i  &\longleftrightarrow& \quad  J_\mu   \,,  \quad \, \mu \in  \check{\cI}_{ 2} \cr
\end{aligned}
 \ee
 
Under F-theory/heterotic duality, the volume moduli of $p^\ast(C_\alpha)$ map to the K\"ahler moduli of the $\rho$-vertical divisors on $Z_3$,
\begin{align}
    \cV_{p^\ast(C_\alpha)} = \eta_{\alpha \beta} t^\beta_\text{het}\,,
\end{align}
where $\eta_{\alpha \beta} = C_\alpha   \cdot_{B_2}  C_\beta$.
The volume moduli of  the blowup divisors map to heterotic 5-brane moduli and the volume of the section divisor $S_-$ is related to the heterotic dilaton, as we will recall now. 
 To this  end note first that for the adiabatic argument for F-theory/heterotic duality to work the volume of the base $B_2$ of the rational fibration has to be much larger compared to the volume of the rational fiber.\footnote{To arrive at a perturbative heterotic string dual one furthermore has to take a stable degeneration limit \cite{Morrison:1996na,Morrison:1996pp} in the F-theory complex structure moduli space which guarantees that the volume of the elliptic fiber of the heterotic compactification space is large and worldsheet instanton corrections from the fiber are suppressed. This requirement is orthogonal to our analysis of the limits in the F-theory K\"ahler moduli space.} In terms of divisor volumes this translates into 
\begin{align}\label{adiabaticlim}
    \cV_{S_i}\gg \cV_{p^\ast(C_\alpha), E_a} \,, 
\end{align}
i.e. the volume of a section divisor of the rational fibration has to be much larger than the volume of a divisor which is vertical w.r.t. the rational fibration or of a blowup divisor.

If we parametrise the asymptotic K\"ahler form as in \eqref{combinedJclassA} or \eqref{Jprime-classB}, to leading order in $\lambda$ the respective volumes  are
given by
\begin{align} \label{vertandsecvolumes}
  \cV_{p^\ast(C_\alpha), E_a} &=& \text{vol} (J_{m\in \{0\}\cup \cI_3}) \precsim \frac{\mu^2}{\lambda} \,,    \qquad \cV_{S_i} &=& \text{vol}(J_{\mu\in \cI_1}) \sim\mu^2\lambda^2\,,  \qquad   \text{for $J$-class A limits} \cr
  \cV_{p^\ast(C_\alpha), E_a}   &=& \text{vol} (J_{m\in \{0\}\cup {\hat\cI}_2}) \precsim \frac{\mu^2}{\lambda}   \,, \qquad \cV_{S_i}  &=& \text{vol} (J_{\mu\in {\check\cI}_2}) \sim \mu^2\lambda^\gamma \,, \qquad   \text{for $J$-class B limits}   \,.
\end{align}  
 We observe that in both cases \eqref{adiabaticlim} is fulfilled in the limit $\lambda\rightarrow \infty$.   To avoid clutter, for the remainder of this review section we focus on the $J$-class A limits.

The relation between the heterotic string scale and the Planck mass is computed in the F-theory frame
by recalling that the heterotic string is associated with a D3-brane wrapping the generic fiber $C_0$ of $B_3$ and hence
 \bea   \label{MIIBvsMhet}
 \frac{M^2_{\rm Pl}}{M^2_{\rm IIB}} = 4 \pi \cV_{B_3}   \,,   \qquad \frac{M^2_{\rm het}}{M^2_{\rm IIB}} = 2 \pi \cV_{C_0}    \,.
\eea
As a result, the heterotic four-dimensional dilaton can be expressed as 
\bea   \label{identifydilaton}
 \text{Re}\,S_\text{het}   := \frac{1}{2}  \frac{M_{\rm Pl}^2}{M_\text{het}^2}  = \frac{\cV_{B_3} }{\cV_{C_0}}   \,.
\eea
In the strict weak coupling limit, this can be approximated by the volume of any of the sections of $B_3$,
\bea
 \text{Re}\,S_\text{het} \stackrel{(\ref{adiabaticlim})}{=} \cV_{S_i} \,.
\eea

 The (complexified) heterotic dilaton is closely related to the heterotic gauge coupling via  the relation, valid at the heterotic tree-level,
 \begin{align}\label{treelevelrelation}
\frac{2 \pi}{g_{{\rm YM}}^2}\big|_{\rm M_\text{het}}= (2m) \, \text{Re}\,S_\text{het} + {\cal O}(S_\text{het}^0) = m \frac{M_{\rm Pl}^2}{M_\text{het}^2} + {\cal O}(S_\text{het}^0)\,.
\end{align}
Here the parameter $m$ appears for an abelian gauge symmetry and fixes the $U(1)$ normalisation; from a heterotic worldsheet perspective, $m$ corresponds to the Kac-Moody level of the $U(1)$ current algebra. 

This relation   receives perturbative loop (and non-perturbative) corrections. At the one-loop level, the gauge coupling, measured at the heterotic string scale ${\rm M_\text{het}}$,  receives threshold corrections which are independent of the heterotic dilaton (see e.g. \cite{Louis:1996ya} and references therein):
\bea   \label{gMhet1}
\frac{2 \pi}{g_{{\rm YM}}^2}\big|_{\rm M_\text{het}}  = (2m) \, \text{Re}\,S_\text{het}  + \tilde\Delta(M,\bar M)  \,.
\eea
The threshold corrections $ \tilde\Delta(M,\bar M)$ obtain two types of contributions,
\bea
 \tilde\Delta =  \tilde\Delta_0 +  \tilde\Delta_1 \,,  
 \eea
 given by
\bea
\tilde\Delta_0 &=&2\pi\, {\rm Re}f^{(1)}(M)    \\
 \tilde\Delta_1 &=& \frac{c}{8\pi} K^H(M, \bar M) + \ldots  \,.
\eea
Here $f^{(1)}(M)$ is the Wilsonian one-loop correction to the gauge kinetic function and depends on the heterotic K\"ahler moduli and NS5-brane moduli, which we collectively denote by $M$.  
The second type of 1-loop corrections, $ \tilde\Delta_1$, includes the contribution from the heterotic tree-level K\"ahler potential, as well as further corrections associated 
with the wave function renormalisation of the charged matter fields and also possible field independent correction terms.
The renormalisation coefficient $c$ depends on the matter content of the gauge theory and can be found e.g. in  \cite{Louis:1996ya}.

On the F-theory side the dual of the perturbative heterotic gauge theory arises from a D7-brane wrapped on a divisor 
\begin{align}
\mathbf{S} = \sum_{m\in \{0\}\cup\cI_3}h^m J_m + \sum_{\mu\in\cI_1}h^\mu J_\mu\,
\end{align}
for which $h^{\mu}\ne 0$ for at least one $\mu\in \cI_1$.\footnote{This includes as special cases the sections $S_-$ and $S_+$ for the perturbative heterotic non-abelian gauge group factors, but also captures the physics of $U(1)$ gauge group factors by identifying ${\bf S}$ with the height pairing, as noted above.} The associated gauge coupling is then given, in the large volume regime on the F-theory side and evaluated at the IIB/F-theory string scale, by 
\be
\label{eq:divS_classical_expansion}
\frac{2 \pi}{g^2_{\rm YM}}\big|_{\rm M_\text{IIB}}  =  \mathcal{V}_{\mathbf{S}}  = \mathcal{V}_{\mathbf{S}}^{(0)} +  \mathcal{V}_{\mathbf{S}}^{(1)}   + \ldots \,,
\ee
where  the superscripts refer to the order in the Type IIB $\alpha'$ expansion.\footnote{More precisely, the leading $\alpha'$-corrections appear at order $(\alpha')^2$.}
In the weak coupling limit on the F-theory side, the classical volume, computed to order  $(\alpha')^0$ in the Type IIB/F-theory frame, can be split it into a leading contribution $\mathcal{V}_{\mathbf{S}}^{(0,0)}$ and a subleading contribution $\mathcal{V}_{\mathbf{S}}^{(0,1)}$:
\bea
\mathcal{V}_{\mathbf{S}}^{(0)} = \mathcal{V}_{\mathbf{S}}^{(0,0)} + \mathcal{V}_{\mathbf{S}}^{(0,1)} +\mathcal{O}\left(1/\lambda^4\right)
\eea
with 
\bea
\mathcal{V}_{\mathbf{S}}^{(0,0)}  &=& (\mu\lambda)^2\frac12 h^\mu k_{\mu m n}\tilde{b}^m\tilde{b}^n   \sim  (\mu\lambda)^2    \\
\mathcal{V}_{\mathbf{S}}^{(0,1)}  &=& \left(\frac{\mu^2}{\lambda}\right)h^\alpha k_{\alpha m \mu}\tilde{b}^m a^\mu   \sim  \left(\frac{\mu^2}{\lambda}\right)    \,.
\eea

For future purposes we will introduce a similar decomposition of the volume of the curve $C_0$ and the F-theory base $B_3$ as
\bea
\mathcal{V}_{C_0} &=& \mathcal{V}^{(0)}_{C_0}  + \mathcal{V}^{(1)}_{C_0}   + \ldots    \\
\mathcal{V}_{B_3} &=& \mathcal{V}^{(0)}_{B_3}  + \mathcal{V}^{(1)}_{B_3}   + \ldots
\eea
where the classical volumes are computed as 
\begin{align}
 \cV^{(0)}_{C_0}&=k_{00\mu}v^\mu \sim \frac{\mu}{\lambda^2}\,, \cr
\label{eq:B3_classical_expansion}
  \mathcal{V}^{(0)}_{B_3}&=\underbrace{\mu^3 \frac12 k_{mn\mu}a^{\mu}\tilde{b}^m\tilde{b}^n}_{\mathcal{V}_{B_3}^{(0,0)}}+\underbrace{\left(\frac{\mu}{\lambda}\right)^{\!3}\frac12 k_{m\mu\nu}\tilde{b}^m a^\mu a^\nu}_{\mathcal{V}_{B_3}^{(0,1)}}+\mathcal{O}\left(1/\lambda^6\right) \,.
\end{align}

To match both expressions (\ref{gMhet1}) and (\ref{eq:divS_classical_expansion})  at the heterotic string scale, we first first run $\frac{2 \pi}{g^2_{\rm YM}}\big|_{\rm M_\text{IIB}}$ from  $M_\text{IIB}$ up to the heterotic string scale 
$M_\text{het}$, which are related via (\ref{MIIBvsMhet}):
\bea
\frac{2 \pi}{g^2_{\rm YM}}\big|_{\rm M_\text{het}}  =\frac{2 \pi}{g^2_{\rm YM}}\big|_{\rm M_\text{IIB}} + \frac{b}{8\pi} \log\left(\frac{M_\text{IIB}^2}{M_\text{het}^2}\right) = \frac{2 \pi}{g^2_{\rm YM}}\big|_{\rm M_\text{IIB}} - \frac{b}{8\pi} \log(2 \pi {\cal V}_{C_0})   \,,
\eea
where $b$ is the $\beta$-function coefficient.
Then, with the help of (\ref{identifydilaton}), 
we obtain
\bea   \label{gMhet2}
\frac{2 \pi}{g^2_{\rm YM}}\big|_{\rm M_\text{het}} = 2m  \text{Re}\,S_\text{het} \frac{\cV_{\bf S}\cV_{C_0} }{2m \cV_{B_3}} - \frac{b}{8\pi} \log(2 \pi {\cal V}_{C_0})   \,.
\eea

We can now expand the ratio of volumes appearing on the RHS in terms of the leading and subleading contributions to the classical volumes as well as the $\alpha'$-corrections as follows:
\bea   \label{ratiodef1}
 \frac{\cV_{\bf S}\cV_{C_0} }{2m \cV_{B_3}} = \frac{\cV^{(0,0)}_{\bf S}\cV^{(0)}_{C_0} }{2m \cV^{(0,0)}_{B_3}} \left(1 + \Delta_0 + \Delta_1   + \ldots  \right)     
\eea
where
\bea   \label{Delta0expre-a}
\Delta_0 = - \frac{\mathcal{V}_{B_3}^{(0,1)}}{\mathcal{V}_{B_3}^{(0,0)}} + \frac{\mathcal{V}_{\mathbf{S}}^{(0,1)}}{\mathcal{V}_{\mathbf{S}}^{(0,0)}} = {\cal O}(1/\lambda^3)   
\eea
appears to order $(\alpha')^0$ and the first $\alpha'$-corrections are assembled in $\Delta_1$.

Then  the heterotic tree-level relation \eqref{treelevelrelation} is equivalent to the geometric relation
\be   \label{VcVSVB}
\frac{\cV^{(0,0)}_{\bf S}\cV^{(0)}_{C_0} }{2m \cV^{(0,0)}_{B_3}}       = 1   \,.
\ee
This relation can indeed be checked  explicitly both for the classical limits of $J$-class A and  $J$-class B, with\footnote{In the special case where ${\bf S}$ is a section $S_i$, we have $2 m =1$ and 
the relation reduces to (\ref{identifydilaton}).}
\be
2 m = {\bf S} \cdot_{B_3}   C_0   \,.
\ee
The proof for classical limits of $J$-class $A$ has already been given in \cite{Lee:2019jan}.
As a new result, we observe here that in the relevant classical $J$-class B limits of the form (\ref{JclassBhetrelabelded}),
(\ref{VcVSVB}) follows from the analysis of Appendix \ref{App_JclassB}, and we discuss the technical steps in Appendix \ref{App_WGCclass-B}.

As a consequence, we find
\bea   \label{ratiodef2}
 \frac{\cV_{\bf S}\cV_{C_0} }{2m \cV_{B_3}} = 1 + \Delta + \ldots = 1 + \Delta_0 + \Delta_1   + \ldots   
\eea
and hence
\bea
\frac{2 \pi}{g^2_{\rm YM}}\big|_{\rm M_\text{het}} = 2m  \text{Re}\,S_\text{het}  \left( 1  +    \Delta_0  +    \Delta_1    \right)    - \frac{b}{8\pi} \log(2 \pi {\cal V}_{C_0})   \,.
\eea
We propose that this matches the structure of the one-loop threshold corrections on the heterotic side,  (\ref{gMhet1}), if we make the following identifications
\bea   
\tilde\Delta_0  &=&   2m \text{Re}\,S_\text{het}  \,   \Delta_0    \label{threshold1} \\                                                        
\tilde\Delta_1   &=& 2m  \text{Re}\,S_\text{het}   \,  \Delta_1   -\frac{b}{8\pi} \log(2 \pi {\cal V}_{C_0})  \,.   \label{threshold2} 
\eea

A detailed interpretation of (\ref{threshold2}) will be given in the next section.
Before we come to this, 
 note that  (\ref{threshold1}) means that the subleading classical corrections to the volumes as encoded in $\Delta_0$ map to the heterotic threshold corrections derived from the holomorphic 1-loop correction
to the gauge kinetic function.
To check   (\ref{threshold1})
at the parametric level,
we observe the scaling behaviour
\bea
2m \text{Re}\,S_\text{het}  \,   \Delta_0     \sim  (\mu^2 \lambda^2)     \frac{1}{\lambda^3}   =      \frac{\mu^2}{\lambda}   \sim  \cV_{p^\ast(C_\alpha), E_a} \,. 
\eea
Here we made use of the scaling behaviour (\ref{vertandsecvolumes}) of the volume of the pullback divisors $p^\ast(C_\alpha)$ or the blowup divisors $E_a$ on the F-theory base $B_3$.
As explained already, these volume moduli map, respectively, to the heterotic base K\"ahler moduli and the heterotic 5-brane moduli.
This perfectly matches the moduli dependence of the 1-loop corrections to the gauge kinetic function on the heterotic side
as computed for instance for heterotic Calabi-Yau compactifications in \cite{Blumenhagen:2005ga,Blumenhagen:2006ux}.
In particular, the corrections of \cite{Blumenhagen:2005ga,Blumenhagen:2006ux} vanish if the dual heterotic instanton is symmetrically embedded into both $E_8$ factors and provided there are no heterotic 5-branes. This means that on the F-theory side, the base $B_3$ is the direct product $B_3 \simeq C_0 \times B_2$ with no blowup divisors $E_a$, in perfect match with the form of the correction terms  $\Delta_0$.

We conclude that already the classical F-theory geometry away from the adiabatic limit ($\lambda\rightarrow \infty$) calculates one-loop corrections to the heterotic gauge coupling which are proportional to the heterotic K\"ahler moduli and the 5-brane moduli and hence leading in a possible $\alpha'_\text{het}$ expansion of the one-loop corrections.

\subsection{Interpretation of F-theory \texorpdfstring{$\alpha'$}{alpha'} corrections in dual heterotic frame}\label{sec:Interpretationalpha}

The remaining task is now to quantify the contribution  $\Delta_1$ to the heterotic threshold corrections (\ref{threshold2}). Recall that $\Delta_1$ was defined in (\ref{ratiodef2}) and involves the $\alpha'$-corrections on the F-theory side
which modify the dependence of the curve, divisor and total threefold volumes on the 
K\"ahler parameters $v^\alpha$.
The main result of this section is that the leading $\alpha'$-corrections appearing in the ratio (\ref{ratiodef2}) indeed have the correct moduli dependence to map to the non-holomorphic threshold corrections
$\Delta_1$ on the heterotic side. This, however, is only possible if we constrain some of the parameters of the $\alpha'$-corrections as in  (\ref{kappa4rel}) and (\ref{zalphaconstraintbeta}) below.
We interpret these as new consistency conditions on the parametrisation of the F-theory $\alpha'$-corrections deduced  by F-theory/heterotic duality.

To compute $\Delta_1$,  recall the expressions
\eqref{VolB3def}, \eqref{ReTmain} and \eqref{CorrectedLinearMultiplets} for the $\alpha'$-corrected quantities. 
The quantum corrected volume of the curve $C_0 = J_0 \cdot J_0 $ is obtained as 
\bea   \label{VC0volume}
\cV_{C_0} = \mathcal{V}_{B_3}   \, k_{00\mu}   \,   L^\mu
\eea
where $k_{00\mu} = J_0 \cdot J_0 \cdot J_\mu$ and the linear multiplets $L^\mu$ are displayed in (\ref{CorrectedLinearMultiplets}).
To evaluate this, we make use of the following two observations: 

First, the expression (\ref{VC0volume}) contains terms involving the matrix $K^{\alpha \beta}$.
In Appendix \ref{App_volC0} we show that these take the form
\begin{equation}
\label{eq:KijC0relation}
  \begin{aligned}
    k_{00\mu}K^{\mu\alpha}&=\frac12\frac{\cV^{(0)}_{C_0} v^\alpha}{\mathcal{V}_{B_3}^{(0)}}   + {\cal O}\left(\frac{1}{\mu\lambda^4}\right)\qquad\qquad&J\textup{-class A}\\
    k_{0\hat{\mu}_0\check{\nu}}K^{\check{\nu}\alpha}&=\frac12\frac{\cV^{(0)}_{C_{\hat{\mu}_0}} v^\alpha}{\mathcal{V}_{B_3}^{(0)}} + {\cal O}\left(\frac{1}{\mu\lambda^4}\right)\qquad\qquad&J\textup{-class B}   
  \end{aligned}
\end{equation}
where $\cV^{(0)}_{C_0}$ is the classical volume of $C_0$ and  $\cV^{(0)}_{C_{\hat{\mu}_0}}$ of the curve $C_{\hat{\mu}_0} = J_0 \cdot J_{\hat{\mu}_0}$ for some reference generator $J_{\hat{\mu}_0}$.\footnote{The precise definition of this generator can be found in Appendix \ref{App_JclassB}.}
Using this one notes that to the given order the terms proportional to $\partial_\alpha \cT_\beta$ in (\ref{CorrectedLinearMultiplets}) all vanish in (\ref{VC0volume}) because after contracting with $K^{\alpha \beta}$ they are proportional to
\bea
v^\alpha \partial_\alpha \cT_\beta =  0   \,.
\eea
Here we used that $\cT_\beta$ is homogeneous of degree zero in $v^\alpha$.
Collecting all remaining terms, including the $\alpha'$-corrections to the pre-factor $\mathcal{V}_{B_3}$ in (\ref{VC0volume}),   gives
\begin{equation}
  \label{eq:C0limit}
  \cV_{C_0}=\cV_{C_0}^{(0)}\left[1+{\alpha^2} \left( \left(\tilde{\kappa}_1+\tilde{\kappa}_2-\kappa_3-\kappa_5\right) \frac{\cal Z}{\cV^{(0)}_{B_3}} +(\tilde{\kappa}_2-\kappa_5)  \frac{\cal T}{\cV^{(0)}_{B_3}} - \delta_1   \right) + {\cal O}\left(\frac{1}{\mu^2 \lambda^2}\right) \right] 
\end{equation}
with 
\be\label{delta1def}
\delta_1 =   \frac{1}{2}\frac{\tilde{\mathcal Z}}{\cV^{(0)}_{B_3}}  = \frac{1}{2} \frac{\tilde{\mathcal Z}_{\alpha} v^\alpha}{ \cV^{(0)}_{B_3}  } = {\cal O} \left(\frac{\lambda}{\mu^2}\right) \,. 
\ee
To understand the scaling of $\delta_1$ recall that $\tilde{\mathcal Z}_{\alpha}$  was defined in (\ref{tildeZdef}) and in the limits (\ref{combinedJclassA}) or (\ref{combinedJclassB}) under consideration $\cV^{(0)}_{B_3} \sim \mu^3$, while the largest K\"ahler modulus scales as $\mu \lambda$.

For the volume of the gauge divisor ${\bf S} = h^\alpha J_\alpha$ and the total volume, one finds
\bea
\cV_{\bf S}  &=& \cV^{(0)}_{\bf S}   \left[ 1 + \alpha^2  \left((\kappa_3 + \kappa_5) \frac{\cal Z}{\cV^{(0)}_{B_3}} + \kappa_5  \frac{\cal T}{\cV^{(0)}_{B_3}}   +  \delta_2 \right)  \right] \,,   \\
\cV_{B_3}   &=&   \cV^{(0)}_{B_3}\left[ 1 + \alpha^2\left( (\tilde{\kappa}_1+\tilde{\kappa}_2) \frac{\cal Z}{\cV^{(0)}_{B_3}}  + \tilde \kappa_2  \frac{\cal T}{\cV^{(0)}_{B_3}}  \right)    \right]   \,,
\eea
where 
\bea \label{delta2def}
\delta_2 =  \frac{ h^\alpha \tilde {\cal Z}_\alpha}{\cV^{(0)}_{\bf S} } {  \rm log}   \cV_{B_3}   + \kappa_6   \frac{h^\alpha \cT_\alpha}{\cV^{(0)}_{\bf S}} + \kappa_7   \frac{h^\alpha {\cal Z}_\alpha}{\cV^{(0)}_{\bf S}}  = {\cal O} \left(\frac{1}{\mu^2 \lambda^2}\right)   \,,  
\eea
with the scaling of $\delta_2$ again referring to the limits (\ref{combinedJclassA}) or (\ref{combinedJclassB}), in which $\cV^{(0)}_{\bf S} \sim \mu^2 \lambda^2$.

We are interested in the ratio of volumes (\ref{ratiodef1}) appearing in the expression (\ref{gMhet2}) for the gauge coupling.
Collecting all terms, we find that 
\be
  \frac{\cV_{\bf S}{ \cV_{C_0}}}{2m {\cV_{B_3}} }    =   \frac{\cV^{(0)}_{\bf S}{ \cV^{(0)}_{C_0}}}{2m {\cV^{(0)}_{B_3}} } \left(1 - \alpha^2 \delta_1   + {\cal O} \left(\frac{1}{\lambda^2\mu^2}\right)  \right) \,.
\ee
This identifies the $\alpha'$-correction $\Delta_1$ defined in (\ref{ratiodef1}) as
\bea
\Delta_1 =    -    \alpha^2 \delta_1 +    {\cal O} \left(\frac{1}{\lambda^2\mu^2}\right)    \,.
\eea

The suppressed subleading terms are of order $1/(\mu\lambda)^2\sim 1/\mathcal{V}_{S_i}$ in the notation of section~\ref{subsec_classical_het-F_duality}. This is the expected scaling of a heterotic one-loop correction to the tree-level relation (\ref{treelevelrelation})
and we will compute these explicitly below.
By contrast,
$\delta_1$ introduced in (\ref{delta1def}) could in principle lead to a correction which scales as 
\be
\delta_1 ={\cal O} \left(\frac{\lambda}{\mu^2}\right) \sim 1/\mathcal{V}_{p^\ast(C_\alpha),E_a} \,.
\ee 
This would be of the same order as the classical result for the emergent string limits~\eqref{JclassAhetrelabelded} and~\eqref{JclassBhetrelabelded} since we have $\mu^2/\lambda\sim 1$ in this case. From the heterotic perspective, such a would-be moduli dependent string tree-level correction to the relation (\ref{treelevelrelation}) between the heterotic dilaton and gauge coupling must not occur; as we discussed in Section \ref{subsec_classical_het-F_duality},
(\ref{treelevelrelation}) is perturbatively only modified at heterotic loop-level, but not by pure heterotic $\alpha'$-corrections.

Consistency of F-theory/heterotic duality hence requires that  the ${\cal O} \left(\frac{\lambda}{\mu^2}\right)$ contribution to  $\delta_1$ vanishes.  In view of (\ref{delta1def}) this implies that 
the product $v^\alpha \kappa'_{4,\alpha}$ (no sum over $\alpha$) must grow slower than at a rate $\mu\lambda$ for all values of $\alpha$. 
Looking back at the definition of the limits of $J$-class A given in (\ref{JclassAhetrelabelded}) and $J$-class B, cf. (\ref{JclassBhetrelabelded}), this leads to the respective constraints that 
\begin{equation}
\label{kappa4rel}
  \begin{aligned}
     \kappa'_{4,\alpha} &\stackrel{!}{=} 0   &\qquad  \forall  \alpha \in \{0\}  \cup  \cI_3      \qquad\qquad                              &J\textup{-class A}\\
     \kappa'_{4,\alpha} &\stackrel{!}{=}  0   & \qquad  \forall  \alpha \in  \{0\}  \cup  \hat\cI_2       \qquad\qquad&J\textup{-class B}   \,.
    \end{aligned}
\end{equation}

As long as these relations are satisfied, we have found that the $\alpha'$-corrected volumes satisfy the relation\footnote{Since we have $\mu=\lambda^{x+1/2}$ with $x\geq 0$ the corrections of the form $1/(\mu\lambda)^2$ are $\cO(1/\lambda^3)$.}
\bea   \label{WGCrelgeomQC}
\frac{\cV_{\bf S}{ \cV_{C_0}}}{2m {\cV_{B_3}} }  = \frac{\cV^{(0)}_{\bf S}{ \cV^{(0)}_{C_0}}}{2m {\cV^{(0)}_{B_3}} } \left(1 +   {\cal O}\left(\frac{1}{\mu^2 \lambda^2} \right) \right) = \left(1 +   {\cal O}\left(\frac{1}{ \lambda^3} \right)+   {\cal O}\left(\frac{1}{\mu^2 \lambda^2} \right) \right)\, ,
\eea
where the  second contribution in the last bracket is due to the classical subleading terms $\Delta_0$ in (\ref{ratiodef1}).

Before scrutinizing the $\alpha'$-correction terms   of order  ${\cal O}(\frac{1}{\mu^2 \lambda^2})$    further,
let us comment on the geometric origin of this relation. 
At the classical level, the K\"ahler potential and K\"ahler coordinates respect the
 no-scale structure 
\bea   \label{noscale}
v^\alpha \text{Re}\,T^{(0)}_\alpha = 3 \cV^{(0)}_{B_3} \,,
\eea
or equivalently
\bea
\big(\cV^{(0)}_{B_3}\big)^{-2} &=& \tfrac{1}{6}  k_{\alpha \beta \gamma} \big(L^{(0)}\big)^\alpha \big(L^{(0)}\big)^\beta  \big(L^{(0)}\big)^\gamma   \,,   \\
\big(\cV^{(0)}_{B_3}\big)^{-2}\,  {\rm Re} T^{(0)}_\alpha &=&  \tfrac{1}{2}   k_{\alpha \beta \gamma} \big(L^{(0)}\big)^\beta \big(L^{(0)}\big)^\gamma   \,.
\eea 
If the no-scale structure continued to hold after including the leading $\alpha'$-corrections, the WGC relation (\ref{WGCrelgeomQC}) would follow immediately from its classical counterpart (\ref{WGCalpha0}).
However, as stressed already in  \cite{Grimm:2017pid,Weissenbacher:2019mef}, the perturbative $\alpha'$-corrections in general break the no-scale condition (\ref{noscale}).
By direct computation,  one finds that no-scale is only restored if 
\bea  \label{noscalekappa4}
\kappa'_{4,\alpha} = 0  \qquad \quad   \forall \, \alpha   
\eea 
and in addition 
\begin{align}
 2 \tilde \kappa_2 - \kappa_6 - 3\kappa_5 =0    \,.   
\end{align}
However, (\ref{noscalekappa4}) would be in conflict with arguments put forward in \cite{Weissenbacher:2020cyf} suggesting the appearance of logarithmic corrections
to the K\"ahler coordinates
in certain Type II orientifold compactifications.
This is to be contrasted with the weaker  constraint (\ref{kappa4rel}), which follows from imposing (\ref{WGCrelgeomQC}) and is hence a prediction of F-theory/heterotic duality.
Indeed, (\ref{kappa4rel}) is compatible with the specific models considered in \cite{Weissenbacher:2020cyf}.
The key point here is our proposal that the coefficients of the $\log \cV_{B_3}^{(0)}$ terms need not be universal among the K\"ahler coordinates in the F-theory effective action, i.e.
the universal parameter $\kappa_4$ appearing in the M-theory quantities (\ref{TiM}) and (\ref{Li}) may
receive non-universal 1-loop corrections under the uplift to F-theory, as parametrised by the non-universal renormalised parameters 
$\kappa'_{4,\alpha}$. 
In particular, we have seen that the relation (\ref{WGCrelgeomQC}) required by F-theory/heterotic duality holds to leading order in $\lambda$ even if the no-scale condition is violated.

To conclude this section, we now give the explicit form of the $\alpha'$-corrections encoded in $\Delta_1$ under the assumption that (\ref{kappa4rel}) indeed holds.
For definiteness we focus here on the $J$-class A limits since the $J$-class B limits behave similarly.
Using the results and definitions from Appendix~\ref{App_volC0}, especially (\ref{C0sublead}),  (\ref{Ssublead}) and (\ref{B3sublead}),
we can collect all correction terms as 
  \begin{equation}
 \begin{gathered}
  \label{eq:subleading_corrections_WGC}
 \Delta_1 =  - \frac{\alpha^2}{\mathcal{V}_{B_3}^{(0,0)}}\left[\left(\tilde{\kappa}_1+\tilde{\kappa}_2-\tfrac32\kappa_3-\tfrac32\kappa_5\right)
  \left(\Delta^\alpha \mathcal{Z}_\alpha+\mathcal{Z}_\mu v^\mu-\tfrac{\mathcal{V}_{B_3}^{(0,1)}}{\mathcal{V}_{B_3}^{(0,0)}}\mathcal{Z}_m v^m\right)  \right.\\
 \left.
  +\left(\tilde{\kappa}_2-\tfrac32 \kappa_5\right)\left(\Delta^\alpha\mathcal{T}_\alpha+\mathcal{T_\mu}v^\mu-\tfrac{\mathcal{V}_{B_3}^{(0,1)}}{\mathcal{V}_{B_3}^{(0,0)}}\mathcal{T}_m v^m\right)+\tfrac12 \tilde{\mathcal{Z}}_\mu v^\mu-\tfrac12 \Delta^\alpha \mathcal{D}_\alpha-\delta_2 \mathcal{V}_{B_3}^{(0,0)}
  \right]+
  \mathcal{O}\left(\frac{1}{(\mu\lambda)^4}\right)\;.
 \end{gathered}
 \end{equation}
These corrections at $\mathcal{O}\left(\alpha'^2\right)$ should now be matched with the non-holomorphic heterotic threshold corrections appearing in (\ref{gMhet1}),
\begin{align}   \label{tildedelta1-2}
 \tilde\Delta_1 = \frac{c}{8\pi} K(M, \bar M) + \ldots =- \frac{c}{8\pi}\log(\mathcal{V}_{B_2}^H) + \dots  = - \frac{c}{8\pi}\log\left(\mathcal{V}^{(0)}_{B_2}  \frac{M^4_{\rm IIB}}{M^4_{\rm het}}\right) + \ldots
\end{align}
via \eqref{threshold2},
\bea \label{threshold2a}
\tilde\Delta_1   \stackrel{!}{=} 2m  \text{Re}\,S_\text{het}   \,  \Delta_1   -\frac{b}{8\pi} \log(2 \pi {\cal V}_{C_0})   \,.
\eea
 In (\ref{tildedelta1-2}), $\mathcal{V}_{B_2}^H$ is the volume of the base of the heterotic Calabi-Yau threefold in heterotic string units,
and after the last equality we have expressed it in terms of the corresponding tree-level volume in Type IIB units.
The only way to obtain such a term in $\Delta_1$ is via the $\log \mathcal{V}_{B_3}$ term that enters through the first term in $\delta_2$ as given in (\ref{delta2def}): 
\begin{align}
2 m \text{Re}\,S_\text{het}    \, \Delta_1 =  2 m \text{Re}\,S_\text{het}    \, \delta_2 + \ldots       =   \alpha^2 h^\alpha \tilde{\mathcal Z}_\alpha\log \mathcal{V}_{B_3}^{(0)}  + \ldots =    \alpha^2 h^\alpha \tilde{\mathcal Z}_\alpha \log\left(\frac{M_\text{het}^2}{M_{\rm IIB}^2} \frac{\mathcal{V}_{B_2}^{(0)}}{2\pi}\right)  + \ldots
\end{align}
After the second equality we used that $2 m \text{Re}\,S_\text{het} = {\cal V}_{\bf S}^{(0)} + \ldots$ according to (\ref{identifydilaton}) and relations such as (\ref{ratiodef2}).
Thus the $\log \mathcal{V}_{B_3}$ term has the same form as a usual one-loop renormalization provided that 
\bea  \label{zalphaconstraintbeta}  
\alpha^2 h^\alpha \tilde{\mathcal{Z}}_\alpha \stackrel{!}{=} -\frac{b}{8\pi} \,.
\eea 
We recall that the parameter $\tilde{\mathcal{Z}}_\alpha$ had been defined in (\ref{tildeZdef}) and that $h^\alpha$ appears in the expansion of the gauge divisor ${\bf S} = h^\alpha J_\alpha$.
Interestingly, (\ref{zalphaconstraintbeta}) hence constrains the parameters $\tilde{\mathcal{Z}}_\alpha$ for those $\alpha$ for which $\tilde{\mathcal{Z}}_\alpha$ is not already zero by means of the earlier consistency condition (\ref{kappa4rel}). This is for all K\"ahler parameters as labelled by $\alpha \in \cI_1$ for limits of $J$-class A and $\alpha \in \check\cI_2$ for limits of $J$-class B.

If this is satisified the RHS of \eqref{threshold2a} reads 
\begin{align}
 - \frac{b}{8\pi} \log\left(\frac{M_\text{het}^2}{M_{\rm IIB}^2} \frac{\mathcal{V}_{B_2}^{(0)}}{2\pi}\right) - \frac{b}{8\pi} \log\left(2\pi \mathcal{V}_{C_0} \right) = - \frac{b}{8\pi} \log\left(\frac{M_\text{het}^4}{M_{\rm IIB}^4} \frac{\mathcal{V}_{B_2}^{(0)}}{2\pi}\right) = - \frac{b}{8\pi} \log\left( \frac{\mathcal{V}_{B_2}^{H}}{2\pi}\right) \,.
\end{align}
In the case of gauge group $G=U(1)$ the beta-function coefficients $b$ and $c$ are in fact equal, $c=b$, and hence the F-theory corrections reproduce the heterotic result. For more general cases one should take into account that group-dependent threshold corrections give additional contributions $\tilde\Delta_1$.

For the simple case of a $U(1)$ gauge group we conclude that heterotic/F-theory duality gives a condition for the F-theory uplift of the parameters $\tilde{\mathcal{Z}}_\alpha$  by relating them to the $\beta$-function on the divisor $\mathbf{S}$ in addition to the constraints encountered in \eqref{kappa4rel}. Similar proposals for the coefficient of the $\log \mathcal{V}_{B_3}$ term have already been made in \cite{Conlon:2009kt,Conlon:2009qa,Conlon:2010ji} and \cite{Weissenbacher:2020cyf} from the perspective of Type IIB orientifolds.

\subsubsection*{Example}

In order to gain some intuition for the threshold corrections at order $1/(\mu\lambda)^2$, let us study a base $B_3=B_2 \times \bbP^1$ that is trivially $\bbP^1$-fibered. The K\"ahler cone of $B_3$ is generated by $J_\star=[B_2]$ and $J_a=[D_a^{\scriptscriptstyle B_2}\times \bbP ^1]$ such that
\begin{equation}
  J_a\cdot J_b\cdot J_\star =k_{ab}\;,\qquad J_\star^2=J_a\cdot J_b\cdot J_c =0\;.
\end{equation}
From the triple intersecions we see immediately that $J_a$ can either be identified with $J_0$ or generators of type $\cI _3$ in a $J$-class $A$ limit, whereas $J_\star$ must be the only generator of type $\cI _1$. It follows that the most general Ansatz for the K\"ahler form is
\begin{equation}
  J=\mu\left(\lambda J_0+\frac{a^\star}{\lambda^2} J_\star+\sum_{r\in \cI _3}b^r J_r\right)\;.
\end{equation}

Even though the trivial $\bbP ^1$-fibration is a rather degenerate case because $K_{\mu\nu}=K_{\star\star}=0=D$, the formulae from appendix~\ref{App_volC0:classA} still apply as the scalar $D$ drops out for any geometry with only a single $\cI _1$ generator. The quantities $\Delta^\alpha$ in~\eqref{eq:KijC0relation_subleading} simplify significantly as a consequence of $\hat{D}=\delta D=\delta B=0$ as well as $k_{00\mu}=k_{00\star}$. We have
\begin{equation}
  \Delta^m=0\;,\qquad\qquad \Delta^\star = \left(\tilde{b}^T A \hat{B}^T\right)^\star=\frac{a^\star \mu}{\lambda^2}=v^\star\;.
\end{equation}

Assuming that the gauge divisor is $\mathbf{S}=J_\star$, it then follows that the only relevant correction terms at order $1/(\mu\lambda)^2$ are proportional to either $\mathcal{Z}_\star$ or $\mathcal{T}_\star$. The quantity $\mathcal{Z}_\star$ depends on the details of the elliptic fibration over $B_3$ through the third Chern class $c_3(Y_4)$\footnote{See table 5 of~\cite{Esole:2018tuz}.}. Independently of the elliptic fibration we can see that
\begin{equation}
  \mathcal{T}_\star\sim J_\star^2\cdot J=0\;.
\end{equation}
The resulting sub-leading $\alpha'$-correction is
\begin{equation}
  \frac{2m\mathcal{V}_{B_3}}{\mathcal{V}_{\mathbf{S}}\mathcal{V}_{C_0}}=1-\frac{\alpha^2\tilde{\mathcal{Z}}_\star}{\mathcal{V}_{\mathbf{S}}^{(0)}}\left[-\frac12 + \log\mathcal{V}_{B_3}^{(0)}\right]\;.
\end{equation}
Note that, as discussed in section~\ref{subsec_classical_het-F_duality} the classical $\mathcal{O}(1/\lambda^3)$ corrections vanish for a trivial fibration.

\section{The Weak Gravity Conjecture with quantum corrections} \label{sec_wgc}

We now apply the analysis of the previous sections  to prove the Weak Gravity Conjecture (WGC) in the infinite distance limits which we have been considering. 
First, we will show the WGC relation for all types of classical limits in the K\"ahler moduli space in which the gauge coupling on a stack of 7-branes tends to zero.
This completes the previous results of \cite{Lee:2019jan} by including also the $J$-class B limits. 
Our main focus is then on the subleading corrections to the WGC, working to 1-loop order in the heterotic frame.
The gauge threshold corrections imply that naively the Weak Gravity Conjecture is no longer satisfied; the WGC can, however, be saved if we allow for a 1-loop mass renormalisation for the states in WGC tower. We make an ansatz for such mass renormalisation effects and constrain it by demanding that the force between two WGC test particles be repulsive.
This determines the required mass renormalisation in terms of the threshold corrections. As a result the WGC is satisfied, but with a modified charge-to-mass ratio (\ref{newWGC-2}) compared to
the leading, classical value.
Interpreting this modified bound as the charge-to-mass ratio of extremal black holes gives a prediction for this quantity in the one-loop corrected four-dimensional effective theory.

\subsection{The WGC relation to leading order in F-theory}   \label{sec_WGCrelQC_recap}

To recapitulate our findings so far, we have identified two types of infinite distance limits in the K\"ahler moduli space of F-theory which are not pure decompactification limits:
The base $B_3$ admits either a torus fibration or a fibration by a rational curve $\mathbb P^1$, whose volume shrinks in each case compared to the base volume. 
In the context of proving the Weak Gravity Conjecture, only the 
 second type of limits plays a role, as it is in one-to-one correspondence with infinite distance limits in which a 7-brane gauge group factor becomes weakly coupled.

Let us briefly recall the argument why this is so from \cite{Lee:2018urn,Lee:2019jan}. If a 7-brane gauge group becomes weakly coupled, the volume of the divisor ${\bf S}$ wrapped by the 7-brane stack in question must tend to infinity.
One can check the consequences of this requirement for both limits of $J$-class A, (\ref{combinedJclassA}), and of $J$-class B, (\ref{combinedJclassB}), subject to the constraints that they are not of pure decompactification type. 
For $J$-class A, this is the limit (\ref{JclassAhetrelabelded}) and for $J$-class B the limit (\ref{JclassBhetrelabelded}).
In both cases, the volume of  ${\bf S}$ tends to infinity in the limit if and only if the intersection product with the fiber is non-zero, ${\bf S} \cdot C_0  > 0$. Furthermore, if ${\bf S}$ supports a 7-brane gauge group, it must satisfy the relation
 ${\bf S} \leq n \bar K_{B_3}$ for some value of $n$, where  $\bar K_{B_3}$ is the anti-canonical divisor on $B_3$.
 This is obvious in the case of a non-abelian gauge symmetry because in this case ${\bf S}$ is a component of the discriminant divisor $\Delta = 12 \bar K_{B_3}$. If the gauge group in question is purely abelian (and not a Cartan subgroup of a non-abelian gauge group factor), then ${\bf S}$ corresponds to the height-pairing associated with the rational section of the elliptic fibration which underlies the existence of the $U(1)$ gauge group factor in F-theory. 
 Even though not  proven in full generality, it is conjectured that such height-pairings likewise satisfy a bound $b \leq n \bar K_{B_3}$ for some $n > 0$ \cite{Lee:2018urn}. 
 Combined with the fact that ${\bf S} \cdot C_0  > 0$, the relation ${\bf S} \leq n \bar K_{B_3}$ implies that ${\bar K}_{B_3} \cdot C_0  > 0$. This singles out $C_0$ as a rational, rather than as a torus, fiber of $B_3$.

Hence in the limits in which $g_{\rm YM} \to 0$ for a 7-brane gauge group in F-theory a unique heterotic string, obtained by wrapping a D3-brane along the rational fiber $C_0$ of $B_3$, becomes tensionless.
This allows us to prove the Weak Gravity Conjecture in such limits as follows \cite{Lee:2018urn,Lee:2019jan}.
For simplicity we focus on a single abelian gauge group factor.
The aim is to show that the excitations of the weakly coupled heterotic string associated with the fiber $C_0= \mathbb P^1$  contain a tower of states of charge $q_k$ and mass $M_k$ which satisfy the WGC relation \label{ArkaniHamed:2006dz}
\bea \label{WGCrel-a}
\frac{g^2_{\rm YM} q_k^2}{M_k^2} \stackrel{!}{\geq} \frac{g^2_{\rm YM} Q^2_{\rm BH}}{M_{\rm BH}^2} \quad   \stackrel{g_{\rm YM} \to 0}{=}  \quad  \frac{1}{M^2_{\rm Pl}}    \,.
\eea
Here $\frac{g^2_{\rm YM} Q^2_{\rm BH}}{M_{\rm BH}^2}$ denotes the charge-to-mass ratio of a suitable extremal black hole 
in the four-dimensional effective theory.
The gauge coupling is to be evaluated at the scale of the WGC tower itself.

In computing the black hole charge-to-mass ratio, one has to take into consideration the effect of all massless scalars in the effective theory
which change the extremal black hole solution \cite{Heidenreich:2015nta}.
In the strict weak coupling limit, our theory asymptotes to
 four-dimensional Dilaton-Einstein-Maxwell theory, and the precise numerical value for this ratio indicated in (\ref{WGCrel-a})
follows from the 
explicit coupling of the dilaton in the four-dimensional heterotic frame \cite{Lee:2019jan}. We will discuss the next-to-leading order corrections to this value momentarily.

An alternative way to understand the Weak Gravity Conjecture bound is to demand that there exists a tower of states which cannot form bound states, taking into account again the contribution of the
massless scalars to the effective interaction between charged massless test particles \cite{Palti:2017elp} \cite{Lee:2018spm} \cite{Heidenreich:2019zkl}.  
This condition amounts to requiring that for the WGC tower
\bea \label{repforce}
|F_{\rm Coulomb}|   & \stackrel{!}{\geq}  &   |F_{\rm Grav}|  +  |F_{\rm Yuk}|   \,,
\eea
where $|F_{\rm Coulomb}|$ denotes the repulsive Coulomb force
while the attractive force includes the universal gravitational interaction  $|F_{\rm Grav}|$ along with a model dependent Yukawa interaction $|F_{\rm Yuk}|$  mediated by massless scalar fields \cite{Palti:2017elp}.
For a single abelian gauge interaction and with the massless real scalar fields $\phi^r$ normalised as in
\bea \label{effectiveaction}
S = \int_{\mathbb R^{1,3}}  \frac{M^2_{\rm Pl}}{2}  ( \sqrt{-g} R  -  g_{r s}   {\rm d} \phi^r \wedge \ast {\rm d} \phi^s )  - \frac{1}{2g^2_{\rm YM}}  F \wedge \ast F \,,
\eea
this criterion requires that the WGC tower satisfies \cite{Palti:2017elp,Lee:2018spm,Heidenreich:2020upe} 
\bea   \label{WGCscalars}
\frac{g^2_{\rm YM} q_k^2}{M_k^2}   &\stackrel{!}{\geq} &    \frac{1}{M^2_{\rm Pl}}   \left(  \frac{d-3}{d-2}|_{d=4} +  \frac{1}{4} \frac{M^4_{\rm Pl}}{M_k^4}  g^{rs}  \partial_r  \left(\frac{M_k^2}{M^2_{\rm Pl}}  \right)\partial_s  \left(\frac{M_k^2}{M^2_{\rm Pl}}\right)  \right)   \,.
\eea
Here we assumed that the ratio $M_k/M_{\rm Pl}$, with $M_k$ the particle mass, depends on the VEV of scalar fields $\phi^r$. In the strict weak coupling limit, the leading contribution to the Yukawa couplings stems from the combination of K\"ahler moduli dual to the heterotic tree-level dilaton.

At least in the infinite distance limit $g_{\rm YM} \to 0$ the two approaches indeed agree \cite{Lee:2018spm,Gendler:2020dfp} (more generally, see \cite{Heidenreich:2019zkl}). 
The point is that, as we now show, asymptotically 
\bea   \label{partialMweaklimit}
\frac{M^4_{\rm Pl}}{M_k^4}  g^{rs}  \partial_r  \left(\frac{M_k^2}{M^2_{\rm Pl}} \right)   \partial_s  \left(\frac{M_k^2}{M^2_{\rm Pl}}\right) = 2   \qquad \text{in the weak coupling limit}   \,.
\eea
Let us see how this works in our particular setup. The real scalar fields $\phi^r$ are the linear multiplets $L^\alpha$, whose inverse kinetic matrix is given by 
\begin{align}\label{galphabeta}
 g^{\alpha \beta}=2 \frac{\partial^2 K}{\partial \text{Re} T_\alpha \partial \text{Re}\,T_\beta} = -2 \frac{\partial L^\alpha}{\partial \text{Re}\,T_\beta} = L^\alpha L^\beta - \frac{2}{{\cal V}_{B_3}}  K^{\alpha \beta}\,.
\end{align} 
Here  $K^{\alpha \beta}$ is the inverse of the matrix $K_{\alpha \beta}$ defined in (\ref{Kalphabeta}).
The massive states on the other hand are given by the excitations of the heterotic string. For a classical, weakly coupled heterotic string, the mass of the states at excitation level $n_k$ is related to the heterotic string scale $M_{\rm het}$ as 
\bea   \label{MkMhetrel}
M_k^2 = 8 \pi M^2_{\rm het} (n_k-1)   \,.
\eea
Together with the second and third relation in
\bea   \label{gMhetMpl}
g^2_{\rm YM}   = \frac{2 \pi}{  {\cal V}_{\bf S} }  \,, \qquad   M^2_{\rm het}   = 2 \pi {\cal V}_{C_0}  \,, \qquad M^2_{\rm Pl}= (4 \pi)  {\cal V}_{B_3}   \,,
\eea
one hence finds
\begin{align}
\frac{M_k^2}{M_{\rm Pl}^2}= 4 \pi (n_k-1)  \frac{{\cal V}_{C_0} }{{\cal V}_{B_3}}   =     4 \pi  (n_k-1)    \sum_{\mu\in \mathcal{I}_1} k_{00\mu} L^\mu  \,. 
\end{align}
In the last equality we applied the relation (\ref{VCalphauantum}) to the curve $C_0 = J_0 \cdot J_0$ associated with the heterotic string, and 
the notation refers to the K\"ahler form in the weak coupling limit of $J$-class $A$. A similar relation holds in the limits of $J$-class B.

In this language the RHS of \eqref{WGCscalars} reduces to
\bea   \label{partialMweaklimit-2}
\frac{1}{M_{\rm Pl}^2}\left[\frac{1}{2} +\frac{1}{4}\left(  \sum_{\mu\in \mathcal{I}_1} k_{00\mu} L^\mu  \right)^{-2}   g^{\mu \nu}   \partial_\mu \left(\sum_{\rho\in \mathcal{I}_1}k_{00\rho} L^\rho \right)\partial_\nu \left(\sum_{\rho\in \mathcal{I}_1} k_{00\rho} L^\rho\right)\right] = \frac{1}{M_{\rm Pl}^2}\left[\frac{1}{2}+ \frac{1}{2}\right]\,,
\eea
where we used that 
\bea
k_{00\mu} g^{\mu \nu} \quad   \stackrel{g_{\rm YM} \to 0}{=}  \quad  2 k_{00\mu} L^\mu L^\nu
\eea
to leading order in 
the strict weak coupling limit. This reproduces the black hole relation \eqref{WGCrel-a}.

To conclude the discussion so far, we must show the existence of a tower of states with the property that to leading order in the weak coupling limit 
\bea   \label{WGCrel-b}
\frac{g^2_{\rm YM} q_k^2}{M_k^2} \stackrel{!}{\geq}   \frac{1}{M^2_{\rm Pl}}     \qquad {\rm as} \quad g_{\rm YM} \to 0 \,. 
\eea
The proof  consists  in two steps:
First one establishes the existence of certain a tower of states in the heterotic spectrum with $U(1)$ charge $q_k$ at excitation level $n_k$ such that 
\bea   \label{tower-ineq}
q_k^2 \geq 4 m   \, n_k    \,,
\eea
where $m= \frac{1}{2} C_0 \cdot {\bf S} > 0$. We will discuss this step in Section \ref{sec_WGCpf}. 

In the second step one deduces from the inequality (\ref{tower-ineq})  the actual asymptotic WGC relation (\ref{WGCrel-b}).  Using \eqref{MkMhetrel} and \eqref{gMhetMpl} we can recast (\ref{tower-ineq}) as
\bea   \label{WGCrecast-gen}
\frac{g^2_{\rm YM} q_k^2 }{M_k^2} \geq   \frac{g^2_{\rm YM}}{4\pi}  \frac{2m}{M^2_{\rm het}} =   \frac{2m {\cV_{B_3}} }{\cV_{\bf S}{ \cV_{C_0}}}   \frac{1}{M^2_{\rm Pl}}    \,.
\eea
This means that the states satisfying (\ref{tower-ineq}) have the property that 
\bea \label{WGCalpha0}
\frac{g^2_{\rm YM} q_k^2 }{M_k^2} \geq   \frac{g^2_{\rm YM}}{4\pi}  \frac{2m}{M^2_{\rm het}} &=&   \frac{2m {\cV_{B_3}} }{\cV_{\bf S}{ \cV_{C_0}}}   \frac{1}{M^2_{\rm Pl}}  = 
   \left(1 -  \Delta + \ldots \right) \frac{1}{M^2_{\rm Pl}}   \,.
\eea
For the last equality we used (\ref{ratiodef2})  and recall that the correction $\Delta$ scales as 
\bea
\Delta = \Delta_0  +  \Delta_1    + \ldots = {\cal O}(1/\lambda^3) +  {\cal O}(1/\mu^2\lambda^2)   \,.
\eea
This relation holds both for limits of $J$-class A and B (see Appendix \ref{App_WGCclass-B}).
Hence to leading order in $\lambda$ the WGC relation (\ref{WGCrel-b}) is indeed satisfied.

\subsection{The WGC at 1-loop order}  \label{sec_WGC1loop}  

We now elaborate on the next-to-leading order terms in the WGC relation (\ref{WGCrel-b}) and 
make a proposal how the WGC is still satisfied taking these into account.

Our point of departure are the subleading corrections $\Delta$ in (\ref{WGCalpha0}). 
In Section \ref{subsec_classical_het-F_duality} we have identified these as one-loop corrections to the classical relation (\ref{treelevelrelation}) between the heterotic gauge coupling $g_{\rm YM}$ and the heterotic dilaton.

If  $g_{\rm YM}$  were the only quantity to receive such subleading loop corrections, this would jeopadize the WGC away from the 
 strict weak-coupling limit. 
 Indeed, the relation  (\ref{WGCalpha0}) obeyed by the would-be superextremal tower
 does not coincide with the bound (\ref{WGCrel-b}) at subleading order, see Figure \ref{fig:WGC_relation_corrections} for an illustration.
 Note that in general the sign of the threshold corrections $\Delta$ is model-dependent, as can be seen both from the geometric expression
 in F-theory and from the dependence of the thresholds on the bundle data in the dual heterotic frame (see e.g. \cite{Blumenhagen:2005ga}).
 The sign chosen in Figure \ref{fig:WGC_relation_corrections} is the one obtained for a specific example discussed at the end of this section.

What saves the day is that also the mass of the string excitations for the solitonic heterotic string in F-theory 
may experience loop corrections modifying (\ref{MkMhetrel}) to subleading order.
The heterotic string itself is BPS in the four-dimensional $N=1$ theory and thus supersymmetry protects the ratio between its tension and its two-form charge. However, the excitations of this string are not BPS and their masses can get renormalised at the same order through string loop effects.
Such a correction to the relation (\ref{MkMhetrel}) will affect both sides of the WGC condition in its form \eqref{WGCscalars}. This can in principle cancel the correction to  $g_{\rm YM}$ such that 
the WGC remains to hold to subleading order.

Ideally one would compute the loop corrections to  (\ref{MkMhetrel}) from first principles and test if they are of the right form such that the WGC \eqref{WGCscalars} continues to hold.
Instead,
we turn tables round and infer the form of the subleading corrections to (\ref{MkMhetrel}) needed so that \eqref{WGCscalars} holds at sub-leading order in $\lambda$. We therefore put the WGC to use in order to make a prediction
 for a physical observable, here the string-loop corrected masses of the string excitations.

The corrections to the mass of the excitations are expected to appear at one-loop level in the heterotic string coupling. Hence in the following we take $M_\text{het}^2/M_{\rm Pl}^2$ as our expansion parameter to single out the one-loop correction and make the ansatz
\begin{align}\label{Mkansatz}
  \frac{M_k^2}{M_{\rm Pl}^2} = 8\pi (n_k-1) \frac{M_\text{het}^2}{M_{\rm Pl}^2}\left(1+ \frac{M_\text{het}^2}{M_{\rm Pl}^2} \mathfrak{m}\left(L\right)\right)  =: 8\pi (n_k-1) \frac{M_\text{het}^2}{M_{\rm Pl}^2}\left(1+ 
   \delta \right)       \,.
\end{align}
The function $\mathfrak{m}(L)$ encodes the corrections to the excitations masses in terms of the linear multiplets denoted collectively as $L$. 
We can now plug this ansatz for the mass correction into the RHS of (\ref{WGCscalars}) and expand it to 
 linear order in $\mathfrak{m}\left(L\right)$. 
 Then
 the WGC relation becomes
\bea \label{newWGC}
\frac{g^2_{\rm YM} q_k^2 }{M_k^2} \stackrel{!}{\geq}   \frac{1}{M^2_{\rm Pl}}  \left(  1 - \delta + \frac{1}{2}  \frac{M^4_{\rm Pl}}{M^4_{\rm het}}  g^{r s} \partial_r \left( \frac{M_\text{het}^2}{M_{\rm Pl}^2} \right)  \partial_s \left( \frac{M_\text{het}^2}{M_{\rm Pl}^2} \delta\right)  + {\cal O}(\delta^2) \right)   \,,
\eea
where we used (\ref{partialMweaklimit}).

We now aim to determine  the mass correction 
\be   \label{deltaexpression}
\delta = \frac{M^2_{\rm het}}{M^2_{\rm Pl}} \mathfrak{m}(L)
\ee 
by imposing  \eqref{newWGC} on the tower of states satisfying (\ref{tower-ineq}).
By combining (\ref{Mkansatz}) and (\ref{WGCalpha0}) we compute the charge-to-mass ratio of these states as
\bea  \label{gMkrelcorr}
\frac{g^2_{\rm YM} q_k^2 }{M_k^2}  \geq \frac{1}{M^2_{\rm Pl}} (1 - \Delta - \delta)  \,.
\eea 
The Weak Gravity Conjecture is therefore satisfied if the RHS of (\ref{gMkrelcorr}) equals  the bound appearing in (\ref{newWGC}). Equating both expressions yields
a differential equation for the mass correction  $\delta$, valid to linear order,
\bea \label{DGLfordelta}
\frac{1}{2}  \frac{M^4_{\rm Pl}}{M^4_{\rm het}}   g^{rs}  \partial_r \left( \frac{M^2_{\rm het}}{M^2_{\rm Pl}} \right)    \partial_s \left( \frac{M^2_{\rm het}}{M^2_{\rm Pl}}  \delta \right)    \stackrel{!}{=} - \Delta \,.
\eea
Since $\delta$ is given by (\ref{deltaexpression}), 
we clearly have 
\bea
\partial_r \left(   \frac{M^2_{\rm het}}{M^2_{\rm Pl}}  \delta \right)   = 2 \partial_r \left( \frac{M^2_{\rm het}}{M^2_{\rm Pl}} \right) \delta +  \frac{M^4_{\rm het}}{M^4_{\rm Pl}} \partial_r \mathfrak{m}(L) \,.
\eea
The combination ${M^2_{\rm het}}/{M^2_{\rm Pl}}$ corresponds to the dilaton, and the derivative acting on it reduces the power of the dilaton for the first term. Hence the second term is subleading
by one power of ${M^2_{\rm het}}/{M^2_{\rm Pl}}$ in the weak coupling limit.
Plugging this back into (\ref{DGLfordelta}) and using
 (\ref{partialMweaklimit}), with $M_k$ traded for $M_{\rm het}$, we arrive at the relation
\bea   \label{deltaDeltarel}
\delta = - \frac{1}{2} \Delta + \ldots   \,,
\eea
where we neglected the terms at subleading power in the heterotic dilaton.
With this input, the WGC condition (\ref{newWGC}) turns into the 
\begin{whitebox}
\bea \label{newWGC-2}
\text{WGC relation at 1-loop}:  \qquad 
\frac{g^2_{\rm YM} q_k^2 }{M_k^2} \stackrel{!}{\geq}   \frac{1}{M^2_{\rm Pl}}  \left(  1  - \frac{1}{2} \Delta \right)   \,,
\eea
\end{whitebox}
where we recall that $\Delta = \Delta_0 + \Delta_1$ is related to the gauge thresholds via (\ref{threshold1}) and (\ref{threshold2}).
This relation is manifestly satisfied by the tower (\ref{tower-ineq}) as long as the mass corrections $\delta$ obey (\ref{DGLfordelta}).

To summarise, 
the 1-loop corrections to the gauge coupling together with the mass renormalisation
effects modify the effective WGC bound on the charge-to-mass ratio
as in (\ref{newWGC-2}). We have come to this conclusion by interpretating the WGC as the repulsive force condition (\ref{repforce}), (\ref{WGCscalars}), and by {\it demanding} that the 
mass renormalisation effects conspire with the gauge threshold corrections such that the repulsive-force condition (\ref{WGCscalars}) is still satisfied.
As reviewed, in the strict weak coupling limit this condition is equivalent to requiring super-extremality with respect to extremal black holes.
If both approaches to the WGC, (\ref{WGCrel-a}) and  (\ref{WGCscalars}), continue to agree including the 1-loop corrections,
this means that the effective value for the charge-to-mass ratio of an extremal black hole appearing on the RHS of (\ref{WGCrel-a}) must likewise be modified at 1-loop level of the effective action, i.e.
\bea
\frac{g^2_{\rm YM} Q_{\rm BH}^2 }{M_{\rm BH}^2} =  \frac{1}{M^2_{\rm Pl}}  \left(  1  - \frac{1}{2} \Delta \right)   \,.
\eea
It would be extremely interesting to investigate this further.

Let us evaluate (\ref{newWGC-2}) in a regime where the classical contributions to $\Delta$ are leading compared to the $\alpha'$-corrections, i.e.
where $\Delta_0 \gg \Delta_1$. In the heterotic duality frame this requires large heterotic volume.
In this regime,
the differential equation (\ref{DGLfordelta}) can be expressed as
\begin{align}\label{horribleequation}
& \frac{1}{4} \frac{ M_{\rm Pl}^4}{M_\text{het}^4} g^{\mu \alpha} \left(k_{00\mu}\frac{\partial}{\partial {L^\alpha}} \left( \frac{M_\text{het}^4}{M_{\rm Pl}^4} \mathfrak{m} \left(L\right) \right)\right) + \mathcal{O}\left(\frac{M_\text{het}^6}{M_{\rm Pl}^6}\right)= \frac{V_{B_3}^{(0,1)}}{V_{B_3}^{(0,0)}}-\frac{V_\mathbf{S}^{(0,1)}}{V_\mathbf{S}^{(0,0)}}\,,
\end{align}
 with $\mu \in \mathcal{I}_1$ (cf. (\ref{JclassA-2}))
and $\alpha$ running over all K\"ahler moduli. Here for definiteness we assumed a weak coupling of $J$-class A, for which we recall that 
\bea
\frac{M_\text{het}^2}{M_{\rm Pl}^2} = \frac{1}{2}    \frac{\cV_{C_0}}{\cV_{B_3}} = \frac{1}{2} k_{00\mu} L^\mu   \,;
\eea
a similar expression holds for the  $J$-class B limits.
To arrive at (\ref{horribleequation}) we furthermore approximated $\Delta$ by $\Delta_0$  and made use of (\ref{Delta0expre-a}).

 To develop an intuition for this differential equation let us consider the simple example where the base of the F-theory fourfold $B_3$ is a rational fibration over $\mathbb{P}^2$,
\begin{align}
 B_3=\mathbb{P}\left(\mathcal{O}_{\mathbb{P}^2}\oplus \mathcal{O}_{\mathbb{P}^2}(-t)\right)\,,\qquad t\in \mathbb{N}_0\,. 
\end{align}
The K\"ahler cone is spanned by the hyperplane class $H$ of $\mathbb{P}^2$ and the section $S_-$, which satisifies 
\begin{align}
 S_- \cdot S_- = t S_- \cdot H\,,
 \end{align}
in agreement with our notation of Section \ref{subsec_classical_het-F_duality}.
The intersection polynomial reads
\begin{align}
 I(B_3)=  t^2 S_-^3 +t S_-^2 H + S_- H^2\,.  
\end{align}
Let us parametrise the K\"ahler form as 
\begin{align}
 J= v^h H + v^s S_-\,,
\end{align}
and denote the associated linear multiplets as $L^h\equiv h$ and $L^s\equiv s$. In this language the heterotic string tension is given by 
\begin{align}
 \frac{M_\text{het}^2}{M_{\rm Pl}^2}= \frac{s}{2}\,,
\end{align}
and therefore we are interested in the terms up to and including $\mathcal{O}(s^2)$. The components of the kinetic metric $g^{\alpha \beta}$ in this example are computed as 
\begin{align}
   g^{ss}=2s^2\,,\qquad g^{sh}=-t s^2 \,,\qquad g^{hh}=h^2 + t hs +\mathcal{O}(s^3)\,. 
\end{align}
As for the corrections to the mass of the string excitations \eqref{Mkansatz} reduces to 
\begin{align}\label{Mkansatz-2}
  \frac{M_k^2}{M_{\rm Pl}^2} = 4\pi (n_k-1) s \left(1+ \frac{s}{2} \,\mathfrak{m}\left(s,h\right)\right) \,.
\end{align}
 
Let us consider a $U(1)$ gauge theory in this geometry. The gauge divisor $\mathbf{S}$ is identified with the height pairing associated with the extra $U(1)$, which we parametrise as
\bea  \label{Sdef}
\mathbf{S}  = x S_- + y H  \,, \qquad \quad  x\geq 0\,, \, \, y \geq - x t \,.
\eea
The range of $x$ and $y$ is chosen such that $\mathbf{S}$ is effective.
Correspondingly 
\be
m = \frac{1}{2}\mathbf{S}   \cdot C_0 = \frac{x}{2} \,.
\ee
For this model, one computes to leading order in $s$ that 
\bea
- \Delta = \frac{V_{B_3}^{(0,1)}}{V_{B_3}^{(0,0)}}-\frac{V_\mathbf{S}^{(0,1)}}{V_\mathbf{S}^{(0,0)}} = - \left(t + 2 \frac{y}{x}\right) \frac{s}{h}  + \ldots  \,,
\eea
such that we get
\begin{align}
\partial_s\left(s^2 \mathfrak{m}(s,h)\right)= -  \frac{2 s}{h}\left(t+ 2 \frac{y}{x}\right)+\mathcal{O}(s^3) \,.
\end{align}
Since we are interested in the terms up to $\mathcal{O}(s^2)$ we can expand $\mathfrak{m}(s,h)=\mathfrak{m}^{(0)}(h) + s\, \mathfrak{m}^{(1)}(h) + \dots$ and read off
\begin{align}
\mathfrak{m}^{(0)}(h)=-\frac{1}{h} \left( t + 2 \frac{y}{x}  \right) \,   \Longrightarrow \delta = -  \frac{s}{2h}  \left( t + 2 \frac{y}{x}  \right) + \ldots = - \frac{1}{2} \Delta + \ldots  \,.
\end{align}
This confirms the general claim (\ref{deltaDeltarel}).
The one-loop corrected excitation mass is accordingly given by 
\begin{align}
 \frac{M_k^2}{M_{\rm Pl}^2} = 4\pi (n_k-1) s\left(1-   \frac{s}{2h}  \left( t + 2 \frac{y}{x}  \right)  \right) + \mathcal{O}(s^3)\,. 
\end{align}

It is also interesting to evaluate the net change in the WGC relation (\ref{newWGC-2}), which becomes
\bea
\frac{g^2_{\rm YM} q_k^2 }{M_k^2} \stackrel{!}{\geq}   \frac{1}{M^2_{\rm Pl}}  \left(1-   \frac{s}{2h}  \left( t + 2 \frac{y}{x}  \right)  \right)   
\eea
to one-loop order. Hence, in this concrete example, the sign of $\Delta$
depends on the choice of $y$ and $x$, which must lie in the range indicated in (\ref{Sdef}).
For instance for a $U(1)$ restricted Tate model, ${\bf S} = 2 \bar K_{B_3} = 4 S_- + 2 (3-t) H$ and therefore, e.g. for $t=1$, the 
bound on the charge-to-mass ratio is \emph{lowered} by the loop corrections, as depicted in Figure \ref{fig:WGC_relation_corrections}.

Finally, let us mention that the WGC for the string itself is not corrected by the subleading terms and hence the string tension as such does not receive loop corrections as expected due to its BPS properties. To see this note that the expressions for the tension and the charge of the BPS string under the 2-from gauge symmetry, obtained by reducing $C_4$ along the rational fiber,
\begin{align}
Q^2= \frac{1}{4} M_{\rm Pl}^2 \, g^{ss}=\frac{1}{2} M_{\rm Pl}^2 s^2 \,,\qquad M_\text{het}^2=\frac{1}{2} s M_{\rm Pl}^2\,,
\end{align}
are not affected by the correction terms and hence satisfy the no-force condition (see e.g. \cite{Lanza:2020qmt}) 
\begin{align}
 M_{\rm Pl}^4g^{ss} \partial_s \frac{M_\text{het}^2}{M_{\rm Pl}^2}\partial_s \frac{M_\text{het}^2}{M_{\rm Pl}^2}= \frac{1}{2} M_{\rm Pl}^4 s^2 = Q^2 M_{\rm Pl}^2 
\end{align}
also in the presence of the corrections due to the non-trivial twist of the fibration.

\subsection{Super-extremal sublattice in the heterotic spectrum} \label{sec_WGCpf}

It remains to show that a tower of states exists which satisfies the inequality~\eqref{tower-ineq},
\bea  \label{tower-ineq2}
q_k^2 \geq 4 m   \, n_k    \,,
\eea
where $q_k$ and $n_k$ are, respectively, the $U(1)$ charge and the mode number of the $k$-th excitation in the tower. As discussed, the inequality~\eqref{tower-ineq2}
is key to proving that the states in the tower are super-extremal, in the sense of (\ref{newWGC-2}).
The results in this section are an extension of the discussion in \cite{Lee:2019jan} in the light of the recent results \cite{Lee:2020gvu,Lee:2020blx}.

We have already noted in Section \ref{sec_FluxedD} the importance of the flux backgrounds $G \in H^4(Y_4)$ to specify an F-theory model. Of particular relevance to
the analysis in this paper are the  gauge fluxes on the internal part of the 7-branes.
As before consider a theory with an abelian gauge group factor $U(1)$ which becomes weakly coupled at infinite distance.
One way to engineer such a theory\footnote{We will comment on a different origin of $U(1)$ symmetries in F-theory on fourfolds at the end of this section.}  is if the elliptic fibration $Y_4$ admits a rational section $S_A$, to which we can associate a non-trivial class $\sigma(S_A)$, the image of the section under the so-called Shioda map (see e.g. the reviews~\cite{Weigand:2018rez,Cvetic:2018bni} for details). The gauge divisor ${\bf S}$ is related to $\sigma(S_A)$ via the relation ${\bf S} = - \pi_\ast(\sigma(S_A)\cdot \sigma(S_A))$, where $\pi$
is the projection of the elliptic fibration.  
On $Y_4$ one can then specify a non-trivial background $U(1)$ flux $F$, which combines with the class $\sigma(S_A)$ into the four-form flux
\beq\label{U(1)flux}
G = \pi^*F \wedge \sigma(S_A) \,, \qquad \text{with}\quad  F\in H^{1,1}(B_3)\,.
\eeq

We will argue that for generic such $U(1)$ fluxes, there exist states obeying~\eqref{tower-ineq2} with $U(1)$ charges 
\beq\label{sublattice}
q_k = 2m k \,, \qquad \forall k \in \mathbb N \,,
\eeq
thereby confirming the Sublattice Weak Gravity Conjecture.

Let us first recall how such a sublattice~\eqref{sublattice} arises in the analogous setup in $6$ dimensions with $N=(1,0)$ supersymmetry. It was established in~\cite{Lee:2018urn, Lee:2018spm} that a subset of the full spectrum, as captured by the elliptic genus of the asymptotically tensionless heterotic string, already provides the desired sublattice particles. Specifically, the elliptic genus of a six-dimensional heterotic string associated with the curve $C_0$, 
\beq\label{exp-2m}
\mathcal Z_{-2,m}\left[C_0\right] (q, \xi) = -q^{-1} \sum_{n,r} N(C_0(n,r))\,q^n\, \xi^r \,,
\eeq
is a meromorphic Jacobi form of weight $-2$ and index $m$. Here, the fugacity parameters $q$ and $\xi$ distinguish states with different $U(1)$ charges and mode numbers, and the expansion coefficients $N(C_0(n,r))$ are interpreted as the degeneracies\footnote{In fact $N(C_0(n,r))$ are the degeneracies of the left-moving excitations counted with signs since the elliptic genus is defined as an index. Nevertheless, non-trivial $N(C_0(n,r))$ with $n\geq 1$ still imply a non-zero number of excitations, and hence, the existence of spacetime particles is guaranteed via level matching.} of the states. 
Then the Jacobi nature of $\mathcal Z_{-2,m}\left[C_0\right]$ was used~\cite{Lee:2018urn} to assure, via its theta expansion~\cite{EichlerZagier}, that the expansion~\eqref{exp-2m} contains super-extremal terms for every charge $q_k$ in the sublattice~\eqref{sublattice}. Furthermore, in the sequel~\cite{Lee:2018spm}, the argument was generalized to theories with multiple $U(1)$ gauge fields; the elliptic genus is then promoted to a higher-rank Jacobi form, for which the index $m$ is a matrix, but the theta expansion is still viable and leads to the Sublattice Weak Gravity Conjecture for multiple $U(1)$s. 

The situation is rather different for four-dimensional compactifications with $N=1$ supersymmetry. After all, the elliptic genus of the four-dimensional heterotic string,
\beq\label{exp-4d}
\mathcal Z_{-1,m} \left[G, C_0\right](q,\xi)= -q^{-1} \sum_{n,r} N_G(C_0(n,r)) \,q^n \,\xi^r  \,,
\eeq
is {\it not} a Jacobi form for a generic four-form flux $G$. The day can be saved, however, by the recent observation~\cite{Lee:2020gvu,Lee:2020blx} that the elliptic genus of a four-dimensional string decomposes as\footnote{This decomposition structure is in fact a manifestation of the quasi-Jacobi nature~\cite{Oberdieck:2017pqm} of four-dimensional elliptic genera. }
\beq\label{dec}
\mathcal Z_{-1,m} \left[G, C_0\right] = Z_{-1,m} + \xi \partial_\xi Z_{-2,m} \,,
\eeq
where $Z_{w, m}$ (for $w=-1, -2$) are individually a Jacobi form of weight $w$ and index $m$. 

For a { \it generic}  flux background, both contributions in (\ref{dec}) are non-vanishing. 
To prove the sublattice WGC it suffices, in fact, for the term of the form $\xi \partial_\xi Z_{-2,m}$ to be non-zero.
Indeed, the non-trivial $Z_{-2,m}$ by itself contains a tower of super-extremal terms obeying~\eqref{tower-ineq2}, which furnish a charge sublattice,
\beq\label{sublattice-generic}
q_k = 2mk \,, \qquad k=1,2,3,\ldots\,, 
\eeq
according to precisely the same logic as in the six-dimensional setup. Then, because the role of the derivative $\xi \partial_\xi$ in~\eqref{dec} is simply to multiply all the expansion coefficients of $Z_{-2,m}$ by the associated $U(1)$ charges, the super-extremal sublattice terms in $Z_{-2,m}$ persist in the full elliptic genus~$\mathcal Z_{-1,m}\left[G,C_0\right]$. 
Note that the super-extremal terms in $\xi \partial_\xi Z_{-2,m}$ cannot be cancelled by $Z_{-1,m}$ since the latter has vanishing expansion coefficients for any $U(1)$ charges in the sublattice~\eqref{sublattice-generic}~\cite{Lee:2019jan}. 
Rather than forming a sublattice, the states encoded in $Z_{-1,m}$ include a
 tower of super-extremal states which form a {\it shifted}\footnote{The shift~\eqref{shift} of the sublattice by $1$ follows from the fact that Jacobi forms of weight $w=-1$ are an odd function in $\xi$. This does not necessarily mean that no {\it unshifted} super-extremal sublattices exist since the elliptic genus captures only a subsector of the full spectrum.} sublattice
\beq\label{shift}
q_k = 2m k + 1\,, \qquad k=0,1,2, \ldots\,,
\eeq 
in the full charge lattice. This proves the WGC in generic flux backgrounds.

For non-generic fluxes, on the other hand, the second, derivative contribution to (\ref{exp-4d}) may be vanishing. 
Such fluxes were called {\it quasi-modular} or {\it modular} fluxes in~\cite{Lee:2019jan}, depending on the properties of the contribution $Z_{-1,m}$.
Hence for such special fluxes, the sublattice version of the WGC cannot be proven based on the elliptic genus $\mathcal Z_{-1,m} \left[G, C_0\right]$ alone because the latter only
guarantees the existence of states in the shifted sublattice (\ref{shift}). This of course is not sufficient to show that no unshifted sublattice of super-extremal states exists because 
 $\mathcal Z_{-1,m} \left[G, C_0\right]$ only computes an index of states, rather than the full partition function.
 An extreme case would be situations without any gauge flux, $F=0$, where the elliptic genus vanishes completely and the theory is non-chiral.
 Chirality is therefore crucial in order to prove the WGC via the route of the elliptic genus.

This proof of the Sublattice Weak Gravity Conjecture, both in six-dimensional $N=(1,0)$ and in chiral four-dimensional $N=1$ setups, relies heavily on the Jacobi behaviour exhibited by the elliptic genus of the heterotic string, propagating in six and four dimensions, respectively. However, the effective physics is much richer in the latter setup, which, to some extent, is a natural consequence of the less amount of supersymmetry. We will end this section by commenting on two distinctive features in four-dimensional $N=1$ compactifications with fluxes:

\subsubsection*{St\"uckelberg mass}

The chirality of the theory in generic flux backgrounds also leads to another important aspect of the abelian gauge theory, namely that its gauge boson acquires a St\"uckelberg mass $M_{st}$.
At scales below $M_{st}$, the $U(1)$ is merely a global symmetry and in particular the Coulomb interaction at those scales is exponentially suppressed.
In view of the interpretation of the WGC via the 
 repulsive-force condition (\ref{repforce}) it is therefore important to compare the scale at which the WGC bound is evaluated and $M_{st}$.
 For the classical finite-volume limit this was already analysed in~\cite{Lee:2019jan}. Since this particular limit is obstructed by $\alpha'$-corrections and we are forced to impose a rescaling in order to stay in the perturbative limit, we will reconsider the evaluation of the St\"uckelberg mass here.
 As we will see, $M_{st}$ lies below the scale of the emergent weakly coupled string tower.
This means that in the emergent string limit the $U(1)$ gauge boson is effectively massless at the mass scale of the emergent string for any choice of gauge flux and hence mediates a long-range force. Since we evaluate the WGC precisely at the emergent string scale, the $U(1)$ is indeed generating the Coulomb potential that enters the repulsive force conjecture used in our previous discussion.

It is well known that the non-trivial anomaly cancelling term $\int B \wedge F$ in a four-dimensional chiral effective theory~\cite{Lerche:1987qk, Lerche:1987sg}
is responsible for the St\"uckelberg mass. 
In our present setup, the two-forms $B^\alpha$ arise from reducing the IIB 4-form $C_4$ along the 2-cycles of $B_3$. Including the $\int B\wedge F$ term into the effective action \eqref{effectiveaction} and dualizing the two-forms $B^\alpha$ and the corresponding linear multiplets $L^\alpha$ into the chiral multiplets $T_\alpha$ we arive at\footnote{We have changed the normalisation of the gauge field compared to \eqref{effectiveaction} in order to directly read off the physical St\"uckelberg mass from the Lagrangian.}
\begin{align}
 S \supset \int_{\mathbb{R}^{1,3}} -\frac{1}{2}F\wedge \ast F + M_{\rm Pl}^2 g^{\alpha \bar \beta} dT_\alpha \wedge \ast d \bar T_{\bar \beta}  + \frac{M_{\rm Pl}^2}{2} f_a^2 \left(da - g_{\rm YM} A \right)\wedge \ast\left(da - g_{\rm YM} A \right)\,.
\end{align} 
The St\"uckelberg axion $a$ is a linear combination of the axions $\text{Im} T_\alpha$  which depends on the chosen fluxes. The St\"uckelberg mass is then 
\begin{align}
 M^2_{st} = g_{\rm YM}^2 f_a^2 M_{\rm Pl}^2\,, 
\end{align}
where the (dimensionless) axion decay constant $f_a$ is determined by the fluxes to be (cf. \cite{Lee:2019jan}) 
\begin{align}
 f_a^2 = g^{\alpha \beta} \Pi_\alpha \Pi_\beta \,.
\end{align} 
Here $g^{\alpha \beta}$ is given by \eqref{galphabeta} and the $\Pi_\alpha$ are flux dependent topological quantities 
\begin{align}
  \Pi_\alpha=\omega_\alpha \cdot \mathbf{S} \cdot F \,. 
\end{align} 
Let us first consider a model without blow-up divisors such that there are no heterotic NS5-brane wrapping curves in the base of the dual heterotic Calabi-Yau. For the most general fluxes, the leading contribution to $f_a^2$ comes from $\Pi_\alpha$ with $\alpha \in \{0\}\cup \mathcal{I}_3$ ($J$-class A, and similarly for $J$-class B limits) and with the corresponding metric element scaling like e.g. 
\begin{align}
 g^{00}\sim \left(L^0\right)^2 \sim \frac{\lambda^2}{\mu^4}\,.
\end{align} 
In this case the St\"uckelberg mass scales like 
\begin{align}
 M_{st}^2 \sim \frac{1}{\mu^2 \lambda^2}\frac{\lambda^2}{\mu^4} M_{\rm Pl}^2 = \frac{1}{\mu^6} M_{\rm Pl}^2 = \frac{\lambda}{\mu^2} M_{\rm KK}^2\,.
\end{align}
Here the KK scale is determined by the volume of the base $B_2$ 
\bea
 \frac{M_{\rm KK}^2}{M_{\rm Pl}^2} = \frac{1}{\mathcal{V}_{B_2}^{1/2}} \frac{M_{\rm IIB}^2}{M_{\rm Pl}^2} \sim \frac{1}{\lambda\mu^4} \,.
\eea 
Thus for emergent string limits, where $\lambda \sim \mu^2$, and generic fluxes the St\"uckelberg mass sits at the KK scale as far as
dependence on the infinite distance parameters $\lambda$ and $\mu$ is concerned.
However, we still have to require
 $\lambda/\mu^2<1$ to keep control of the perturbative $\alpha'$-corrections, i.e. 
 $M_{st}$ is numerically suppressed with respect to $M_{\rm KK}$. In particular this also means that the St\"uckelberg mass is numerically suppressed with respect to the emergent string scale.

Let us analyse what happens for more special choices for fluxes. For so-called \textit{modular fluxes} in the nomenclature of \cite{Lee:2019jan}, $\Pi_\alpha\ne 0$ only for $\alpha \in \mathcal{I}_1$ (in the language of $J$-class A limits) such that we get 
\begin{align}
  M_{st}^2 =  \frac{M_{\rm Pl}^2}{(\lambda \mu)^6} = M_{\rm Pl}^2 g_{\rm YM}^6 \,,
\end{align}
where we used $g^{\mu \nu} \sim L^\mu L^\nu$ if $\mu, \nu \in \mathcal{I}_1$. This is consistent with the  dual heterotic picture since for modular fluxes the St\"uckelberg axion is just the universal heterotic axion with axion decay constant $f_a^2 = g_{\rm YM}^4 M_{\rm Pl}^2$. Finally let us consider the case that $B_3$ has blow-up divisors. As reviewed in \cite{Lee:2019jan} in this case the St\"uckelberg axion on the heterotic side receives contributions from the axions dual to the two-forms of NS5-brane worldvolume theory. Wrapping the NS5 brane on a curve $\Gamma_a^H$ in the base of the heterotic Calabi-Yau threefold yields for the effective action of the 2-form $B_a$
\begin{align}
 S_a = \frac{1}{f_a^2 M_{\rm Pl}^2} \int_{\mathbb{R}^{1,3}} dB_a \wedge \ast d B_a = \frac{\mathcal{V}_{\Gamma_a^H}}{M_{\rm Pl}^2} \int_{\mathbb{R}^{1,3}} dB_a \wedge \ast d B_a \,.
\end{align}
As discussed in section \ref{subsec_classical_het-F_duality} we have to identify $\mathcal{V}_{\Gamma_a^H}$ on the F-theory side with the volume of the corresponding vertical divisor such that we get 
\begin{align}
 f_a^2 =  \frac{1}{\mathcal{V}_{\Gamma_a^H} } \sim \frac{\lambda}{\mu^2} \, \qquad \Rightarrow \qquad M_{st}^2 = \frac{1}{\lambda \mu^4} M_{\rm Pl}^2 = M_{\rm KK}^2\,. 
\end{align}
Thus even if the St\"uckelberg axion receives contributions from the NS5-brane axions the St\"uckelberg mass of the $U(1)$ gauge boson is of the order of the KK-scale and thus numerically suppressed with respect to the emergent string scale.

\subsubsection*{Flux-induced breaking to $U(1)$}
While $U(1)$ gauge bosons can be realised via a non-trivial Mordell-Weil lattice of sections in six-dimensional F-theory, fourform fluxes provide yet another route to realise them in four dimensions via flux-induced breaking. 
Let us briefly convince ourselves that the proof of the WGC can be extended to such constructions.

Suppose the theory contains a non-abelian gauge algebra $\mathfrak{g}$, 
which is realised by a stack of 7-branes along a divisor ${\bf S}$ on $B_3$.
There are largely two qualitatively different types of vertical $\mathfrak{g}$ fluxes one may consider in such geometries. 
Cartan fluxes take the form
\beq
\pi^*F \wedge E_{i} \,, \qquad \text{with}\quad  F\in H^{1,1}(B_3)\,,
\eeq
where $E_i$ is the exceptional divisor on the fourfold $Y_4$ associated to the Cartan $\mathfrak{u}(1)_i$ of $\mathfrak g$. With such a flux turned on, the Lie algebra is by construction broken as
\beq\label{Cartan-flux-breaking}
\mathfrak g \to \mathfrak h_i \oplus \mathfrak{u}(1)_i \,, 
\eeq
where $\mathfrak h_i$ is the commutant of $\mathfrak{u}(1)_i$ within $\mathfrak{g}$. Then, the elliptic genus refined with respect to $\mathfrak u(1)_i$ once again decomposes as~\eqref{dec}. Hence, precisely the same logic applies as for $U(1)$ symmetries engineered via rational sections, leading to (at worst shifted) sublattices of super-extremal particles. 

One can also consider turning on
gauge fluxes which can induce chirality with respect to a non-abelian gauge group factor without breaking it to its Cartan by itself (see e.g \cite{Weigand:2018rez} and references therein).
The elliptic genus in such a flux background would still have a derivative sector as in~\eqref{dec}, where $Z_{w,m}$ are $\mathfrak{g}$-invariant Jacobi forms, as opposed to weak Jacobi forms. One may then turn on in addition a Cartan flux to break $\mathfrak{g}$ as in~\eqref{Cartan-flux-breaking}. Then, the theta expansion of the final elliptic genus with both fluxes combined would in the end lead to the desired super-extremal particles.

All in all, we conclude that the elliptic genus guarantees the presence of a desired super-extremal tower also for such flux-induced $U(1)$s. 

\section{Conclusions and outlook}

In this paper we have elucidated open questions arising in the {\it Swampland Program} for four-dimensional ${N}=1$ supersymmetric theories.
The computational framework in which we have embedded the discussion is F-theory compactified on a Calabi-Yau fourfold, for which initial investigations of the Swampland Distance Conjecture (SDC) \cite{Ooguri:2006in} and the Weak Gravity Conjecture (WGC)  \cite{ArkaniHamed:2006dz} have appeared in~\cite{Lee:2019jan}.
As for theories with eight supercharges, infinite distance limits either lead to a decompactification or feature emergent critical strings \cite{Lee:2019oct}, whose excitations are candidates for the WGC and SDC towers of states. The ubiquity of perturbative and non-perturbative quantum corrections to the K\"ahler potential and superpotential makes the four-dimensional minimally supersymmetric setting an interesting playground for sharpening our understanding of the rich interplay between swampland conjectures, string dualities and the quantum geometry of moduli spaces.

 The main new ingredient that we have taken into account in the present work are the leading $\alpha'^2$-corrections to the F-theory effective action, as computed in \cite{Grimm:2013gma, Grimm:2013bha,Grimm:2017pid,Weissenbacher:2019mef, Weissenbacher:2020cyf}. Already at this level we have seen that the inclusion of quantum corrections is crucial for the consistency of the theory in that they censor pathological limits which would lead to a four-dimensional critical string theory with the compactification space decoupled. This mirrors the situation in theories with eight supersymmetries \cite{Marchesano:2019ifh,Baume:2019sry}. Although our results are promising, due to the low amount of supersymmetry we necessarily have to rely on a limited amount of information and a fully non-perturbative understanding of the fate of these limits is currently out of reach.

The limits that we have considered are for the K\"ahler parameters of the three-dimensional base $B_3$ of the elliptically fibered fourfold on which we compactify F-theory. The proper $N=1$ coordinates in this case are the four-cycle volumes of the base. Tensionless emergent strings can appear in case the base is itself fibered by a rational curve (heterotic string) or a torus (Type II string). Since the limits involve a shrinking of the fiber to zero size, one has to carefully take into account stringy $\alpha'$-effects which could obstruct it. Although we find that the shrinking of the fiber itself does not appear to pose a problem, the implied shrinking of vertical divisors can lead to diverging $\alpha'$-corrections. This is because of the nature of the perturbative $\alpha'$-expansion as an expansion in inverse divisor volumes as opposed to curve volumes.

Delving into some more detail, we have found that, unlike in six dimensions \cite{Lee:2018urn}, the naive finite volume infinite distance limits of~\cite{Lee:2019jan} generically do not survive the leading perturbative $\alpha'$-corrections. 
In order to make the limits work, we had to superimpose them with an additional homogeneous rescaling of the F-theory base. We have then demonstrated that a subset of the resulting two-parameter limits is able to evade the constraint imposed by the $\alpha'$-corrections. Interestingly, the censored limits are precisely those that would lead to an inconsistent decoupling of the KK tower from the emergent string, i.e. $T_{\rm str}/M_{\rm KK}^2\to 0$, while the parametric scaling $T_{\rm str}\sim M_{\rm KK}^2$ is marginally allowed.

To be more precise, the leading $\alpha'$-correction depends on certain characteristic integrals such as $\int c_3\wedge \pi^*(J_\alpha)$ on the Calabi-Yau fourfold \cite{Grimm:2013gma, Grimm:2013bha,Grimm:2017pid,Weissenbacher:2019mef, Weissenbacher:2020cyf}. In the case of a $\mathbb{P}^1$-fibration these are generically non-zero and we have shown this explicitly at least in $J$-class B limits of a smooth Weierstrass model over a smooth, projective, almost Fano base $B_3$. Moreover, in the latter class of models the sign is such that in the pathological limits perturbative control is lost already at \emph{finite distance}. In contrast, if the vanishing fiber of $B_3$ is a $T^2$ and $Y_4$ is a smooth Weierstrass model the characteristic integrals vanish so there can be no obstruction at leading order although we have no reason to expect this to be the case for higher orders in the $\alpha'$ expansion. It would certainly be desirable, once available, to include also higher order $\alpha'$-corrections into the analysis. In addition, a more exhaustive and systematic study of the sign of the leading order correction is possible but left for future work.

Another puzzle of the remaining consistent limits was that there could in principle exist several non-contractible curves in $B_3$ that shrink at the same rate in the limit, leading to a theory with multiple light fundamental strings. In fact it is not hard to come up with an example such as $B_3=(\mathbb{P}^1)^3$ ($J$-class B limit). By refining the analysis of~\cite{Lee:2019jan} of the intersection properties of the relevant divisors in $B_3$ and the scaling of the divisor volumes we find that such limits must always correspond to a decompactification $M_{\rm KK}^2/T_{\rm str}\to 0$. For the truly emergent string limits $T_{\rm str}\sim M_{\rm KK}^2$ the string is always unique!

Regarding non-perturbative corrections we have  explained why the volume of the shrinking fiber does not receive any corrections. First, compactifying on a torus to two dimensions one can use fourfold mirror symmetry to study world-sheet instanton corrections to the volume of the fiber. We have found that neither are these important in the limit, nor do they uplift to F-theory. Second, D3-brane instantons on four-cycles can give rise to obstructions through their contribution to the K\"ahler potential or superpotential. For those limits which survive the perturbative $\alpha'$-corrections we have found that even though the D3-instanton induced superpotential can be (at most) constant the scalar potential is suppressed due to the universal $1/\mathcal{V}_{B_3}$ factor. By contrast, the scalar potential can in principle diverge in those limits in which classically $T_{\rm str}/M_{\rm KK}^2\to 0$. In the case of a rationally fibered $B_3$ we have seen that this is in fact expected and including also the perturbative $\alpha'$-corrections the superpotential blows up at finite distance so that the solitonic heterotic string runs into a strong coupling singularity.

Quite differently, in the case of a genus-one fibration where we already observed the vanishing of the leading perturbative $\alpha'$-correction we have also found no indication of an obstruction by a superpotential. Although the fact that we have not observed any necessary obstructions for limits with a shrinking genus one fiber is surprising, this does not present an inconsistency as these limits never lead to the offending hierarchy $T_{\rm str}/M_{\rm KK}^2\to 0$. Nevertheless it would be extremely interesting to study this further.

After clarifying the existence and nature of the infinite distance limits we have addressed the interpretation of the excitations of the emergent string as the tower of states satisfying the sublattice WGC. At leading order in the limit and ignoring the $\alpha'$-corrections we have shown that the relation $\mathcal{V}_{\mathbf{S}}\cdot\mathcal{V}_{C_0}=2m\mathcal{V}_{B_3}$, which was shown for the $J$-class A limits already in~\cite{Lee:2019jan}, holds also in the $J$-class B limits. This relation between the volumes of the gauge divisor ${\bf S}$, shrinking fiber $C_0$ and $B_3$ when translated into a relation between the gauge coupling, emergent string tension and Planck mass is one of the crucial ingredients in proving the sublattice WGC. The other ingredient is the spectrum of the emergent heterotic string. Extending the results of~\cite{Lee:2019jan} and making use of the recent observations of~\cite{Lee:2020gvu,Lee:2020blx} we have managed to identify a sublattice of charged states satisfying the WGC, at least for generic flux backgrounds.

Beyond the strict weak coupling limit we have proposed a dictionary between corrections to the relation $\frac{1}{g_{\rm YM}^2}\sim\mathcal{V}_{B_3}/\mathcal{V}_{C_0}$ in F-theory and loop corrections to the gauge coupling on the heterotic side. In F-theory the corrections arise both from deviations from the relation $\mathcal{V}_{\mathbf{S}}\cdot\mathcal{V}_{C_0}=2m\mathcal{V}_{B_3}$ in classical K\"ahler geometry away from the strict infinite distance limit as well as from the perturbative $\alpha'$-corrections. Here the analysis of the $\alpha'$-corrections was limited to the case of a $U(1)$ gauge group. For different gauge groups, a more careful analysis of the group-dependent threshold corrections would be required.

Including these corrections we have found a deviation from the classical repulsive force condition for the tower of states just discussed, which could only be repaired by assuming that the classical relation between the heterotic string tension $T_{\rm het}$ and the excitation masses $M_n^2\sim n T_{\rm het}$ is corrected. 
Requiring that the same sublattice of states which satisfies the classical WGC bound still complies with the WGC at the 1-loop level hence leads to a prediction for 
the mass renormalisations of the string excitations. 

Furthermore, by demanding that the repulsive force condition is still identical with the black hole extremality bound we have turned this into a prediction for the corrected black hole masses. It would be exciting to check the consistency of our proposal with a proper calculation of the one-loop corrected black hole charge-to-mass ratio in heterotic supergravity or of the 1-loop mass renormalisation effects for the string excitations. Another interesting question for future work is whether we can make the index of the charge sublattice arbitrarily large. 

Clearly, the topic of swampland conjectures in low-dimensional theories with minimal or no supersymmetry continues to be an exciting area of research. 

\subsubsection*{Acknowledgements}

We thank Wolfgang Lerche, Fernando Marchesano, Irene Valenzuela and Eran Palti for helpful comments and correspondence.
SL and TW are particularly grateful to Wolfgang Lerche for many important discussions and collaboration on related matters. The work of MW  is supported by the Spanish Research Agency (Agencia Estatal de Investigaci\'on) through the grant IFT Centro de Excelencia Severo Ochoa SEV-2016-0597, and by the grant PGC2018-095976-B-C21 from MCIU/AEI/FEDER, UE. The work of MW also received the support of a fellowship from ”la Caixa” Foundation (ID 100010434) with fellowship code LCF/BQ/DI18/11660033 and funding from the European Union’s Horizon 2020 research and innovation programme under the Marie Sklodowska-Curie grant agreement No. 713673. The work of SJL is supported by IBS under the project code, IBS-R018-D1.

\appendix

\section{Uniqueness results for classical infinite distance limits}

In this appendix we prove the uniqueness of the fastest-shrinking fibers in finite volume infinite distance limits of $J$-class A and $J$-class B, respectively.

\subsection{\texorpdfstring{$J$}{J}-class A}   \label{App_JclassA}

We consider a limit of the form (\ref{JclassA-1}). In addition to the intersection numbers (\ref{JclassA-2}), we recall from \cite{Lee:2019jan} that 
\bea\label{0r}
J_0 \cdot J_r &=& n_r J_0 \cdot J_0 \,,\quad n_r >0 \,,\\ \label{rs}
J_r \cdot J_s &=& n_{rs} J_0 \cdot J_0 \,, \quad n_{rs} \geq 0\,.\nn
\eea
Let us furthermore introduce the notation
\bea
k_{\alpha\beta\gamma}   = J_{\alpha}   \cdot J_\beta  \cdot J_\gamma   \,.
\eea
Then (\ref{0r}) together with  (\ref{JclassA-2}) implies $k_{0rs} = 0$ and $k_{rst}=0$.

As reviewed in Section \ref{subsec_JAclassical}, the base $B_3$ admits a rational or genus-one fibration whose generic fiber is given by the curve $C_0 = J_0\cdot J_0$ and whose volume shrinks in the limit $\lambda \to \infty$.
Furthermore, if $B_3$ admits another fibration by a curve $\tilde C_0$, there must exist a nef divisor $\tilde D$ with the properties (\ref{D2}) and (\ref{D3}).
Let us assume that $\tilde C_0$ is a shrinking curve, i.e. that
\beq
{\cal V}'_{\tilde D^2} \to 0     \quad \text{for}~\l \to \infty\,.
\eeq
As we will see, this leads to a contradiction.

The proof is an application of the proof leading to Proposition 1 in \cite{Lee:2019oct}, which studied limits of $J$-class A for Calabi-Yau threefolds, rather than for a K\"ahler threefold $B_3$. 
Since $\tilde D$ is a nef divisor, it can be expanded in terms of the K\"ahler cone generators of $B_3$ as 
\beq
\tilde D = p^0 J_0 + \sum_{\m \in \cI_1} p^\m J_\m + \sum_{r \in \cI_3} p^r J_r \,, \quad p^0,\, p^\m,\, p^r \geq 0 \,.
\eeq
Furthermore, since by assumption $\tilde D^2$ is proportional as a class to $C_0 = J_0 \cdot J_0$, at least one $p^\m$ must be strictly positive (otherwise, by the intersection properties~\eqref{0r} and~\eqref{rs}, $\tilde D^2$ would be proportional to $C_0$). We pick one index 
\beq
\a_0 \in\cI_1   \quad  \text{such that }   \quad p^{\a_0} >0   \,.
\eeq
An obvious consequence is that $p^0 =0$; if not, $\tilde D^3$ contains a positive contribution,
\beq
(p^0)^2 p^{\a_0} k_{00\a_0}    \neq 0 \,,
\eeq
which contradicts the vanishing~\eqref{D3}. 
It also follows that 
\beq
p^r =0  \quad \forall r \in \cI_3 \,,
\eeq
since otherwise ${\cal V}'_{\tilde D^2} = J' \cdot \tilde D^2$ contains a non-trivial $\cO(\l)$ contribution given by
\beq
\l p^{\a_0} p^{r_0} k_{0r_0\a_0} = \l p^{\a_0} p^{r_0} n_r k_{00\a_0}\,,
\eeq
where $r_0$ is a putative index in $\cI_3$ for which $p^{r_0} >0$, and $n_r$ is a constant of proportionality in~\eqref{0r} which is strictly positive. We thus have
\beq
\tilde D = \sum_{\a \in \cI_1} p^\m J_\m \,.
\eeq

Let us now choose a pair $(\m_0, \n_0) \in \cI_1$ for which 
\beq
J_{\m_0} \cdot J_{\n_0} \neq 0 \quad \text{and}\quad p^{\m_0}, p^{\n_0} >0 \,.
\eeq
Such pair must exist in order for the class $\tilde D^2$ to be non-vanishing. Then, 
\beq\label{0a0b0}
{\cal V}'_{\tilde D^2} = J' \cdot \tilde D^2 \geq \l p^{\m_0} p^{\n_0} k_{0\m_0\n_0}   \,, 
\eeq
and the asymptotic vanishing of ${\cal V}'_{\tilde D^2}$ implies that 
\be   \label{doab=0}
 k_{0\m_0 \n_0} = 0    \,.
\ee
Now, Lemma 5 in Appendix B.1 of \cite{Lee:2019oct}  states that 
\beq
J_0 \cdot J_{\m_0} = \k J_{\m_0} \cdot J_{\n_0} \,, \quad \k >0 \,. 
\eeq
Using this result along with (\ref{doab=0})   leads to 
\beq
k_{00\m_0} = \k k_{0 \m_0 \n_0} = 0 \,,
\eeq
in contradiction with the fact that $k_{00\m} = J_0 \cdot J_0 \cdot J_{\m}  \neq 0$ for all $\m \in \cI_1$ for limits of $J$-class A.
This proves that no $B_3$ cannot admit any other fibration by a genus-one or a rational curve which also shrinks in the limit $\lambda \to \infty$.

\subsection{\texorpdfstring{$J$}{J}-class B}     \label{App_JclassB}

We now turn to the $J$-class B limit as defined in  (\ref{Jprime-classB})   subject to (\ref{Jprime-classB-int}).
As explained in the main text, in order for the limit not to correspond to a (partial) decompactification, there must exist a leading-shrinking  fiber $C_0$ of genus $0$ or $1$, obtained as~\eqref{D2-B}, whose volume behaves as
\beq\label{JB-C0-para}
{\cal V}'_{C_0} \simeq {\cal V}'_{D^2} \sim \l^{-\g}\,, \quad \text{with}~~\g \geq 1   \,.
\eeq

Let us start by expanding the nef divisor $D$, associated to $C_0$ by~\eqref{D2-B}, as
\beq  
D = p^0 J_0 + \sum_{\hat \m \in \hat \cI_2} p^{\hat \m}_+ J_{\hat \m} \,, \quad \text{with}~~p^0 \geq 0, ~p^{\hat \m}_+ >0\,,
\eeq
where $\hat \cI_2 \subset \cI_2$ is a non-empty index subset. It is then natural to decompose $\cI_2$ as
\beq
\cI_2 = \hat \cI_2 \cup \bar \cI_2 \cup \check \cI_2 \,,
\eeq
such that 
\bea
D^2 \cdot J_{\bar \m} &=& 0 \,,\quad \forall \bar \m \in \cI_2  \,,\\
D^2 \cdot J_{\check \m} &>& 0 \,, \quad \forall \check \m \in \cI_2 \,.
\eea
We will now show  that without loss of generality $\bar \cI_2 = \emptyset$ and also that the K\"ahler coefficients $b^{\prime \mu}$ are subject to the following constraints:
\bea\label{muhat}
b^{\prime \hat \m} &\precsim& \l^{\g-1} \quad \forall \hat\m \in \hat \cI_2\,, \quad \text{and the inequality saturates for at least one}~~\hat \mu \,, \\ \label{mucheck}   
b^{\prime \check \m} &\precsim& \l^{-\g}\quad  \, \,  \forall\check\m \in \check \cI_2\,, \quad  \text{and the inequality saturates for at least one}~~\check \mu\,. 
\eea
Note that~\eqref{muhat} will imply in particular that $\gamma \leq 2$ due to the general form~\eqref{Jprime-classB} of the $J$-class B limit. 

First, to justify $\bar \cI_2 = \emptyset$, we observe that the nef divisor $\bar D$ defined as
\beq
\bar D = D + \sum_{\bar \mu \in \bar \cI_2} p^{\bar \m}_+ J_{\bar \m} \,, \quad \text{with}~~ p^{\bar \m}_+ > 0 
\eeq
leads to the same curve class (up to scaling) as $C_0$, that is,\footnote{See the arguments from (B.103) to (B.114) in \cite{Lee:2019oct} for details. The key is to observe that $D^2$, $D \cdot J_{\bar \m}$, and $J_{\bar \m} \cdot J_{\bar \n}$ are proportional to one another as curve classes for all $\bar \m, \bar \n  \in \bar \cI_2$.}
\be
\bar D^2 = \k D^2 = \k n C_0 \,,\quad \text{with}~~\k >0 \,.
\ee
The scaling behavior of ${\cal V}'_{C_0}$ is then computed as
\be
{\cal V}'_{C_0} \sim {\cal V}'_{D^2} = J'\cdot D^2 \sim J' \cdot J_0 \cdot J_{\hat \mu} = \sum_{\check \nu \in \check \cI_2} b^{\prime\check \n} k_{0\check \n \hat \m} \sim \sum_{\check \n} b^{\prime\check \n} \,,
\ee
where $\hat \m \in \hat \cI_2$ is an arbitrarily chosen index. Here, we have made use of the fact that $D^2$ and $J_0 \cdot J_{\hat\mu}$ are proportional classes, and also of the following intersection properties\footnote{\label{proprels}One observes that $D^2$,  $D \cdot J_0$, $J_0 \cdot J_{\hat \m}$, and $J_{\hat \m} \cdot J_{\hat \n}$ are all proportional to one another as curve classes for all $\hat \m, \hat \n \in \hat \cI_2$. This follows from the vanishing of $V_{C_0}$ in the limit $\l\to \infty$, which implies $k_{0\hat \m \hat \n}=0$ $\forall \hat \m, \hat\n \in \hat \cI_2$.}: 
\beq
k_{00\hat \m} = 0\,,\quad k_{0\hat \n \hat \m}=0 \,,\quad k_{0 \check \n \hat \m} >0 \, \qquad \forall \hat \n \in \hat \cI_2, ~\forall \check \n \in \check \cI_2 \,.
\eeq
If ${\cal V}'_{C_0}$ scales as in~\eqref{JB-C0-para}, we thus see that 
\beq\label{subleading}
b^{\prime\check \nu} \precsim \l^{-\g} \,\quad \forall\check\n\in\check\cI_2\,,
\eeq
which furthermore has to saturate for at least one $\check \n$. 
Now, the volume of $B_3$ is computed as
\beq
{\cal V}'_{B_3} = \frac16J'^3 \sim \l b^{\prime\hat \m} b^{\prime\check \n} k_{0\hat \m \check\n} \,,
\eeq
where the subleading contributions have been neglected based on~\eqref{subleading}. Finiteness of ${\cal V}'_{B_3} $ implies that
\beq\label{bhm}
b^{\prime\hat \m_0} \sim \l^{\g-1}  \,\quad \text{for some}~~{\hat\m_0}\in\hat\cI_2\,,
\eeq
and also that 
\beq
b^{\prime\hat \m} \precsim \l^{\g-1} \, \quad \forall\hat\m \in \hat\cI_2 \,,
\eeq
as claimed. 

We will now show that there cannot exist any other such leading-shrinking curve fiber $\tilde C_0$. To this end we suppose that another such fiber $\tilde C_0$ exists with the volume behavior, 
\beq\label{tC0}
{\cal V}'_{\tilde C_0} \sim {\cal V}'_{C_0} \sim \l^{-\g} \,.
\eeq
Necessarily the fiber $\tilde C_0$ is associated with a nef divisor $\tilde D$ obeying
\bea
\tilde D^2 &=& \tilde n \tilde C_0 \,, \quad \tilde n \in \mathbb N\\ 
\tilde D^3 &=& 0 \,.
\eea
We then have
\beq\label{tDDD}
\tilde D \cdot D^2 = \k \tilde D \cdot D \cdot J_0 = \k' \tilde D \cdot J_0 \cdot J_{\hat \mu_0} = \k'' \tilde D^2 \cdot J_{\hat \m_0} \,, \qquad \k, \k', \k'' >0 \,,
\eeq
for $\hat \m_0 \in\hat \cI_2$ appropriately chosen to have the property~\eqref{bhm}. Here, we have used the fact that, for a nef divisor $D$ associated with $C_0$, the intersection products $D^2$, $D \cdot J_0$, $J_0\cdot J_{\hat \m_0}$ are proportional classes, and similarly for $\tilde D$ (see footnote~\ref{proprels}). Let us now note that the last intersection in~\eqref{tDDD} must vanish since, otherwise, we have  
\beq
{\cal V}'_{\tilde C_0} \sim {\cal V}'_{\tilde D^2} = J'\cdot \tilde D^2  \succsim b^{\prime\hat \m_0}  \tilde D^2 \cdot J_{\hat \m_0} \sim b^{\prime\hat \m_0} \,,
\eeq
which, given the scaling behaviors~\eqref{bhm} and~\eqref{tC0}, implies $-\g \geq \g-1$ and hence, $\g \leq \frac12$, contradicting~\eqref{JB-C0-para}. Therefore,
\beq
\tilde D \cdot D^2 = 0 \,,
\eeq
and hence
\beq
\tilde D \cdot D  = \k_D D^2 \,, \quad \text{with}~~\k_D >0\,, 
\eeq
by Lemma 5 of \cite{Lee:2019oct}. Of course, precisely the same logic leads to the proportionality
\beq
\tilde D \cdot D = \k_{\tilde D} \tilde D^2 \,, \quad \text{with}~~\k_{\tilde D} >0 \,, 
\eeq
and therefore, we learn that $D^2$ and $\tilde D^2$ are proportional to each other as curve classes, and hence, $C_0$ and $\tilde C_0$ are one and the same fiber.

\section{The classical weak gravity bound in \texorpdfstring{$J$}{J}-class B limits}   \label{App_WGCclass-B}

In this appendix, we consider $J$-class B limits and prove the geometric relation required for the sublattice weak gravity tower to asymptotically saturate the classical super-extremality bound: 
\beq\label{bound-C0}
\cV'_{\bf S} \cV'_{C_0} = 2m \cV'_{B_3} \,.
\eeq

We work in the unrescaled limit (\ref{Jprime-classB-int}) such that, following the notation of Section \ref{subsec_JBclassical}, $\cV'_{\ast}$ refers to the classical volume of $\ast$, ignoring subleading corrections in the limit.

The proof of (\ref{bound-C0}) is analogous in spirit to the case of $J$-class A provided in \cite{Lee:2019jan} and goes as follows. Note first that since $m=\frac12 C_0 \cdot \bf S$ is linear in the curve class $C_0$, we may replace $C_0$ in~\eqref{bound-C0} by $C_{\hat \mu_0}=J_0 \cdot J_{\hat \m_0}$ for a fixed $\hat \mu_0 \in \hat \cI_2$ (recall that $D^2$ and $J_0 \cdot J_{\hat \m}$ are both proportional to $C_0$ as curve classes; see footnote~\ref{proprels}). We will thus prove that 
\beq
\cV'_{\bf S} \cV'_{C_{\hat \mu_0}} = 2m_{\hat \mu_0} \cV'_{B_3} \,,
\eeq
where $m_{\hat \mu_0}$ is defined as
\beq\label{rescalem}
m_{\hat \mu_0} = \frac12 C_{\hat \mu_0} \cdot {\bf S}\,.
\eeq

To show this, we first expand the gauge divisor $\bf S$ as
\beq\label{Sexp}
{\bf S} = h^0 J_0 + \sum_{\hat \mu \in \hat \cI_2} h^{\hat \m} J_{\hat \mu} + \sum_{\check \nu \in \check\cI_2} h^{\check \nu} J_{\check \n}\,,
\eeq
and compute in turn the parametric behaviours for the volumes of $C_{\hat \mu_0}$, $\bf S$ and $B_3$ in the limit $\l \to \infty$:
\bea
\cV'_{C_{\hat \mu_0}} &=& J'\cdot J_0 \cdot J_{\hat \mu_0} = \sum_{\check\nu\in\check\cI_2} b^{\prime\check\n} k_{0\hat\mu_0\check\n}  \,,\\
\cV'_{\bf S} &=& \frac12 J'^2 \cdot {\bf S} = \frac12 (\l J_0 + \sum_{\hat\m\in\hat\cI_2} b^{\prime\hat\m} J_{\hat \m})^2 \cdot {\bf S} + \cdots  \\
&=& \frac12 (2 \l \sum_{\hat\m \in \hat\cI_2} b^{\prime\hat \m} \k_{\hat\m} + \sum_{\hat\m,\hat\n\in\hat\cI_2}b^{\prime\hat\m}b^{\prime\hat\n} \k_{\hat\m\hat\n} ) J_0 \cdot J_{\hat\m_0} \cdot \sum_{\check \r \in \check\cI_2} h^{\check \r} J_{\check \r} +\cdots \\ 
&=& \frac12 \left(\sum_{\check \r \in \check\cI_2} h^{\check\r} k_{0\hat\m_0 \check\r} \right)\left( \sum_{\hat \m,\hat \n\in \hat \cI_2} b^{\prime\hat \m} (2\l \k_{\hat \m} + b^{\prime\hat \n} \k_{\hat \m \hat \n})\right) +\cdots \\
\cV'_{B_3} &=& \frac16 J'^3 = \frac12 (\l J_0 + \sum_{\hat \m \in \hat \cI_2} b^{\prime\hat \m} J_{\hat \m} )^2 \cdot \sum_{\check \r\in\check\cI_2} b^{\prime\check\r} J_{\check\r} + \cdots \\
&=& \frac12 \left(\sum_{\check \r \in \check\cI_2} b^{\prime\check\r} k_{0\hat\m_0 \check\r} \right) \left( \sum_{\hat \m,\hat \n\in \hat \cI_2} b^{\prime\hat \m} (2\l \k_{\hat \m} + b^{\prime\hat \n} \k_{\hat \m \hat \n})\right) +\cdots \\
\eea
Note that, in deriving the above expressions, the vanishing of the triple intersections
\beq
k_{000}=k_{00\hat \m} = k_{0\hat\m\hat\n} = k_{\hat\m\hat\n\hat\r}=0\,,\quad \forall \hat\m, \hat\n, \hat\r \in \hat\cI_2 \,,
\eeq
the non-vanishing of the triple intersections, 
\beq
k_{0\hat \m \check \n} >0 \,, \quad \forall \hat \m \in\hat\cI_2, ~\forall \check \n \in \check\cI_2\,, 
\eeq
as well as the following proportionality relations 
\bea
J_{\hat \m} \cdot J_{\hat \n} &=& \k_{\hat \m \hat\n} J_0 \cdot J_{\hat \mu_0}\,, \qquad \text{with}~~{\k_{\hat \m \hat \n} >0}\,, \quad \forall \hat \m, \hat \n \in \hat \cI_2 \,,\\
J_0 \cdot J_{\hat \m} &=& \k_{\hat\m} J_0 \cdot J_{\hat \m_0} \,, \qquad \text{with}~~{\k_{\hat \m} >0}\,,\quad \forall \hat\m \in \hat\cI_2 \,,
\eea
have been used multiple times. Furthermore, in singling out the leading order contributions, we have relied on the constraints on the K\"ahler parameters~\eqref{muhat} and~\eqref{mucheck}. We also assume that the coefficients in the expansion of $\bf S$ in~\eqref{Sexp} should obey $\sum_{\check \r \in \check\cI_2} h^{\check\r} k_{0\hat\m_0 \check\r}>0$ since otherwise ${\bf S}$ cannot correspond to a gauge divisor of our interest in the limit.   

Finally, note that the rescaled elliptic index~\eqref{rescalem} is expressed as
\beq
m_{\hat \m_0} = \frac12 J_0 \cdot J_{\hat \m_0} \cdot {\bf S}  = \frac12 \sum_{\check \r \in \check\cI_2} h^{\check\r} k_{0\hat\m_0 \check\r} \,.
\eeq
Altogether, therefore, we conclude that for $\l \to \infty$: 
\beq
\cV'_{\bf S} \cV'_{C_{\hat \m_0}} =  \frac12 \left(\sum_{\check \r \in \check\cI_2} h^{\check\r} k_{0\hat\m_0 \check\r} \right)\left( \sum_{\hat \m,\hat \n\in \hat \cI_2} b^{\prime\hat \m} (2\l \k_{\hat \m} + b^{\prime\hat \n} \k_{\hat \m \hat \n})\right) \left(\sum_{\check\sigma\in\check\cI_2} b^{\prime\check\sigma} k_{0\hat\mu_0\check\sigma}  \right)
= 2m_{\hat \m_0} \cV'_{B_3} \,. 
\eeq

\section{The linear multiplets \texorpdfstring{$L^\alpha$}{Lalpha} in the effective action} \label{App_Lalpha}

In this appendix, we review the calculation of the real scalar field $L^\alpha$ in the linear multiplets in the M-theory and F-theory effective action. 
We  start in the three-dimensional $N=2$ theory obtained by reducing M-theory on a Calabi-Yau fourfold. The K\"ahler potential and the K\"ahler coordinates in this theory have been calculated in \cite{Weissenbacher:2019mef}, building on and extending earlier works \cite{Grimm:2013gma,Grimm:2013bha,Grimm:2014xva} (see also \cite{GarciaEtxebarria:2012zm}). They are given by 
\begin{align}
    K=-3\log\left(\mathcal{V}_{Y_4}^{(0)}+\alpha^2 \left((\kappa_1+\kappa_2) \mathcal{Z}_i v^i + \kappa_2 \mathcal{T}_i v^i\right)\right)
\end{align}
and the K\"ahler coordinates by
\begin{align}   \label{TiM}
    \text{Re}\, T_i = \frac{K_i}{3!} + \alpha^2\left((\kappa_3+\kappa_5)\frac{K_i \mathcal{Z}_kv^k}{3! \mathcal{V}_{Y_4}^{(0)}} + \kappa_5 \frac{K_i \mathcal{T}_kv^k}{3! \mathcal{V}_{Y_4}^{(0)}}+\kappa_4\mathcal{Z}_i\log \mathcal{V}_{Y_4}^{(0)}+\kappa_6 \mathcal{T}_i+\kappa_7 \mathcal{Z}_i\right)\,,
\end{align}
where 
\begin{align}
    \alpha^2=\frac{1}{3^2 2^{13}} \,,
\end{align}
and the $\mathcal{Z}_i$ and $\mathcal{T}_i$ are the M-theory precursors of the F-theory quantities introduced in section~\ref{sec_corr}.\footnote{For the definition of $\cT_i$ we pulled out the $\mathcal{Z}_i$ contribution that arises in the definition of the divisor integrals in \cite{Weissenbacher:2019mef}.} The dual coordinates to $\text{Re}\,T_i$ can be determined by looking at 
\begin{align}
    L^i=-\frac{\partial K}{\partial v^j}\frac{\partial v^j}{\partial \text{Re}\,T_i}\,.
\end{align}
To calculate this one has to take into account that the $\mathcal{T}_i$ now also depend on the $v^i$. We thus get 
\begin{align}  
\nonumber    L^i=\frac{v^i}{\mathcal{V}_{Y_4}^{(0)}}+\frac{\alpha^2}{\mathcal{V}_{Y_4}^{(0)}}&\left[6\kappa_2 v^k (\del_m \mathcal{T}_k) \mathcal{K}^{mi} -2\kappa_6v^l\partial_l \mathcal{T}_m \mathcal{K}^{mi} - 8 \kappa_5 v^k\partial_m\mathcal{T}_k\mathcal{K}^{mi}+\left(6\kappa_2-8\kappa_5\right)\mathcal{K}^{ij} \mathcal{T}_j \right.\\ \label{Li}
    +&\left(6(\kappa_1+\kappa_2)-8(\kappa_3+\kappa_5)\right)\mathcal{K}^{ij} \mathcal{Z}_j-\frac{v^i}{3\mathcal{V}_{Y_4}^{(0)}}\left(3(\kappa_1+\kappa_2)-(\kappa_3+\kappa_5)+\kappa_4\right)v^k\mathcal{Z}_k\\
\nonumber    - &\left.\frac{v^i}{3\mathcal{V}_{Y_4}^{(0)}}\left(3\kappa_2-\kappa_5\right)v^k\mathcal{T}_k\right]\,. 
\end{align}
By comparison with the one-modulus case studied in \cite{Grimm:2017pid}, \cite{Weissenbacher:2019mef} fixed 
\begin{align}
    \kappa_1+\kappa_2=512\,,\qquad \kappa_4=-1152\,,\qquad 4\kappa_3+4\kappa_5+\kappa_6=0\,. 
\end{align}
If we demand that for $\mathcal{T}_i\rightarrow 0$ we reproduce the results in \cite{Grimm:2013gma}, this fixes in addition
\begin{align}
    -\kappa_7=\kappa_3+\kappa_5=768\,. 
\end{align}
Let us now come to the F-theory uplift: The uplift of the K\"ahler potential and of $\Re\, T_\alpha$ is readily obtained as
\begin{align} \label{KFapp}
K^F=-2\log\left(\mathcal V_{B_3}^{(0)} +\alpha^2\left(\tilde \kappa_1+\tilde \kappa_2) \mathcal{Z}_\alpha v^\alpha +\tilde \kappa_2 \mathcal{T}_\alpha v^\alpha \right)\right)\,,
\end{align}
where $\tilde \kappa_i=\frac{3}{2}\kappa_i$ and 
\begin{align}\label{ReTapp-App}
    \text{Re}\, T_\alpha= \frac{K_\alpha}{2} + \alpha^2\left((\kappa_3+\kappa_5)\frac{K_\alpha\mathcal{Z}_\beta v^\beta}{2 \mathcal{V}_{B_3}^{(0)}} + \kappa_5 \frac{K_\alpha \mathcal{T}_\beta v^\beta}{2 \mathcal{V}_{B_3}^{(0)}}+  \tilde{\mathcal{Z}}_\alpha \log \mathcal{V}_{B_3}^{(0)}+\kappa_6 \mathcal{T}_\alpha+\kappa_7\mathcal{Z}_\alpha \right)\,.
\end{align}
Here
\bea
\tilde{\mathcal{Z}}_\alpha  = \kappa'_{4,\alpha}\mathcal{Z}_\alpha    \qquad (\text{no sum over} \, \alpha)
\eea
and the parameters
 $\kappa'_{4,\alpha}$ are still to be determined. Note that
  as explained in \cite{Grimm:2017pid,Weissenbacher:2019mef}, they can differ from their M-theory counterpart $\kappa_4$  because the F-theory uplift might receive 1-loop quantum corrections.
Differently from \cite{Weissenbacher:2019mef}, however, we allow for the possibility that the renormalised parameters are non-universal among the divisor classes, i.e. upon uplifting the chiral coordinates to F-theory quantities $ \text{Re}\, T_\alpha$ we replace
\bea
\kappa_4    \stackrel{\rm 1-loop}\longrightarrow   \kappa'_{4,\alpha}   \,.
\eea
All quantities above are understood to be related to the base $B_3$ which we indicate by the Greek letters. 

There are now two ways to obtain the F-theory expression for the $L^\alpha$: The first method directly uplifts the expression in \eqref{Li} by using \cite{Grimm:2013gma}
\begin{align}
 \mathcal{K}^{ij}\rightarrow \frac{1}{2}\left(\mathcal{K}^{\alpha \beta}-\frac{1}{6}\frac{v^\alpha v^\beta}{\mathcal{V}_{B_3}^{(0)}}\right)\,,
\end{align}
and replacing $\kappa_{1/2}$ by $\tilde\kappa_{1/2}$. Again we allow for a non-universal renormalisation of $\kappa_4$, but this time as 
\bea
\kappa_4    \longrightarrow \hat\kappa_{4,\beta}  \,,
 \eea
 where the parameters $\hat\kappa_{4,\beta}$ appearing in the uplift of the $L^i$ to $L^\beta$ may in principle differ from the parameters $\kappa'_{4,\alpha}$ in (\ref{ReTapp-App}).
 This gives 
\begin{align}  \label{eq:LalphafromM}
     L^\alpha=\frac{v^\alpha}{ \mathcal{V}_{B_3}^{(0)}} + &\frac{\alpha^2}{2 \mathcal{V}_{B_3}^{(0)}} \left(\left(6\tilde \kappa_2 v^\gamma\partial_\beta \mathcal{T}_\gamma -2\kappa_6 v^\gamma \partial_\gamma \mathcal{T}_\beta-8\kappa_5 v^\gamma\partial_\beta \mathcal{T}_\gamma
    \right)\left(\mathcal{K}^{\beta \alpha}-\frac{1}{6}\frac{v^\beta v^\alpha}{\mathcal{V}_{B_3}^{(0)}}\right)\right.\\\nonumber 
    &+ \left(6\tilde \kappa_2-8\kappa_5\right) \left(\mathcal{K}^{ \alpha\beta}-\frac{1}{6}\frac{v^\alpha v^\beta}{\mathcal{V}_{B_3}^{(0)}}\right) \mathcal{T}_\beta+\left(6(\tilde \kappa_1+\tilde \kappa_2)-8(\kappa_3+\kappa_5)\right) \left(\mathcal{K}^{ \alpha\beta}-\frac{1}{6}\frac{v^\alpha v^\beta}{\mathcal{V}_{B_3}^{(0)}}\right) \mathcal{Z}_\beta\\
     &\left. -\frac{2v^\alpha }{3\mathcal{V}_{B_3}^{(0)}}\left(3(\tilde \kappa_1+\tilde \kappa_2)-(\kappa_3+\kappa_5)\right)\mathcal{Z}- \frac{2v^\alpha }{3\mathcal{V}_{B_3}^{(0)}} \widehat{\mathcal{Z}}_\beta v^\beta-\frac{2v^\alpha }{3\mathcal{V}_{B_3}^{(0)}}\left(3\tilde \kappa_2-\kappa_5\right)\mathcal{T}\right)\,,\nonumber
\end{align}
where
\bea
\widehat{\mathcal{Z}}_\beta =   {\mathcal{Z}}_\beta   \hat\kappa_{4,\beta}   \qquad (\text{no sum over} \, \beta)   \,.
\eea

The second strategy is to compute the $L^\alpha$ directly from (\ref{KFapp}) and  (\ref{ReTapp-App}), yielding
\begin{align}
\label{eq:LalphaFTheory}
    \nonumber L^\alpha=\frac{v^\alpha}{ \mathcal{V}_{B_3}^{(0)}} + &\frac{\alpha^2}{ \mathcal{V}_{B_3}^{(0)}} \left(\left(2\tilde \kappa_2 v^\gamma\partial_\beta \mathcal{T}_\gamma-\kappa_6 v^\gamma \partial_\gamma \mathcal{T}_\beta-3\kappa_5 v^\gamma\partial_\beta \mathcal{T}_\gamma
    \right)\mathcal{K}^{\beta \alpha}\right.\\
    &+ \left(2\tilde \kappa_2-3\kappa_5\right)\mathcal{K}^{\beta \alpha} \mathcal{T}_\beta+\left(2(\tilde \kappa_1+\tilde \kappa_2)-3(\kappa_3+\kappa_5)\right) \mathcal{K}^{\beta \alpha}\mathcal{Z}_\beta\\
     &\left. -\frac{v^\alpha }{\mathcal{V}_{B_3}^{(0)}}\left((\tilde \kappa_1+\tilde\kappa_2)-\frac{\kappa_3+\kappa_5}{2}\right)\mathcal{Z} -\frac{1}{2} \frac{v^\alpha }{\mathcal{V}_{B_3}^{(0)}} \tilde{\mathcal{Z}_\beta} v^\beta-\frac{v^\alpha }{\mathcal{V}_{B_3}^{(0)}}\left(\tilde \kappa_2-\frac{\kappa_5}{2}\right)\mathcal{T}\right)\,. \nonumber 
\end{align}
Here the potential quantum corrections to the uplift are encapsulated in the same parameters $\kappa'_{4,\alpha}$ entering the definition of $\tilde{\mathcal{Z}_\beta}$ already in (\ref{ReTapp-App}).

Consistency of the F-theory uplift requires that both approaches agree. This allows us to fix the remaining open parameters as follows: First matching the $\mathcal{K}^{\alpha \beta}\mathcal{T}_\beta$ implies 
\begin{align} \label{upliftconstraint}
\tilde \kappa_2 = \kappa_5 \qquad \Rightarrow \qquad \tilde \kappa_1 = \kappa_3\,. 
\end{align}
Similarly, the terms involving $\widehat{\mathcal{Z}}_\beta v^\beta$ in (\ref{eq:LalphafromM}) and $\tilde{\mathcal{Z}_\beta} v^\beta$ in  (\ref{eq:LalphaFTheory}) must also match. This is to be read as a consistency condition on the 
quantum corrections to the uplift. 
Under these two conditions (\ref{eq:LalphafromM}) and (\ref{eq:LalphaFTheory}) agree.

Let us stress that the conclusion (\ref{upliftconstraint}) is of course only valid  under the assumption that the uplift of the  terms in the $L^i$ other than those related to $\kappa_4$ does not obtain quantum corrections.
In any event, we do not impose (\ref{upliftconstraint}) in the main text but continue to compute with the most general parametrisation, using the form (\ref{eq:LalphaFTheory}).
As we discuss in  Section \ref{sec:Interpretationalpha}, the WGC relation (\ref{WGCrelgeomQC}) holds regardless of whether or not (\ref{upliftconstraint}) is imposed to leading order in $\lambda$ - provided the quantum corrected parameters $\kappa'_{4,\alpha}$ satisfy (\ref{kappa4rel}).
It is nonetheless interesting to state that if in addition
(\ref{upliftconstraint}) holds, the relation  (\ref{WGCrelgeomQC})  becomes particularly simple: In this case the leading order corrections to $\cV_{C_0}$ vanish
and the leading order corrections to $\cV_{\bf S}$ and $\cV_{B_3}$ on both sides of the equation (\ref{WGCrelgeomQC}) are identical.

\section{Computation of \texorpdfstring{$\mathcal{T}_0$}{T0}  and \texorpdfstring{$\mathcal{Z}_0$}{Z0} in classes of geometries}  \label{app_T0Z0}

In this appendix we exemplify the computation of the invariants $\mathcal{T}_0$  and $\mathcal{Z}_0$ for rationally and elliptically fibered base spaces.
These invariants govern the perturbative $(\alpha')^2$ corrections as discussed in Section \ref{sec_alphapcorr}.

\subsection{Rational fibration}

We begin with the case where the infinite distance limit leads to the shrinking of the rational fiber of the F-theory base $B_3$.
Of particular importance is the combination ${\cal W}_0$ in (\ref{W0def}). As discussed in Section \ref{sec_alphapcorr}, if ${\cal W}_0 < 0$ 
we reach a strong coupling singularity at finite distance in the parameter regime where $x <0$.

First we will exemplify situations in which the combination ${\cal W}_0$ in (\ref{W0def}) is indeed negative. 
Establishing such examples requires control of both $\mathcal{T}_0$  and $\mathcal{Z}_0$. 

The correction $\mathcal{T}_0$ to the K\"ahler coordinates depends on the unknown coefficients $\alpha_2$ and $\kappa_5$ in \eqref{Talpha} and \eqref{ReTmain}, respectively, which makes it difficult to quantify.
Despite this problem, let us evaluate $\mathcal{T}_0$ for the simple class of F-theory geometries 
given by a smooth Weierstrass model over $B_3$. In this case the gauge group is trivial.
On the dual heterotic side, this corresponds  to a gauge bundle that breaks the perturbative gauge group completely. We can now use \eqref{Talpha} to calculate $\mathcal{T}_0$. We have to differentiate between $J$-class A and B limits due to the different intersection properties of $J_0$ in this case. For $J$-class A, we obtain 
\begin{align}
   \nonumber \text{$J$-class A:}\qquad \mathcal{T}_0&=-18(1+\alpha_2) \frac{1}{\text{Re}\, T^\text{cl}_0}\left(\int_{B_3} J_0 \wedge J_0\wedge  J\right)\left(\int_{B_3} J \wedge J_0 \wedge c_1(B_3)\right) \\
    &\sim -18(1+\alpha_2) \frac{1}{\sum_{\mu\in \mathcal{I}_1} k_{00\mu} v^0 v^\mu} \left(\sum_{\mu \in \mathcal{I}_1} k_{00\mu} v^\mu\right)\left(2 v^0\right) = -36(1+\alpha_2)\,,
\end{align}
where we used $\bar K\cdot J_0\cdot J_0=2$. For $J$-class A limits, we thus arrive at a model-independent and non-vanishing contribution to $\Re\,T_\alpha$ whose precise pre-factor depends on the value for $\alpha_2$ and $\kappa_5$. For $J$-class B limits we also find a model independent result, but this time the leading contribution in $\lambda$ is vanishing, \footnote{Note that in the case that there exists a K\"ahler cone divisor  $J_\alpha$ with $\alpha\in \hat{\mathcal{I}}_2$  and $J_\alpha \cdot J_\alpha \ne 0$ and we choose to scale $v^\alpha\sim \lambda$ the $J$-class B limit effectively turns into a $J$-class A limit upon replacing $\alpha \leftrightarrow 0$.}
\be
 \text{$J$-class B:}\qquad \mathcal{T}_0=0\,,
\ee 
since the first integral in the expression for $\mathcal{T}_0$ vanishes due to $J_0\cdot J_0=0$. Thus, for smooth Weierstrass models, the contribution from $\mathcal{T}_0$ does not scale with $\lambda$ and is therefore of the same order as any contribution coming from the constant topological term $\mathcal{Z}_0$.

Next, we will show that 
 in the case of a smooth Weierstrass model with projective, smooth, and almost Fano base $B_3$ we find $\mathcal{Z}_0<0$.

To see this, note first that for smooth Weierstrass models the topological term $\mathcal{Z}_0$ reduces to 
\be\label{chismoothweierst}
\mathcal{Z}_0= -60 \int_{B_3} c_1(B_3)^2 \wedge J_0\,. 
\ee
As reviewed in the main text in Section \ref{subsec_classical_het-F_duality}, among  F-theory models with a rationally fibered base $p: B_3\rightarrow B_2$ there exist the class of models that are dual to a perturbative heterotic string theory. In particular if the heterotic dual is perturbative there cannot be any exceptional divisors due to blow-ups of curve in $B_2$ that would give rise to E-strings. 
In this case $c_1(B_3)$ is given by 
\begin{align}\label{eq:c1pertmodels}
    c_1(B_3)=p^*c_1(B_2) + 2S_+ - t\,. 
\end{align}
Here $t=c_1({\cal L})$ characterises the twist of the fibration and $S_+$ is the exceptional section at infinity of the fibration $p$. Using $S_+(S_+ -t )=0$ we find 
\begin{align}
      \mathcal{Z}_0=-240\int_{B_2} c_1(B_2) \wedge p_* J_0\,. 
\end{align}
The integral in the expression above is strictly positive for $B_2$ either a del Pezzo surface or a (blow-up in r points) of a Hirzebruch surface. Thus for these models, which have a perturbative heterotic dual and no non-abelian or abelian gauge factors, $\mathcal{Z}_0$ is strictly negative. We can generalise this slightly by considering smooth Weierstrass models over bases $B_3$ that are projective, smooth, and almost Fano.  This in particular includes cases where some curves along the bases $B_2$ of the geometries considered before are blown up to exceptional divisors of $B_3$. However, performing too many independent blow-ups will destroy the almost Fano condition. Under these assumptions $-K_{B_3}$ is known to be nef and big and through \eqref{chismoothweierst} we are interested in the quantity
\be 
J_0 \cdot (-K_{B_3})^2= (-K_{B_3}|_{J_0})^2 \,.
\ee
Now $-K_{B_3}|_{J_0}$ is still nef as this property is preserved under restriction and since $J_0$ is a K\"ahler cone divisor and therefore ample the restriction of $-K_{B_3}$ to $J_0$ is also big. These two properties of $-K_{B_3}|_{J_0}$ imply that $(-K_{B_3}|_{J_0})^2$ is strictly positive. Thus in the case of a smooth Weierstrass model with projective, smooth, and almost Fano base $B_3$ we also find $\mathcal{Z}_0<0$.

Using \eqref{k3plusk5} we note that the correction to $\Re\,T_\alpha$ due to $\mathcal{Z}_0$ is strictly negative in these models but unlike the contribution due to $\mathcal{T}_0$, it is model-dependent. 
However in the $J$-class A limit, a strictly negative $\mathcal{Z}_0$ for smooth Weierstrass models over almost Fano bases does not yet imply a strictly negative $\cW_0$ due to the non-zero contribution coming from the $\mathcal{T}_0$ term in \eqref{W0def}. In case $\kappa_5(1+\alpha_2)<0$, $\mathcal{T}_0$ contributes with a positive pre-factor to $\cW_0$ which could therefore yield to $\cW_0\ge 0$.

Note that since $\mathcal{T}_0$ is model-independent and just depends on whether we take a $J$-class A or B limit, the case where $\cW_0=0$ is very special as it can only be reached in specific models with certain values of $\mathcal{Z}_0$. We thus expect that in generic $N=1$ models we have $\cW_0\ne 0$.\footnote{A special case is e.g. a smooth Weierstrass model over a base $B_3$ which is rationally fibered over the Enriques surface $K3/\mathbb{Z}_2$ which has $\mathcal{Z}_0=0$. In case a $J$-class B limit for these models exists, the obstruction to leading order vanish since $\cW_0=0$. However, these models are closely related to models with higher supersymmetry which might be responsible for the vanishing of the leading order correction and just yield subleading corrections. These subleading corrections arise since $K3/\mathbb{Z}_2$ has additional curves $J_\alpha$, $\alpha\ne 0$, with $J_\alpha^2\ne0$ that uplift to divisors of $B_3$. Since for these divisors we have $\mathcal{T}_\alpha \ne 0$ the limits of vanishing $\Re\,T_\alpha$ is obstructed which then also obstructs the original J-class B limit on $K3/\mathbb{Z}_2$.} It would be further interesting to see whether $\cW_0>0$ is actually possible or whether we always run into a strong coupling singularity as it is the case for $J$-class B limits for smooth Weierstrass models over almost Fano bases.   

As a first step into that direction we can go beyond smooth Weierstrass models over almost Fano bases  $B_3$, for which we had shown that  $\mathcal{Z}_0<0$. First of all, we can perform a large number of blow-ups of curves in the class of bases $B_2$ considered around \eqref{eq:c1pertmodels}. In this case the exceptional blow-up divisors $E_a$ will contribute to $c_1(B_3)$ as 
\begin{align}
    c_1(B_3)=p^*c_1(B_2) + 2 S_+ - t +\sum_a c_a E_a \,,
\end{align}
where $c_a$ are constants. This leads to an additional term for $\cW_0$: 
\be 
\mathcal{Z}_0=-240\int_{B_2} c_1(B_2) \wedge p_* J_0 - 60 \sum_{ab}c_a c_b \int_{B_3} E_a \wedge E_b \wedge J_0\,. 
\ee 
If the blow-ups are all independent $E_a\cdot E_b=\delta_{ab}$, the second term in the expression for $\cW_0$ above gives a strictly positive contribution such that $\cW_0<0$ cannot be ensured any more if the number of independent blow-ups is large enough. 

Furthermore note that in the presence of a non-abelian gauge symmetry ${\cal Z}_0$ will not be given by \eqref{chismoothweierst} any more. If we assume that the singularity corresponding to a single gauge group factor realised on a divisor ${\cal S}$ is crepantly resolvable, we have \cite{Esole:2018tuz}
\be
\mathcal{Z}_0= \int_{B_3} \left(-60 c_1(B_3)^2 + a c_1(B_3) {\cal S} - b {\cal S}^2\right)J_0\,, \qquad a,b > 0\,. 
\ee
This reflects the observation of \cite{Grimm:2013bha} that $\mathcal{Z}$ is closely related to the gauge theory in F-theory. Starting from an almost Fano base $B_3$ and a suitably chosen gauge group it may in principle be possible to reach $\mathcal{Z}_0>0$, but we have not verified whether this actually occurs in explicit examples.

\subsection{Genus-one fibration}
Let us now turn to the case of genus-one fibrations that, as discussed above, yield a tensionless fundamental Type IIB string at infinite distance. In this case, we can again look at the $\alpha'$-corrections related to $\mathcal{Z}$ and $\cT$. Assuming a smooth Weierstrass model, we use \eqref{chismoothweierst} to evaluate $\mathcal{Z}_0$. However, now that we have a $T^2$-fibration $p: B_3 \rightarrow B_2$, $c_1(B_3)$ is given by  
\be
c_1(B_3)= p^*c_1(B_2)-\frac{1}{12} p^* \Delta(B_2)\,.
\ee 
Since $J_0$ is also a pull-back divisor from $j_0\in H^2(B_2)$, i.e. $J_0=p^*(j_0)$, we find
\begin{align}
    \mathcal{Z}_0 = \int_{B_3} c_1(B_3)^2 \wedge J_0=0 \,,
\end{align}
for any $T^2$-fibered base $B_3$. As for the $\cT$ contribution we note that since $\bar K\cdot C_0=0$ for elliptic fibrations we find to leading order in $\lambda$
\begin{align}
    \mathcal{T}_0=0 \,,
\end{align}
unlike in the case of rational fibrations for both $J$-class A and $J$-class B limits.

\section{Quantum volume in Type IIA on \texorpdfstring{$Y_4$}{Y4}}   \label{app_QV}

In this appendix we consider Type IIA string theory compactified on an elliptically fibered fourfold
$Y_4$ and
investigate quantum effects in its $\sigma$-model K\"ahler moduli space.
Our goal is to determine if there exists a non-perturbative kinematical obstruction
to taking infinite distance limits in which the rational fiber $C_0$ of the base $B_3$ shrinks.

To illustrate the general idea we consider a simple example of $J$-type A as in \eqref{combinedJclassA} with $\mathcal{I}_1=\{1\}$ and $\mathcal{I}_3=\{2\}$ such that the K\"ahler form is given by
\begin{align}
    J= t^0 J_0 + t^1 J_1 + t^2J_2\,. 
\end{align}
Such a system is realised for instance if $B_3$ is a rational fibration over a Hirzebruch surface. In this case $t^1$ gives the $\sigma$-model volume of the rational fiber of $B_3$, $t^0$ the volume of the fiber of the Hirzebruch surface and $t^2$ the volume of its base. The question is now whether the point of vanishing $\sigma$-model volume $t^1$ is part of the moduli space. For the moment let us forget about the $\mathcal{I}_3$ modulus $t^2$ and focus on the two-modulus system associated to $t^0$ and $t^1$. We follow the example presented in \cite{Mayr:1996sh} where the Picard-Fuchs equations for $Y_4$, when reduced to the moduli  $t^0$ and $t^1$, yield the differential equations describing the Hirzebruch surface $\mathbb{F}_2$. For this system the discriminant locus is given by
\begin{align}\label{DeltaF2}
    \Delta_{\mathbb{F}_2}= (1-4z_0)^2\times \left((1-4z_1)^2-64z_1^2z_0\right)=0\,,
\end{align}
where, in this particular example, the $z_{0,1}$ are related to $t^{0,1}$ in the large volume limit via 
\begin{align}
    t^{0,1}=\frac{1}{2\pi i} \log\left(\frac{1-2z_{0,1}-\sqrt{1-4z_{0,1}}}{2z_{0,1}}\right)\,, \qquad \text{for} \qquad z_{1,0}=0\,. 
\end{align}
We thus see that e.g. at $z_0=0$ corresponding to $t^0=\infty$, the $\sigma$-model volume of the rational fiber of $B_3$ vanishes if $z_1=1/4$. And indeed the point $(z_0,z_1)=(0,1/4)$ lies on the divisor $\Delta_{\mathbb{F}_2}=0$. However, once we fix $0<z_0\ll 1/4$, $z_1=1/4$ is not on the discriminant divisor anymore as the singularity splits for finite $z_0$. In this example the split of the singularity induces a quantum volume for the rational fiber of $B_3$ due to its coupling to the curve in the base $B_2$. On the other hand, the point $(z_0,z_1)=(1/4, z_1)\in \Delta_{\mathbb{F}_2}$ for any $z_1\ge 0$ such that the curve in $B_2$ does not receive a minimal non-zero quantum volume due to the $\mathbb{P}^1$ fiber of $B_3$. 

The important question is now whether this effect is big enough to obstruct any limit in which the rational fiber shrinks asymptotically to zero size. From \eqref{DeltaF2} the effect of switching on $z_0>0$ is to shift the singularity as
\begin{align}
    z_1^\text{sing.}=\frac{1}{4} \pm \mathcal{O}(z_0)\,.
\end{align}
Thus, for $z_0\ll 1$ the minimal corrected $\sigma$-model volume $t^1$ scales as 
\begin{align}
    t^1_\text{min.}(z_1^\text{sing.})\sim - \log\left(1-\mathcal{O}(z_0)\right) \sim \mathcal{O}(z_0) \sim e^{-2\pi t^0}\,.
\end{align}
From here we learn that the minimal quantum volume of the rational fiber is exponentially suppressed in the volume of the base curve $t^0$. 

If we now impose the scaling \eqref{combinedJclassA} we see that the corrected scaling due to the quantum volume is 
\begin{align}
    t^1\sim \frac{\mu}{\lambda^2} + e^{-2 \pi \mu \lambda} \,.
\end{align}
Accordingly, the effect of the quantum volume is negligible in this limit.

In fact the exponential suppression of the quantum volume as compared to the classical scaling holds in general. Of course also the $\cI_3$ modulus $t^2$, which so far we have only treated as a spectator, might be non-trivially coupled to $t^1$ and introduce an additional contribution to the quantum volume. However, this effect will be at least suppressed as 
\begin{align}
    \delta t^1 \sim e^{-2\pi \lambda^{1/2+x}} \rightarrow 0 \,,\qquad \text{for}\qquad x>-1/2 \,. 
\end{align}

Similar results as for the rationally fibered case also hold if $B_3$ is genus-one fibered. In summary, non-perturbative corrections as captured by fourfold mirror symmetry cannot further obstruct limits with vanishing fiber limits.

\section{\texorpdfstring{$\alpha'$}{alpha'}-corrections to the volume of \texorpdfstring{$C_0$}{C0}}
\label{App_volC0}
Here we prove the relations
\begin{equation}
\label{eq:KijC0relation_app}
  \begin{aligned}
    k_{00\mu}K^{\mu\alpha}&=\frac12\frac{\mathcal{V}^{(0)}_{C_0}v^\alpha}{\mathcal{V}_{B_3}^{(0)}} + \mathcal{O}\left(\frac{1}{\mu \lambda^4}\right)
    \qquad\qquad&J\textup{-class A}\\
    k_{0\hat{\mu}_0\check{\nu}}K^{\check{\nu}\alpha}&=\frac12\frac{\mathcal{V}^{(0)}_{C_{\hat{\mu}_0}}v^\alpha}{\mathcal{V}^{(0)}_{B_3}} + \mathcal{O}\left(\frac{1}{\mu \lambda^4}\right) \qquad\qquad&J\textup{-class B}
  \end{aligned}
\end{equation}
 in the limits~\eqref{combinedJclassA} and~\eqref{combinedJclassB}. The identities presented above are needed in order to evaluate the $\alpha'$-corrections to the volume of the distinguished curve $C_0$. In conjunction with~\eqref{eq:LalphaFTheory} they imply that the volume of this curve is only corrected by a multiplicative factor as in~\eqref{eq:C0limit}.

\subsection{\texorpdfstring{$J$}{J}-class A}
\label{App_volC0:classA}
Consider a  limit of $J$-class A,
\be \label{combinedJclassA_app}   
J = \mu \Big(\lambda J_0+\sum_{\mu\in\mathcal{I}_1}\frac{a^\mu}{\lambda^2}J_\mu+\sum_{r\in\mathcal{I}_3} b^{\prime r} J_r\Big)   \,, 
\ee
with intersection numbers as in (\ref{JclassA-2}).
We are interested in the volume of the curve $C_0=J_0\cdot J_0$, for which we find
\begin{equation}
\mathcal{V}^{(0)}_{C_0}=\frac{\mu}{\lambda^2} k_{00\mu}a^\mu\equiv \frac{\mu}{\lambda^2}k^Ta\, ,   \quad \quad k_{00\mu} = J_0 \cdot J_0\cdot J_\mu \,.
\end{equation}
We evaluate the righthand side of~\eqref{eq:KijC0relation_app} as follows
\bea
\label{eq:correction_factor_firstterm}
\renewcommand{\arraystretch}{1.2}
  \frac12\frac{\mathcal{V}_{C_0}^{(0)}v^\alpha}{\mathcal{V}_{B_3}^{(0)}}
&  =&\frac12\frac{\frac{\mu}{\lambda^2}k^Ta}{\tfrac12 \mu^3\, k^Ta\left(1+2\tfrac{n_rb^{\prime r}}{\lambda}+\tfrac{n_{rs}b^{\prime r}b^{\prime s}}{\lambda^2}\right)}
  \begin{pmatrix}\mu\lambda\\
  \mu b^{\prime r}\\
  \frac{\mu}{\lambda^2}a^\mu\end{pmatrix} +\mathcal{O}\left(\frac{1}{\lambda^4}\right)  \\
 & =&\frac{1}{1+2\tfrac{n^Tb'}{\lambda}+\tfrac{b^{\prime T}Nb'}{\lambda^2}}
  \frac{1}{\mu\lambda}
  \begin{pmatrix}1\\
  \tfrac{b^{\prime r}}{\lambda}\\0^\mu\end{pmatrix}
  +\mathcal{O}\left(\frac{1}{\lambda^4}\right)\;.
\eea
Here we have defined the vectors 
\be
n=(n_r)_{r\in\mathcal{I}_3} \,,  \qquad b'=(b^{\prime r})_{r\in\mathcal{I}_3}
\ee
 and matrix $N=(n_{rs})_{r,s\in\mathcal{I}_3}$ that encode the linear dependencies between the intersection numbers~\eqref{0r}.

The next step is to determine the asymptotic form of the kinetic matrix $K_{\alpha\beta}$ defined in (\ref{Kalphabeta}) and its inverse $K^{\alpha\beta}$, which appears on the lefthand side of~\eqref{eq:KijC0relation_app}.
It is convenient to group the index set as $(\{0\}\cup\mathcal{I}_3)\cup\mathcal{I}_1$. In the following, indices $m,n$ will range over $\{0\}\cup\mathcal{I}_3$. We find the block matrix structure
\begin{equation}
\label{eq:kinetic_matrix}
\renewcommand{\arraystretch}{1.1}
  \begin{gathered}
    X\equiv (K_{\alpha\beta})_{\alpha,\beta\in\{0\}\cup\mathcal{I}_3\cup\mathcal{I}_1}=
    \left(
    \begin{array}{c:c}
      A & B^T\\
      \hdashline
      B & D
    \end{array}
    \right)\;,\\
    A=\frac{\mu}{\lambda^2}a^Tk
    \left(
    \begin{array}{c:c}
      1 & n^T\\
      \hdashline
      n & N
    \end{array}
    \right)\;,
    \qquad\qquad
    B=\mu\lambda\; k\otimes
    \left(
    \begin{array}{c:c}
      1+\tfrac{n^Tb'}{\lambda}&
      n^T+\tfrac{b^{\prime T}N}{\lambda}
    \end{array}
    \right)\;,\\
    D=\mu\lambda \left(k_{0\mu\nu}+k_{r\mu\nu}\frac{b^{\prime r}}{\lambda}\right)_{\mu,\nu\in\,\mathcal{I}_1}\;.
  \end{gathered}
\end{equation}
We recall the definition 
\be
\tilde{b}=\begin{pmatrix}1&b'/\lambda\end{pmatrix}
\ee
and define the block components of the inverse of $X$ as
\begin{equation}
\renewcommand{\arraystretch}{1.2}
  X^{-1}=
  \left(
  \begin{array}{c:c}
  \hat{A}&\hat{B}^T\\
  \hdashline
  \hat{B}&\hat{D}
  \end{array}
  \right)\;.
\end{equation}
The lefthand side of~\eqref{eq:KijC0relation_app} can now be brought in the form
\begin{equation}
\label{eq:correction_factor_secondterm}
  \begin{aligned}
    k_{00\mu}K^{\mu m}&=\frac{1}{\mu\lambda(1+2\tfrac{n^Tb'}{\lambda}+\tfrac{b^{\prime T}Nb'}{\lambda^2})}\cdot(\tilde{b}^TB^T\hat{B})^m=\mu\lambda\frac{\mathcal{V}_{C_0}^{(0)}}{2\mathcal{V}_{B_3}^{(0)}}\cdot(\tilde{b}^TB^T\hat{B})^m\;,\\
    k_{00\mu}K^{\mu \nu}&=\frac{1}{\mu\lambda(1+2\tfrac{n^Tb'}{\lambda}+\tfrac{b^{\prime T}Nb'}{\lambda^2})}\cdot(\tilde{b}^TB^T\hat{D})^\nu=\mu\lambda\frac{\mathcal{V}_{C_0}^{(0)}}{2\mathcal{V}_{B_3}^{(0)}}\cdot(\tilde{b}^TB^T\hat{D})^\nu\;.
  \end{aligned}
\end{equation}

Provided the sub-matrices $A,D$ are invertible, the inverse kinetic matrix is determined as follows. The matrix block $\hat{A}$ is a rank one update of $A^{-1}$
\begin{equation}
  \hat{A}=A^{-1}-\frac{\tilde{b}\otimes\tilde{b}^T}{\tilde{b}^TA\tilde{b}}\;,
\end{equation}
where the inverse of $A$ is
\begin{equation}
\renewcommand{\arraystretch}{1.2}
  A^{-1}=\frac{\lambda^2}{\mu}\frac{1}{k^Ta(1-n^TN^{-1}n)}
  \left(
  \begin{array}{c:c}
    1 & -n^TN^{-1}\\
    \hdashline
    -N^{-1}n & (1-n^TN^{-1}n)N^{-1}+N^{-1}nn^TN^{-1}
  \end{array}
  \right)\;,
\end{equation}
provided $N$ is invertible.

The block $\hat{B}$ is implicitly determined in terms of the inverse of $D$ as
\begin{equation}
  \hat{B}=\frac{(D^{-1}k)\otimes\tilde{b}^T}{(k^TD^{-1}B\tilde{b})}\;.
\end{equation}
Finally, $\hat{D}$ is again a rank one update of $D^{-1}$
\begin{equation}
  \hat{D}=D^{-1}-\frac{(D^{-1}k)\otimes(k^TD^{-1})}{k^TD^{-1}k}\;.
\end{equation}

Using these formulae it is straightforward to verify that~\eqref{eq:correction_factor_secondterm} reduces to
\begin{equation}
\label{eq:correction_factor_secondterm_simplified}
  \begin{aligned}
    k_{00\mu}K^{\mu m}&=\mu\lambda\frac{\mathcal{V}^{(0)}_{C_0}}{2\mathcal{V}^{(0)}_{B_3}}\cdot\tilde{b}^m\;,\\
    k_{00\mu}K^{\mu \nu}&=0^\nu\;.
  \end{aligned}
\end{equation}
Combining~\eqref{eq:correction_factor_secondterm_simplified} with~\eqref{eq:correction_factor_firstterm} we find that to leading order in the scaling parameter $\lambda$ the relation~\eqref{eq:KijC0relation_app} holds.

\subsubsection*{Subleading Corrections}

Away from the limit~\eqref{JclassAhetrelabelded} the kinetic matrix $K_{\alpha\beta}$ receives corrections of the form
\begin{equation}
\label{eq:kinetic_matrix_corrections}
\renewcommand{\arraystretch}{1.1}
  \begin{gathered}
    \delta X=
    \left(
    \begin{array}{c:c}
      0 & \delta B^T\\
      \hdashline
      \delta B & \delta D
    \end{array}
    \right)\;,\\
    \delta B=\frac{\mu}{\lambda^2}
    \left(
      k_{m\mu\nu} a^\nu
    \right)_{\mu,m}\;,
    \qquad\qquad
    \delta D=\frac{\mu}{\lambda^2} (k_{\rho\mu\nu}a^\rho)_{\mu,\nu}\;.
  \end{gathered}
\end{equation}
The corresponding first subleading corrections to the inverse are given by
\begin{equation}
\label{eq:inverse_kinetic_matrix_corrections}
\renewcommand{\arraystretch}{1.3}
\begin{gathered}
    \delta X^{-1}=
    \left(
    \begin{array}{c:c}
      \delta \hat{A} & \delta \hat{B}^T\\
      \hdashline
      \delta \hat{B} & \delta \hat{D}
    \end{array}
    \right)\;,\\
    \begin{aligned}
    \delta \hat{A}&=-\hat{A}\delta{B}^T\hat{B}-\hat{B}^T\delta B\hat{A}-\hat{B}^T D\hat{B}+\mathcal{O}\left(1/\lambda^4\right)\;,\\
    \delta \hat{B}&=-\hat{B}\delta{B}^T\hat{B}-\hat{D}\delta B\hat{B}+\hat{B}A\hat{B}^T(\delta B\hat{A}+D\hat{B})+\hat{D}\delta B\hat{B}^T(D\hat{B}+\delta{B}\hat{A})+\mathcal{O}\left(1/\lambda^7\right)\;,\\
    \delta \hat{D}&=-\hat{B}A\hat{B}^T-\hat{D}\delta{D}\hat{D}-\hat{D}\delta B\hat{B}^T-\hat{B}\delta B^T\hat{D}+\mathcal{O}\left(1/\lambda^7\right)\;.\\
  \end{aligned}
  \end{gathered}
\end{equation}
Using the definitions~\eqref{eq:B3_classical_expansion} and the fact that $B^T\hat{D}=0$ it follows that
\begin{equation}
\label{eq:KijC0relation_subleading}
\begin{gathered}
  \begin{aligned}
    \frac{\mathcal{V}_{B_3}^{(0)}}{\mathcal{V}_{C_0}^{(0)}}k_{00\mu}K^{\mu m}&=\frac12 \left[\left(1+\frac{\mathcal{V}_{B_3}^{(0,1)}}{\mathcal{V}_{B_3}^{(0,0)}}\right)v^m-\Delta^m\right]+\mathcal{O}\left(\frac{1}{\lambda^4}\right)\;,\\
    \frac{\mathcal{V}_{B_3}^{(0)}}{\mathcal{V}_{C_0}^{(0)}}k_{00\mu}K^{\mu \nu}&=-\frac12 \Delta^\nu+\mathcal{O}\left(\frac{1}{\lambda^4}\right)\;,
  \end{aligned}\\
\Delta^m=\left(\tilde{b}^T\delta B^T \hat{B}-\tilde{b}^TA\hat{B}^T(\delta B \hat{A}+D\hat{B})\right)^m\qquad\qquad
    \Delta^\mu=\left(\tilde{b}^TA\hat{B}^T+\tilde{b}^T\delta B^T\hat{D}\right)^\mu\;.
\end{gathered}
\end{equation}

Note that the leading order term $\tfrac12 v^m$ in equation~\eqref{eq:KijC0relation_subleading} scales as $\mu\lambda$, while the other, sub-leading terms are all of order $\mu/\lambda^2$ and hence relatively suppressed by $1/\lambda^3\sim \cV _{p^*(C_\alpha),E_a}/\cV _{S_i}$.

We can now go back to the expressions for $\mathcal{V}_{C_0}$, $\mathcal{V}_{B_3}$ and $\mathcal{V}_{\bf S}$ and evaluate these to $\mathcal{O}\left(\frac{1}{(\mu\lambda)^4}\right)$.
In particular, the expression for ${\mathcal{V}}_{C_0}$, (\ref{VC0volume}), which depends on the  linear multiplets $L^\mu$ given in (\ref{CorrectedLinearMultiplets}), explicitly involves the 
subleading terms in (\ref{eq:KijC0relation_subleading}):
  \begin{multline}
    \frac{\mathcal{V}_{C_0}}{\mathcal{V}_{C_0}^{(0)}}
  =1+\frac{\alpha^2}{\mathcal{V}_{B_3}^{(0,0)}}
  \Bigg[(\tilde{\kappa}_1+\tilde{\kappa}_2-\kappa_3-\kappa_5)\mathcal{Z}_m v^m
  -(\tilde{\kappa}_1+\tilde{\kappa}_2-\tfrac32 \kappa_3-\tfrac32 \kappa_5)\Delta^\alpha\mathcal{Z}_\alpha\\
  -(\tilde{\kappa}_2-\tfrac32 \kappa_5)\Delta^\alpha \mathcal{T}_\alpha
  +\tfrac12 \Delta^\alpha D_\alpha+(\tilde{\kappa}_2-\kappa_5)\mathcal{T}_m v^m+\tfrac12 \kappa_5\left(\mathcal{T}_\mu v^\mu-\tfrac{\mathcal{V}_{B_3}^{(0,1)}}{\mathcal{V}_{B_3}^{(0,0)}}\mathcal{T}_m v^m \right)    \label{C0sublead}\\
  +\tfrac{1}{2}(\kappa_3+\kappa_5)\left(\mathcal{Z}_\mu v^\mu-\tfrac{\mathcal{V}_{B_3}^{(0,1)}}{\mathcal{V}_{B_3}^{(0,0)}}\mathcal{Z}_m v^m\right)
  -\tfrac12 \tilde{\mathcal{Z}}_\mu v^\mu
  \Bigg]+\mathcal{O}\left(\frac{1}{(\mu\lambda)^4}\right)\;,
  \end{multline}
  \begin{multline}
    \frac{\mathcal{V}_{\mathbf{S}}}{\mathcal{V}_{\mathbf{S}}^{(0)}}
    =1+\frac{\alpha^2}{\mathcal{V}_{B_3}^{(0,0)}}\left[(\kappa_3+\kappa_5)\left(\mathcal{Z}_\alpha v^\alpha-\tfrac{\mathcal{V}_{B_3}^{(0,1)}}{\mathcal{V}_{B_3}^{(0,0)}}\mathcal{Z}_m v^m\right)+\delta_2\mathcal{V}_{B_3}^{(0,0)}  \right.\\
    \left.+\kappa_5 \left(\mathcal{T}_\alpha v^\alpha-\tfrac{\mathcal{V}_{B_3}^{(0,1)}}{\mathcal{V}_{B_3}^{(0,0)}}\mathcal{T}_m v^m\right)\right]
    +\mathcal{O}\left(\frac{1}{(\mu\lambda)^4}\right)\;,  \label{Ssublead}  \\
  \end{multline}
  \begin{multline}
    \frac{\mathcal{V}_{B_3}}{\mathcal{V}_{B_3}^{(0)}}
    =1+\frac{\alpha^2}{\mathcal{V}_{B_3}^{(0,0)}}\left[(\tilde{\kappa}_1+\tilde{\kappa}_2)\left(\mathcal{Z}_\alpha v^\alpha-\tfrac{\mathcal{V}_{B_3}^{(0,1)}}{\mathcal{V}_{B_3}^{(0,0)}}\mathcal{Z}_m v^m\right)\right.\\
    +\left.\tilde{\kappa}_2\left(\mathcal{T}_\alpha v^\alpha-\tfrac{\mathcal{V}_{B_3}^{(0,1)}}{\mathcal{V}_{B_3}^{(0,0)}}\mathcal{T}_m v^m\right)\right]
  +\mathcal{O}\left(\frac{1}{(\mu\lambda)^4}\right)\;.   \label{B3sublead}
  \end{multline}
Here we have collected the derivative terms that appear in the $\alpha'$-corrections to $\mathcal{V}_{C_0}$ in the quantity $D_\alpha=2\tilde \kappa_2 v^\gamma\partial_\alpha \mathcal{T}_\gamma-\kappa_6 v^\gamma \partial_\gamma \mathcal{T}_\alpha-3\kappa_5 v^\gamma\partial_\alpha \mathcal{T}_\gamma$.

\subsection{\texorpdfstring{$J$}{J}-class B}
\label{App_volC0:classB}

The proof for the $J$-class B limits proceeds in an analogous fashion. We recall that the curve $C_0\sim D^2$ is proportional to $C_{\hat{\mu}_0}=J_0\cdot J_{\hat{\mu}_0}$ for some fixed $\hat{\mu}_0\in \hat{\mathcal{I}}_2$.
Using the results from Appendix~\ref{App_WGCclass-B}, the right hand side of~\eqref{eq:KijC0relation} evaluates to
\begin{equation}
  \frac12\frac{\textup{vol}(C_{\hat{\mu}_0})v^\alpha}{\mathcal{V}^0}\simeq\frac{1}{\mu^2\left(2\lambda\kappa_{\hat{\mu}}b^{\prime\hat{\mu}}+\kappa_{\hat{\mu}\hat{\nu}}b^{\prime\hat{\mu}}b^{\prime\hat{\nu}}\right)}
  \begin{pmatrix}
    \mu \lambda\\
    \mu b^{\prime\hat{\mu}}\\
    \mu b^{\prime\check{\mu}}
  \end{pmatrix}
  \equiv
  \frac{1}{2\frac{\kappa^T\hat{b}'}{\lambda}+\frac{\hat{b}^{\prime T} \mathbb{K} \hat{b}'}{\lambda^2}}
  \frac{1}{\mu\lambda}
  \begin{pmatrix}
    1\\
    \hat{b}'/\lambda\\
    \check{b}'/\lambda
  \end{pmatrix}\;,
\end{equation}
where we have defined $\hat{b}'=(b^{\prime\hat{\mu}})_{\hat{\mu}\in\hat{\mathcal{I}}_2}$, $\check{b}'=(b^{\prime\check{\mu}})_{\check{\mu}\in\check{\mathcal{I}}_2}$, $\kappa=(\kappa_{\hat{\mu}})_{\hat{\mu}\in\hat{\mathcal{I}}_2}$ as well as $\mathbb{K}=(\kappa_{\hat{\mu}\hat{\nu}})_{\hat{\mu},\hat{\nu}\in\hat{\mathcal{I}}_2}$.

Using the constraints on the intersection numbers from Appendix~\ref{App_WGCclass-B} we find the following kinetic matrix
\begin{equation}
\label{eq:kinetic_matrix_typeB}
\renewcommand{\arraystretch}{1.1}
  \begin{gathered}
    X\equiv (K_{\alpha\beta})_{\alpha,\beta\in\{0\}\cup\hat{\mathcal{I}}_2\cup\check{\mathcal{I}}_1}=
    \left(
    \begin{array}{c:c}
      A & B^T\\
      \hdashline
      B & D
    \end{array}
    \right)\;,\\
    A=\mu\;\check{k}^T\check{b}'
    \left(
    \begin{array}{c:c}
      0 & \kappa^T\\
      \hdashline
      \kappa & \mathbb{K}
    \end{array}
    \right)
    \sim \mu\lambda^{-\gamma}
    \qquad\qquad
    B=\mu\lambda\; \check{k}\otimes
    \left(
    \begin{array}{c:c}
      \tfrac{\kappa^T\hat{b}'}{\lambda}&
      \kappa^T+\tfrac{\hat{b}^{\prime T}\mathbb{K}}{\lambda}
    \end{array}
    \right)
    \sim\mu\lambda^{\gamma-1}\;,\\
    D=\mu\lambda\left(k_{\check{\mu}\check{\nu}}+k_{\check{\mu}\check{\nu}\hat{\rho}}\frac{b^{\prime\hat{\rho}}}{\lambda}\right)_{\check{\mu},\check{\nu}\in\check{\mathcal{I}}_2}
    \sim\mu\lambda\;.
  \end{gathered}
\end{equation}
Here we have introduced the vector $\check{k}=(k_{0\hat{\mu}_0\check{\nu}})_{\check{\nu}\in\check{\mathcal{I}}_2}$. The rest of the proof follows mutatis mutandis.

\bibliography{papers}
\bibliographystyle{JHEP}

\end{document}